\documentclass[a4paper,11pt]{article}
\pdfoutput=1 

\usepackage{jcappub} 

\usepackage[T1]{fontenc} 
\usepackage{ifthen}
\usepackage{mathtools}
\usepackage{physics}
\usepackage{cleveref}
\usepackage{verbatim}
\usepackage[dvipsnames]{xcolor}
\usepackage[export]{adjustbox}
\usepackage{arydshln}
\usepackage{subcaption}
\usepackage{multirow}
\usepackage[normalem]{ulem}
\usepackage{bm}

\usepackage[compat=1.0.0]{tikz-feynman}
\tikzfeynmanset{warn luatex=false}

\usepackage{tcolorbox}
\tcbuselibrary{minted,breakable,xparse,skins}

\definecolor{bg}{gray}{0.95}
\DeclareTCBListing{mintedbox}{O{}m!O{}}{%
  breakable=true,
  listing engine=minted,
  listing only,
  minted language=#2,
  minted style=default,
  minted options={%
    linenos,
    gobble=0,
    breaklines=true,
    breakafter=,,
    fontsize=\small,
    numbersep=8pt,
    #1},
  boxsep=0pt,
  left skip=50pt,
  right skip=50pt,
  left=25pt,
  right=0pt,
  top=3pt,
  bottom=3pt,
  arc=5pt,
  leftrule=0pt,
  rightrule=0pt,
  bottomrule=1pt,
  toprule=1pt,
  colback=bg,
  colframe=orange!70,
  enhanced,
  overlay={%
    \begin{tcbclipinterior}
    \fill[orange!20!white] (frame.south west) rectangle ([xshift=20pt]frame.north west);
    \end{tcbclipinterior}},
  #3}

\newcommand{\bth}{\bm{\theta}}
\newcommand{\bD}{\bm{D}}
\newcommand{\calP}{\mathcal{P}}
\newcommand{\calL}{\mathcal{L}}
\newcommand{\GeV}{\mathrm{GeV}}
\newcommand{\dneff}{\Delta N_\mathrm{eff}}
\newcommand{\logfa}{\log f_a}

\definecolor{python_c0}{HTML}{1F77B4}
\definecolor{python_c1}{HTML}{FF7F0E}
\definecolor{python_c2}{HTML}{2CA02C}
\definecolor{python_c3}{HTML}{D62728}

\title{Beyond thermal approximations: Precise cosmological bounds on Axion-Like Particles}

\author[*,a,b]{Nicola Barbieri,}
\author[*,c]{Luca Caloni,}
\author[b]{Martina Gerbino,}
\author[b]{Massimiliano Lattanzi,} 
\author[d,e]{and Luca Visinelli}

\affiliation[a]{Instituto de Física Corpuscular (IFIC), CSIC‐Universitat de València, Parc Científic UV, c/ Catedrático José Beltrán, 2, E-46980 Paterna (València), Spain}
\affiliation[b]{Istituto Nazionale di Fisica Nucleare (INFN), Sezione di Ferrara, Via G. Saragat 1, I-44122 Ferrara, Italy}
\affiliation[c]{Faculdade de Ci\^{e}ncias e Tecnologia and CFisUC, Departamento de F\'isica, Universidade de Coimbra, Rua Larga P-3004-516 Coimbra, Portugal}
\affiliation[d]{Dipartimento di Fisica ``E.R.\ Caianiello'', Universit\`a degli Studi di Salerno,\\ Via Giovanni Paolo II, 132 - 84084 Fisciano (SA), Italy \looseness=-1}
\affiliation[e]{Istituto Nazionale di Fisica Nucleare - Gruppo Collegato di Salerno - Sezione di Napoli,\\ Via Giovanni Paolo II, 132 - 84084 Fisciano (SA), Italy \looseness=-1}
\affiliation[*]{{\bf These authors contributed equally to this work}}

\emailAdd{barbieri@ific.uv.es}
\emailAdd{luca.caloni@uc.pt}
\emailAdd{gerbinom@fe.infn.it}
\emailAdd{lattanzi@fe.infn.it}
\emailAdd{lvisinelli@unisa.it}

\newcommand{\round}[2][]{%
  \mathopen{}\left( #2 \right)%
  \ifthenelse{\equal{#1}{}}{}{^{#1}}%
  \mathclose{}%
}

\newcommand{\sround}[2][]{%
  \mathopen{}\left[ #2 \right]%
  \ifthenelse{\equal{#1}{}}{}{^{#1}}%
  \mathclose{}%
}

\abstract{
We derive updated cosmological bounds on light axion-like particles (ALPs) coupled to leptons or photons, using a full phase-space treatment of their production from the primordial thermal plasma.
The ALP phase-space distribution, obtained by solving the momentum-dependent Boltzmann equation for the relevant production processes, is consistently propagated into the computation of cosmological observables, allowing us to assess the impact of non-thermal spectral distortions on the effective number of relativistic species, $\Delta N_{\rm eff}$.
Using state-of-the-art measurements of the cosmic microwave background from \textit{Planck}, the Atacama Cosmology Telescope, and the South Pole Telescope, complemented with Big Bang Nucleosynthesis determinations of primordial deuterium and helium abundances, we obtain the following 95\% credible limits on the ALP decay constant:  $f_a > 1.63 \times 10^6 \, {\rm GeV}$, $9.41 \times 10^6 \, {\rm GeV}$ and $8.06 \times 10^4 \, {\rm GeV}$ for ALPs coupled to electrons, muons and taus, respectively. For the ALP-photon coupling we find $g_{a\gamma} < 1.98 \times 10^{-8} \, {\rm GeV}^{-1}$.
Including baryon acoustic oscillation data from the Dark Energy Spectroscopic Instrument mildly relaxes the constraints, in line with previous analyses of extra relativistic degrees of freedom. Finally, we present forecasts for the {\it LiteBIRD}$+$Simons Observatory and {\it LiteBIRD}$+$CMB-HD configurations, discussing the importance of an exact phase-space treatment for robust cosmological bounds on ALP interactions.
}

\begin{document}
\maketitle
\flushbottom


\section{Introduction}

Axions and axion-like particles (ALPs) arise in a wide range of extensions of the Standard Model (SM) of particle physics. The QCD axion was originally proposed in the context of the Peccei-Quinn (PQ) solution to the strong CP problem~\cite{Peccei:1977hh}, predicting a light pseudo Nambu-Goldstone boson whose mass and couplings are inversely proportional to the PQ symmetry breaking scale, $f_a$~\cite{Weinberg:1977ma, Wilczek:1977pj}. Notably, it constitutes a compelling dark matter candidate through the misalignment mechanism and the decay of topological defects~\cite{Preskill:1982cy, Abbott:1982af, Dine:1982ah, Davis:1986xc, Battye:1994au, Marsh:2015xka}. More general ALPs appear in many frameworks of high-energy physics, such as string compactifications~\cite{Witten:1984dg, Svrcek:2006yi, Conlon:2006tq, Grimm:2007hs, Arvanitaki:2009fg} and models with additional spontaneously broken global symmetries~\cite{Srednicki:1985xd, Georgi:1986df, Kim:2015yna, DiLuzio:2020wdo}. In contrast to the QCD axion, the ALP mass and couplings are not determined by a single scale and can be treated as independent parameters.
This opens up the parameter space available for ALPs and motivates a systematic exploration of their observational signatures~\cite{Cadamuro:2011fd, Bauer:2017ris, Alonso-Alvarez:2019ssa}. 

Given their couplings to SM particles, axions and ALPs are naturally produced in the early Universe through scatterings and decays in the primordial plasma. For the QCD axion, production at high temperatures is dominated by interactions in the quark–gluon plasma~\cite{Masso:2002np, Graf:2010tv, Salvio:2013iaa, Ferreira:2018vjj}, while below the QCD crossover, pion- and nucleon-mediated reactions govern thermal production~\cite{Turner:1986tb, Chang:1993gm, Hannestad:2005df, DiLuzio:2021vjd, Carenza:2021ebx}. Predictions for the resulting axion abundances within selected QCD axion models have also been compiled~\cite{Ferreira:2020bpb, DEramo:2022nvb}. Axions and ALPs with electromagnetic or leptonic couplings can also be produced through Primakoff and lepton-scattering processes, leading to additional thermal production mechanisms beyond QCD interactions. The thermal bath is an unavoidable source of relativistic axions when they interact with SM particles in the early universe~\cite{Langhoff:2022bij, DEramo:2023nzt, Cheng:2025cmb, Yin:2025amn}.

The resulting population of thermally produced ALPs can be probed through early-Universe signatures such as Big Bang Nucleosynthesis (BBN) and the Cosmic Microwave Background (CMB), which are sensitive to even small amounts of additional radiation contributing to the effective number of relativistic species, $\Delta N_{\rm eff}$. Predictions for light element abundances within standard BBN are in very good agreement with observational data~\cite{Steigman:2007xt, Pisanti:2007hk, Iocco:2008va, Pospelov:2010hj, Consiglio:2017pot, Pitrou:2018cgg}, placing tight bounds on any additional dark radiation component during this epoch~\cite{Baumann:2016wac, Yeh:2020mgl, Yeh:2022heq}.
Similarly, extra radiation modifies the anisotropy pattern of the CMB~\cite{Bashinsky:2003tk,Hou:2011ec,Planck:2018vyg, SPT-3G:2025bzu, AtacamaCosmologyTelescope:2025blo} and, for unstable species, can also affect the reionization history~\cite{Langhoff:2022bij, Cheng:2025cmb}.
Looking ahead, future CMB surveys are expected to reach a percent sensitivity to $\Delta N_{\rm eff}$ and below~\cite{SimonsObservatory:2018koc, LiteBIRD:2022cnt, Sehgal:2019ewc, CMB-HD:2022bsz}.

Cosmological constraints on ALPs have traditionally been derived either under the instantaneous decoupling approximation or employing an integrated Boltzmann equation treatment, which assumes an equilibrium (thermal) distribution function for the ALPs. Examples include the model-independent limits on axion-photon and axion-gluon couplings that we obtained in ref.~\cite{Caloni:2022uya}, as well as earlier studies such as refs.~\cite{Hannestad:2005df, Melchiorri:2007cd, Cadamuro:2011fd, Archidiacono:2013cha, DEramo:2018vss, Ferreira:2020bpb}. 
These approaches implicitly rely on the premise that possible departures from a thermal distribution have a negligible impact on the cosmological observables from which the bounds are inferred.
However, with the increasing precision of small-scale CMB measurements from current and forthcoming ground-based experiments, this assumption requires a careful re-examination.
Recent work has emphasized that the full momentum dependence of the ALP phase-space distribution function (PSD) can in fact play a crucial role in accurately predicting its cosmological impact. Departures from a thermal shape can significantly alter the resulting relic abundance and the associated contribution to dark radiation, thus modifying the inferred constraints on ALP interactions with SM particles~\cite{DEramo:2024jhn,Badziak:2024qjg}.

In this work we take a step further. We perform a fully consistent cosmological analysis in which the exact ALP phase-space distribution, obtained by solving the momentum-dependent Boltzmann equation, is propagated into the computation of CMB observables through the Boltzmann solver \texttt{CLASS}~\cite{Blas:2011rf}, and consistently incorporated in the Markov chain Monte Carlo analysis with \texttt{Cobaya}~\cite{Torrado:2020dgo}. We focus on ALP interactions with leptons and photons, and consider light ALPs with mass $m_a = 10^{-3} \, {\rm eV}$, such that they are ultra-relativistic both during production and at the epoch relevant for CMB observations, and can be treated as effectively massless throughout this work. Extending this analysis to heavier ALPs, including the assessment of mass effects, is left for future work. 

We consider different combinations of state-of-the-art datasets. For the CMB, this includes the low-$\ell$ temperature and polarization data along with high-$\ell$ TT, TE, EE spectra, and the lensing reconstruction from {\it Planck} datasets~\cite{Planck:2018vyg, Planck:2019nip}, the Atacama Cosmology Telescope (ACT) in its data release 6 (DR6)~\cite{AtacamaCosmologyTelescope:2025blo}, and the South Pole Telescope (SPT-3G) D1~\cite{SPT-3G:2025bzu, Balkenhol:2024sbv}. We also include baryon acoustic oscillation (BAO) measurements from the Dark Energy Spectroscopic Instrument (DESI) DR2~\cite{DESI:2025zgx}. BBN data include the abundances of deuterium and helium, bringing along up-to-date constraints on the expansion rate and additional radiation during the BBN epoch~\cite{Baumann:2016wac, Yeh:2020mgl, Yeh:2022heq}.

The paper is organized as follows. In section~\ref{sec:phase-space}, we present the essential formulas describing the phase-space treatment of thermal ALP production, establishing the framework used here. In section~\ref{sec:constraints} we present the constraints imposed by current cosmological data and discuss the regions of parameter space already excluded. In section~\ref{sec:forecasts}, we provide forecasts for the sensitivity of forthcoming CMB surveys, demonstrating the prospects for substantial discovery potentials. Our conclusions are summarized in section~\ref{sec:conclusions}. Throughout this paper, we adopt natural units with $c = \hbar = 1$, and use the reduced Planck mass $M_{\rm Pl} = (8\pi G)^{-1/2} \simeq 2.43 \times 10^{18} \, {\rm GeV}$.


\section{Phase-space analysis of ALP production}
\label{sec:phase-space}

\subsection{ALP model}
We work under the assumption that, below the energy scale $f_a$ at which the global symmetry is spontaneously broken, the ALP field $a$ is described by an effective field theory (EFT). The low-energy interactions that we consider are captured by the Lagrangian~\cite{Georgi:1986df, Jaeckel:2010ni}
\begin{equation}
    \label{eq:lagrangian}
	\mathcal{L}_a \supset \frac{1}{2}(\partial^\mu a) (\partial_\mu a) - \frac{1}{2}m_a^2 a^2 + \sum_\ell\,c_\ell \frac{\partial_\mu a}{2 f_a} \bar{\ell} \gamma^\mu \gamma^5 \ell + \frac{1}{4} g_{a\gamma} a F_{\mu\nu}\tilde{F}^{\mu\nu} \, ,
\end{equation}
where the first term corresponds to the kinetic term of the ALP, and the mass term $m_a$ softly breaks the shift symmetry associated with the ALP. The interactions with SM leptons, $\ell = \{e, \mu, \tau\}$, are parameterized in terms of the the derivative dimensionless couplings $c_\ell$, while the ALP-photon coupling at tree-level is expressed by the quantity $g_{a\gamma}$. We neglect higher-dimension couplings to hadronic currents, which are absent at the UV scale. We also assume that no additional light degree of freedom is present and mixes with the ALP, so we isolate the production processes we focus on in this work, namely the Primakoff and leptonic channels.

In addition to leptons, ALPs may also couple directly to quarks through analogous derivative interactions. We do not include such couplings in our analysis, as the evaluation of the collision term below the QCD confinement scale ($T \lesssim 1 \, {\rm GeV}$) is highly uncertain. Nevertheless, these interactions could still play a significant role in ALP production, even when restricting to temperatures above the GeV scale, where the collision term can be computed with theoretical control. Such contributions are particularly relevant for future CMB observations sensitive to light relic production at $T \gtrsim 1 \, {\rm GeV}$ (see, e.g., refs.~\cite{Ferreira:2018vjj,DEramo:2024jhn}).

Note that, even in the absence of a tree-level ALP-photon coupling, an effective interaction with photons is generically induced at loop level through radiative mixing between operators~\cite{Srednicki:1985xd, Georgi:1986df}. More in detail, triangle diagrams with charged fermions induce an ALP–photon coupling even if $g_{a\gamma}$ vanishes at tree-level, as
\begin{equation}
	g_{a\gamma}^{\rm 1-loop} \sim \frac{\alpha_{\rm EM}}{2\pi\,f_a}\,\sum_i\,c_\ell^{(i)}\,Q_\ell^2\,F\left(\frac{4m_\ell^2}{m_a^2}\right)\,,
\end{equation}
in terms of a triangle-loop form factor $F$~\cite{Adler:1969gk, Marciano:1977wx}, which reduces to $F\to 1$ for the regime $m_a \ll m_\ell$ of interest for dark radiation.  Since the loop-induced photon coupling generated from leptonic couplings is parametrically suppressed for light ALPs, we are allowed to vary one coupling at a time in our analysis.

\subsection{Boltzmann equation for the ALP phase-space distribution}
We consider an ALP field $a$, whose PSD is denoted by $\mathcal{F}_a(k,t)$. Here, $k$ denotes the physical momentum of the ALP and $t$ is the cosmic time. 
The time evolution of $\mathcal{F}_a$ is governed by the Boltzmann equation
\begin{equation}
    \label{eq:Boltzmann-t}
    \frac{{\rm d}\mathcal{F}_a(k,t)}{{\rm d}t} = \mathcal{C}_a(k,t) \left[ 1 - \frac{\mathcal{F}_a(k,t)}{\mathcal{F}_a^{\rm \, eq} (k,t)} \right] \, ,
\end{equation}
where $\mathcal{C}_a(k,t)$ is the collision rate for single particle production processes. The equilibrium distribution function for the ALP reads
\begin{equation}
    \label{eq:Fequilibrium}
    \mathcal{F}_a^{\rm \, eq}(k, t) = \frac{1}{e^{E_k/T(t)} - 1} \, ,
\end{equation}
depending on the cosmic time $t$ through the temperature of the SM plasma, $T$. Here $E_k = \sqrt{k^2 + m_a^2}$, and the minus sign encodes Bose-Einstein distribution. The factor $1 - \mathcal{F}_a(k,t)/\mathcal{F}_a^{\rm \, eq}(k,t)$ in eq.~\eqref{eq:Boltzmann-t} comes from the detailed-balance identity in the full quantum Boltzmann collision term from freeze-in~\cite{Gondolo:1990dk}, for derivation see e.g.\ refs.~\cite{Hall:2009bx, Graf:2010tv, Arias:2012az, Blennow:2012de, Salvio:2013iaa}. While this expression neglects backreaction on the SM plasma, this approximation is well justified, especially in the freeze-in regime, where the ALP abundance remains always subdominant (see e.g.~\cite{DEramo:2023nzt}).
In this work, we focus on the production of ALPs through $2 \rightarrow 2$ scatterings of the type $1 + 2 \rightarrow 3 + a$, arising from ALP–lepton and ALP–photon interactions (see sections~\ref{sec:ALP-lepton} and~\ref{sec:ALP-photon}, respectively). The collision rate term for such processes is
\begin{equation}
    \label{eq:collision-term}    
    \mathcal{C}_a(k,T) = \frac{1}{2 E_k} \int \, {\rm d}\Pi_1 {\rm d}\Pi_2 {\rm d}\Pi_3 \round[4]{2 \pi} \delta^{\round{4}} \round{P_1 + P_2 - P_3 - K} \abs{\overline{\mathcal{M}}_{12\rightarrow 3a} \round{s,t,u} }^2 f_1 f_2 \round{1 \mp f_3} \, ,
\end{equation}
where $s$, $t$ and $u$ are the Mandelstam variables, $P_i$, $K$ are four-momenta, the plus (minus) sign corresponds to particle 3 being a boson (fermion), and the phase-space measure is
\begin{equation}
    {\rm d}\Pi_i \equiv \frac{g_i}{\round[3]{2\pi}} \frac{{\rm d}^3 \mathbf{p}_i}{2E_{p_i}} \; ,
\end{equation}
with $g_i$ the number of internal degrees of freedom of the species $i$.
Here, $\abs{\overline{\mathcal{M}}}^2$ denotes the squared matrix element averaged over both initial and final polarization states. 
The integral in eq.~\eqref{eq:collision-term} can be reduced to a four-dimensional form for any generic $2 \to 2$ scattering process (see appendix~\ref{app:collision-integral} for details). 
For each production channel considered in the following sections, we evaluate the integral~\eqref{eq:coll_step3_red} numerically with Monte Carlo methods, using the \texttt{VEGAS} algorithm~\cite{Lepage:1977sw} through the \texttt{PyCuba} library~\cite{Hahn:2004fe}.

Once the collision integral has been computed, the next step is to integrate the Boltzmann equation~\eqref{eq:Boltzmann-t}.
For convenience in the numerical analysis, we introduce the dimensionless evolution variable $x \equiv M/T$, with $M$ a fixed reference energy scale, whose choice has no physical impact, and define the dimensionless comoving momentum of the ALP
\begin{equation}
    \label{eq:comoving-momentum}
    q \equiv \frac{a(T)}{a(T_{\rm in}) T_{\rm in}} k = \frac{k}{T} \left( \frac{g_{*s}(T)}{g_{*s}(T_{\rm in})} \right)^{-1/3} \, ,
\end{equation}
where $g_{*s}(T)$ denotes the effective number of degrees of freedom in entropy and $T_{\rm in}$ is the initial temperature from which we start to evolve the equation. 
In terms of these variables, the Boltzmann equation~\eqref{eq:Boltzmann-t} can be recast as (see, e.g., ref.~\cite{DEramo:2023nzt})
\begin{equation}
    \label{eq:Boltzmann-x}
    \frac{{\rm d}\mathcal{F}_a(q,x)}{{\rm d}\log x} = \frac{\mathcal{C}_a(q,x)}{H(x)} \left(1 - \frac{1}{3}\frac{{\rm d}\log g_{*s}}{{\rm d}\log x} \right) \left[ 1-\frac{\mathcal{F}_a(q,x)}{\mathcal{F}_a^{\rm \, eq}(q,x)} \right] \, ,
\end{equation}
where the Hubble parameter during the radiation-dominated era is given by
\begin{equation}
    H(x) = \frac{\pi}{\sqrt{90}}\frac{g_*(x)^{1/2}}{M_{\rm Pl}} \left(\frac{M}{x}\right)^2 \, ,
\end{equation}
with $g_*(x)$ the effective number of degrees of freedom in energy density.\footnote{In our analysis we adopt the parametrization of $g_*(T)$ and $g_{*s}(T)$ given in~\cite{Saikawa:2018rcs}.} We integrate equation~\eqref{eq:Boltzmann-x} numerically in momentum space, with 64 logarithmically-spaced values of $q$ chosen such that $k/T \in [0.005,50]$. 
We assume the ALP population to be vanishing at the initial temperature, $T_{\rm in}$, and integrate the equation down to the final temperature $T_f = 10^{-5} \, {\rm GeV}$. This corresponds to a temperature below the electron mass, at which point $\Delta N_{\rm eff}$ stops evolving. 
In the following sections, we apply this framework to the specific cases of ALP-lepton and ALP-photon interactions, deriving the corresponding collision terms and PSDs. For ALP-lepton interactions, the production rate is IR-dominated and the resulting $\Delta N_{\rm eff}$ is insensitive to the exact value of the initial temperature; in this case, we fix $T_{\rm in} = 10^3\,{\rm GeV}$ as a representative choice. Conversely, ALP production via the Primakoff process is UV-dominated and therefore sensitive to the upper limit of integration. Therefore, for ALP-photon couplings we vary $T_{\rm in}$ within [$10^{-1}, 10^3$]\,GeV, in order to assess the dependence of $\Delta N_{\rm eff}$ on this parameter.

All the numerical tools, encompassing both the computation of collision integrals and the subsequent solution of the Boltzmann equations for particle production, have been integrated into a single computational framework, which we plan to make publicly available in a subsequent stage of this work. In the following, we will refer to this setup as the \textit{PSD solver}.


\subsection{Production via ALP-lepton interactions} \label{sec:ALP-lepton}

\begin{figure}[t!]
  \centering
  \resizebox{\textwidth}{!}{%
  \begin{tikzpicture}
    \begin{feynman}

      \vertex (t_i1) at (0,0.9) {$\ell^-$};
      \vertex [below=2cm of t_i1] (t_i2) {$\ell^+$};

      \vertex [right=2cm of t_i1, dot] (t_v1) {};
      \vertex [right=2cm of t_i2, dot] (t_v2) {};

      \vertex [right=2cm of t_v1] (t_f1) {$\gamma$};
      \vertex [right=2cm of t_v2] (t_f2) {$a$};

      \diagram*{
        (t_i1) -- [fermion] (t_v1),
        (t_i2) -- [fermion] (t_v2),
        (t_v1) -- [fermion] (t_v2),
        (t_v1) -- [photon] (t_f1),
        (t_v2) -- [scalar] (t_f2),
      };

      \vertex (u_i1) at (5.2,0.9) {$\ell^-$};
      \vertex [below=2cm of u_i1] (u_i2) {$\ell^+$};

      \vertex [right=2cm of u_i1, dot] (u_v1) {};
      \vertex [right=2cm of u_i2, dot] (u_v2) {};

      \vertex [right=2cm of u_v1] (u_f1) {$\gamma$};
      \vertex [right=2cm of u_v2] (u_f2) {$a$};

      \diagram*{
        (u_i1) -- [fermion] (u_v1),
        (u_i2) -- [fermion] (u_v2),
        (u_v1) -- [fermion] (u_v2),
        (u_v2) -- [photon] (u_f1),
        (u_v1) -- [scalar] (u_f2),
      };

    \vertex [dot] (s_v1) at (11.7,0.0) {};  
    \vertex [dot] (s_v2) at (13.2,0.0) {};

    \vertex (s_Lin)  at (10.6,1.2); 
    \node[below left=-4pt] at (s_Lin) {$\ell^\pm$};
 
    \vertex (s_Lout) at (14.2,1.2);  
    \node[below right=-4pt] at (s_Lout) {$\ell^\pm$};

    \vertex (s_Gin)  at (10.6,-1.2); 
    \node[above left] at (s_Gin) {$\gamma$};  
    
    \vertex (s_Aout) at (14.2,-1.2); 
    \node[above right=-2pt] at (s_Aout) {$a$};        

    \diagram*{
      (s_Lin)  -- [fermion] (s_v1),
      (s_v1)   -- [fermion] (s_v2),
      (s_v2)   -- [fermion] (s_Lout),
    
      (s_Gin)  -- [photon] (s_v1),
      (s_v2)   -- [scalar] (s_Aout),
    };
    
    \def\xC{15.75}

    \vertex [dot] (c_vL) at (\xC,0.0) {};        
    \vertex [dot] (c_vR) at (\xC+2.0,0.0) {};    

    \vertex (c_lL) at (\xC-0.65, 1.3);
    \node[left] at (\xC+0.2, 1.1) {$\ell^\pm$};

    \vertex (c_lR) at (\xC+2.55, 1.3);
    \node[right] at (\xC+1.8, 1.1) {$\ell^\pm$};

    \diagram*{
      (c_lL) -- [fermion] (c_vL),
      (c_vR) -- [fermion] (c_lR),
      (c_vL) -- [fermion] (c_vR),
    };

    \vertex (c_G) at (\xC+0.,-1.6) {$\gamma$};
    \vertex (c_A) at (\xC+2.0,-1.6) {$a$};

    \diagram*{
      (c_vR) -- [photon] (c_G),   
      (c_vL) -- [scalar] (c_A),   
    };

    \end{feynman}
  \end{tikzpicture}
  }
  \caption{Feynman diagrams for ALP production via interactions with leptons. From left to right: lepton pair annihilation ($t$ and $u$-channel) and Compton-like scattering ($s$ and $u$-channel).}
  \label{fig:Feynman_leptons}
\end{figure}
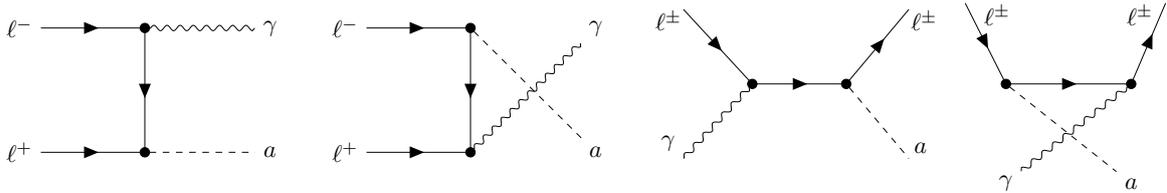

In this section we consider ALPs coupled to leptons through flavor-diagonal interactions described by the Lagrangian 
\begin{equation}
    \mathcal{L}_{a\ell} = c_\ell \frac{\partial_\mu a}{2 f_a} \bar{\ell} \gamma^\mu \gamma^5 \ell  \, ,
\end{equation}
with $\ell = \{ e, \mu, \tau \}$. 
ALP production from thermal leptons proceeds mainly via two processes: lepton-antilepton annihilation, $\ell^+ \ell^- \rightarrow a \gamma$, and Compton-like scattering, $\ell^\pm \gamma \rightarrow \ell^\pm a$ (see \cref{fig:Feynman_leptons} for the corresponding Feynman diagrams). 
We compute the squared matrix elements for these processes, averaged over both initial and final states, in their most general form retaining the full dependence on the ALP mass. Although our analysis is carried out for an ALP with fixed mass $m_a = 10^{-3} \, {\rm eV}$, chosen such that the particle is effectively massless at BBN and recombination epochs, we write down the full mass-dependent expressions for completeness. These are particularly relevant in scenarios where the ALP mass is not negligible compared to the temperature at which production occurs (i.e., for $m_a \sim m_\ell$).
The resulting expressions are:
\begin{align}
    \nonumber
    |\overline{\mathcal{M}}_{\ell^+ \ell^- \rightarrow a \gamma}|^2 &= 
    \frac{c_\ell^2 e^2}{2f_a^2} \frac{m_\ell^2}{(m_\ell^2 - t)^2 (m_\ell^2 - u)^2}
    \Big\{ s^2 (m_\ell^2 - u)(-m_\ell^2 + s + u) - m_a^6 (3 m_\ell^2 - u) \\
    & - m_a^4 \big[m_\ell^4 - m_\ell^2 (5s + 2u) + u(s + u)\big] 
     - m_a^2 s^2 (3 m_\ell^2 - u) \Big\} \, , \\
    \nonumber
    |\overline{\mathcal{M}}_{\ell^\pm \gamma \rightarrow \ell^\pm a}|^2 &= 
    \frac{c_\ell^2 e^2}{2f_a^2} \frac{m_\ell^2}{(m_\ell^2 - s)^2 (m_\ell^2 - u)^2} 
    \Big\{-t^2 (m_\ell^2 - u)(-m_\ell^2 + t + u) + m_a^6 (3 m_\ell^2 - u) \\
    & + m_a^4 \big[m_\ell^4 - m_\ell^2 (5t + 2u) + u(t + u)\big] + m_a^2 t^2 (3 m_\ell^2 - u) \Big\} \, ,
\end{align}
where $s$, $t$ and $u$ are the Mandelstam variables. In the limit $m_a \rightarrow 0$, we recover the expressions of refs.~\cite{DEramo:2018vss,DEramo:2024jhn}.\footnote{Note that ref.~\cite{DEramo:2018vss} reports the matrix elements averaged over the initial states but summed over the final ones. As a consequence, their expressions differ by an overall factor of 1/2 compared to ours and to those presented in ref.~\cite{DEramo:2024jhn}.}
\begin{figure}[t!]
    \centering
    \includegraphics[width=0.5\linewidth]{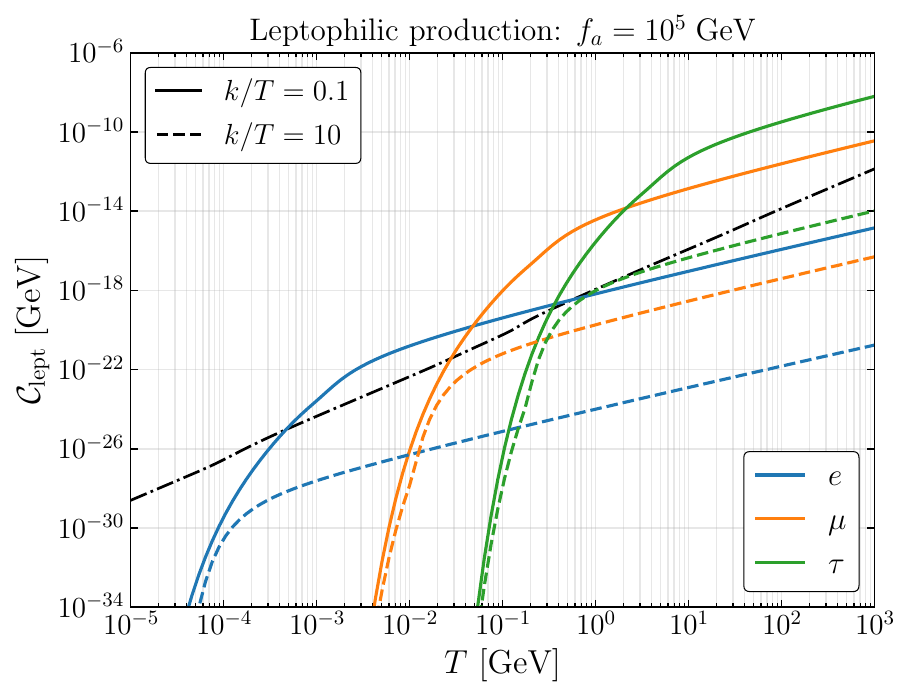}
    \includegraphics[width=0.485\linewidth]{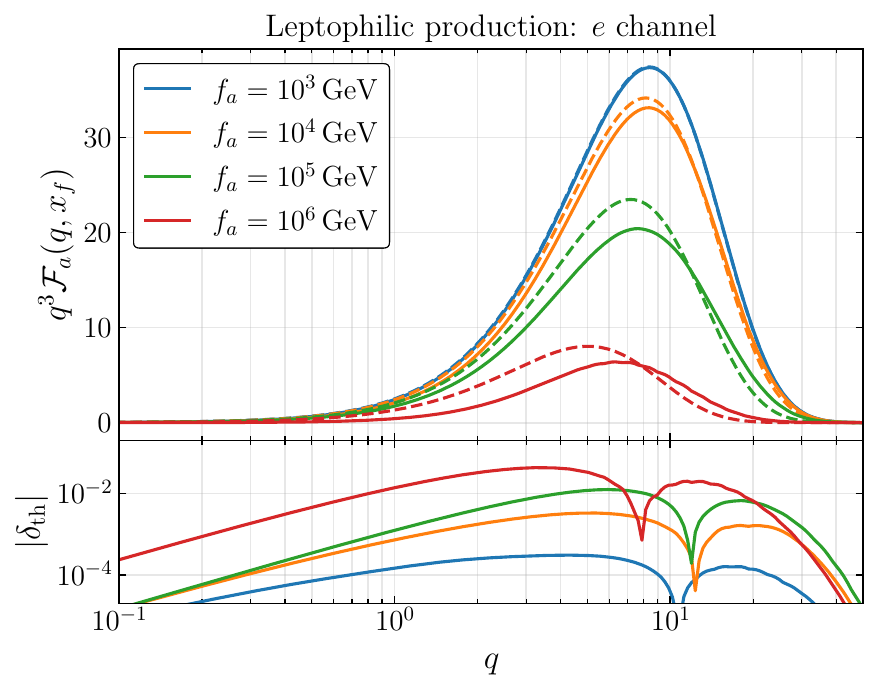}
    \caption{\textit{Left}: collision rate for production of ALPs through leptonic interactions with electrons (blue), muons (orange), and taus (green), as a function of the temperature of the primordial plasma. The curves correspond to two fixed values of the momentum, $k/T = 0.1$ (solid) and $k/T = 10$ (dashed). The ALP decay constant is set to $f_a = 10^5 \, {\rm GeV}$, with $c_\ell = 1$ by convention. The dot-dashed line denotes the Hubble expansion rate.
    \textit{Right}: Exact ALP phase-space distributions for the electron channel (solid lines) compared with Bose-Einstein distributions with temperatures rescaled to yield the same $\Delta N_{\rm eff}$ (dashed lines), for different values of $f_a$. The lower sub-panel shows the difference between the exact and thermal distribution functions, quantified by the parameter $\delta_{\rm th}$ defined in eq.~\eqref{eq:delta_th}.}
    \label{fig:lepton-collision-psd}
\end{figure}
The collision term is then obtained integrating eq.~\eqref{eq:coll_step3_red} using the matrix elements above. In \cref{fig:lepton-collision-psd} (left panel) we show the result for the different leptonic production channels and for two representative momenta, $k/T = 0.1$ and $k/T = 10$. In this plot, the ALP decay constant is fixed to $f_a = 10^5 \, {\rm GeV}$. Here and in the following, we set $c_\ell = 1$ by convention. This choice does not entail any loss of generality, since the limits presented can be straightforwardly reinterpreted as constraints on the ratio $f_a / c_\ell$.
Note that heavier leptons become Boltzmann suppressed at higher temperatures, causing their corresponding collision term to drop at earlier times.

The right panel of \cref{fig:lepton-collision-psd} shows the ALP phase-space distributions obtained by solving the Boltzmann equation~\eqref{eq:Boltzmann-x} for the electron channel, for different values of $f_a$.
For comparison, we also show the corresponding thermal (Bose-Einstein) distributions with temperatures rescaled to yield the same values of $\Delta N_{\rm eff}$.
To quantify this deviation from a thermal spectrum, we define the quantity 
\begin{equation}
    \label{eq:delta_th}
    \delta_{\rm th} \equiv \frac{q^3 \left[ \mathcal{F}_a(q,x_f) - \mathcal{F}_a^{\rm \, eq}(q,\bar{x}) \right]}{ \displaystyle \int {\rm d}q \, q^3 \mathcal{F}_a^{\rm \, eq}(q,\bar{x})} \, ,
\end{equation}
which is a dimensionless function of the comoving momentum $q$. Here $\mathcal{F}_a^{\rm \, eq}$ denotes the equilibrium distribution defined in eq.~\eqref{eq:Fequilibrium}, and $\bar{x}$ is the value of $x$ corresponding to the rescaled temperature that reproduces the same $\Delta N_{\rm eff}$.
As can be seen in \cref{fig:lepton-collision-psd}, larger values of $f_a$ (corresponding to smaller couplings) result in more pronounced departures from a thermal distribution, since the ALP population interacts too feebly with the primordial plasma to achieve thermal equilibrium.


\subsection{Production via ALP-photon interactions}
\label{sec:ALP-photon}
\begin{figure}[t!]
  \centering
  \begin{tikzpicture}
    \begin{feynman}
      \vertex (i1) {$\gamma$}; 
      \vertex [right=2cm of i1, dot] (v1) {};
      \vertex [right=2cm of v1] (f1) {$a$};
      \vertex [below=2cm of i1] (i2) {$q^\pm$};
      \vertex [below=2cm of v1, dot] (v2) {};
      \vertex [right=2cm of v2] (f2) {$q^\pm$};
      \diagram* {
        (i1) -- [photon] (v1) -- [scalar] (f1),
        (v1) -- [photon, edge label=$\gamma$] (v2),
        (i2) -- [fermion] (v2) -- [fermion] (f2),
      };
    \end{feynman}
  \end{tikzpicture}
  \caption{$t$-channel diagram for Primakoff production of an ALP.}
  \label{fig:Feynamn_Primakoff}
\end{figure}
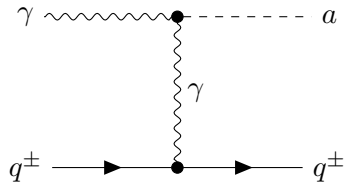

\begin{figure}
    \centering
    \includegraphics[width=0.53\linewidth]{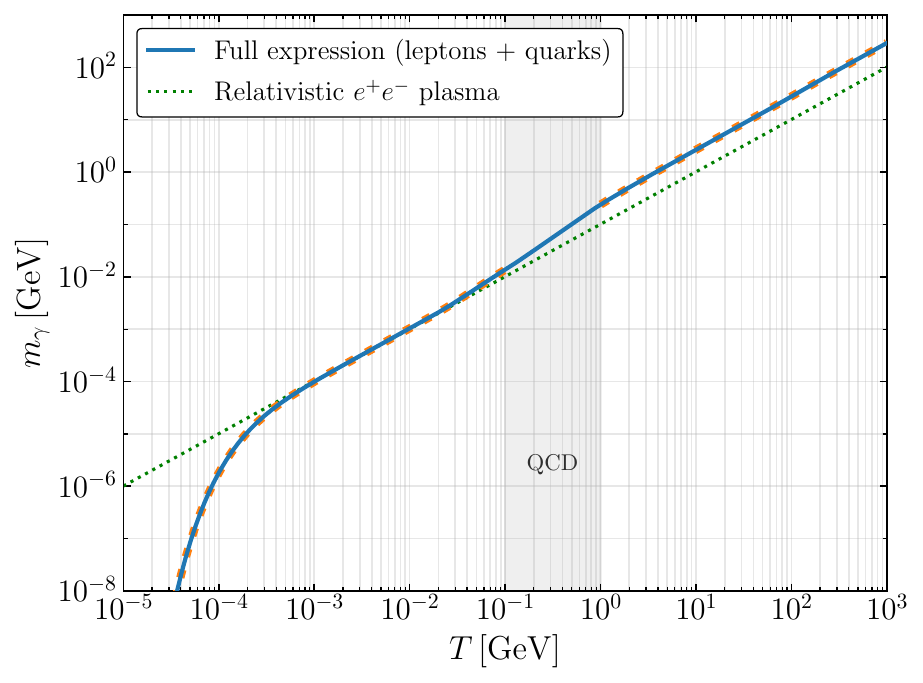}
    \caption{Photon plasma mass, $m_\gamma$, as a function of the temperature of the primordial plasma. The orange markers denote values computed using eq.~\eqref{eq:plasma_mass}, including only leptons in the sum for $T \leq 100 \, {\rm MeV}$, and both leptons and quarks for $T \geq 1 \, {\rm GeV}$. The solid blue curve represents the smooth interpolation between these two regimes across the QCD phase transition epoch. 
    The green dotted curve corresponds to the photon mass in a relativistic $e^+e^-$ plasma, $m_\gamma \simeq e T / 3$ (see, e.g.,~\cite{Carrington:1997sq, Bolz:2000fu}). This is well recovered from our full expression of $m_\gamma$ for temperatures $m_e \ll T \ll m_\mu$.}
    \label{fig:m_gamma}
\end{figure}

The interaction between ALPs and photons is described by the dimension-five operator
\begin{equation}
    \label{eq:ALP-photon}
    \mathcal{L}_{a\gamma} = \frac{1}{4} g_{a\gamma} a F_{\mu\nu}\tilde{F}^{\mu\nu} \, ,
\end{equation}
where $g_{a\gamma}$ denotes the ALP-photon coupling, $F_{\mu\nu}$ is the electromagnetic field strength tensor, and $\tilde{F}^{\mu\nu} \equiv \epsilon^{\mu\nu\alpha\beta} F_{\alpha\beta}/2$ its dual. The dominant production channel for light ALPs is the Primakoff process,\footnote{ALP production from fermion-antifermion annihilations mediated by a photon is indeed subdominant~\cite{Langhoff:2022bij, Jain:2024dtw,Cima:2025zmc}, and is therefore neglected in our analysis. Analogously, production via inverse decays ($\gamma + \gamma \rightarrow a$) is kinematically not allowed until the photon plasma drops below half of the ALP mass, so it is negligible for light ALPs.} whereby a photon is resonantly converted into an ALP in the presence of the charged particles in the primordial plasma (see \cref{fig:Feynamn_Primakoff}).
For a species $i$ with electric charge $Q_i$ (in units of the elementary charge) and mass $m_i$, the squared matrix element for Primakoff production (``Prim''), averaged over both initial and final states, is given by~\cite{Cadamuro:2010cz,Jain:2024dtw}\footnote{Refs.~\cite{Cadamuro:2010cz,Jain:2024dtw} report the expression for the squared matrix element summed over both initial and final states. Here we instead quote the expression averaged over them, as appropriate for our definition of the collision term.}
\begin{equation}
    \label{eq:MatrixPrimakoff}
    |\overline{\mathcal{M}}_{\rm Prim}^i|^2 = \frac{\pi\alpha Q_i^2 g^2_{a\gamma}}{ 2N_i^c (t-m^2_\gamma)^2} \bigg\{-2m_i^2 m_a^4 - t\Big[m_a^4+2(s-m_i^2)^2-2m_a^2(s+m_i^2)^2\Big] -2t^2(s-m_a^2) -t^3 \bigg\} \, ,
\end{equation}
where $\alpha \equiv e^2 / (4\pi) \simeq 1/137$ is the electromagnetic fine-structure constant, and the factor $N_i^c$ accounts for color multiplicity, with $N_i^c = 3$ for quarks and $N_i^c = 1$ for leptons. 
Plasma effects are taken into account by introducing the photon plasma mass, $m_\gamma$, in the propagator of the Primakoff process, as in refs.~\cite{Bolz:2000fu,Cadamuro:2010cz,Jain:2024dtw}. This effective mass serves to regularize the logarithmic divergence arising due to the exchange of a massless photon. 
We compute the plasma mass as (see~\cite{Jain:2024dtw})
\begin{equation}
    \label{eq:plasma_mass}
    m_\gamma^2 = 4\pi\alpha \sum_i Q_i^2 \frac{n_{q_i}}{\langle \omega_{q_i} \rangle} \, ,
\end{equation}
where $n_{q_i}$ is the number density of the $i$-th charged species in the plasma (including the contribution from the corresponding anti-particle) and 
\begin{equation}
    \langle \omega_{q_i} \rangle \equiv \frac{\displaystyle \int {\rm d}^3 p \, \omega_{q_i} f_{q_i}(\omega_{q_i})}{\displaystyle \int {\rm d}^3 p \, f_{q_i}(\omega_{q_i})} = \frac{\rho_{q_i}}{n_{q_i}}
\end{equation}
denotes its average energy. 
The sum in eq.~\eqref{eq:plasma_mass} runs over all electrically charged fermions of the Standard Model. In practice, we include only leptons for temperatures $T \le 100 \, {\rm MeV}$, and both leptons and quarks for $T \ge 1 \, {\rm GeV}$. The full temperature dependence of the plasma mass is then reconstructed by interpolating between these two regimes. This interpolation, shown in \cref{fig:m_gamma}, smoothly bridges across the QCD phase transition, during which quarks and gluons confine into hadrons and a direct calculation of $m_\gamma$ becomes challenging. At temperatures below the electron mass, the number density of electrons is Boltzmann suppressed, so does the plasma mass in eq.~\eqref{eq:plasma_mass} which depends on the charged particle density. On the other hand, above the QCD phase transition the photon self energy gets contributions from relativistic quark degrees of freedom.

\begin{figure}
    \centering
    \includegraphics[width=0.487\linewidth]{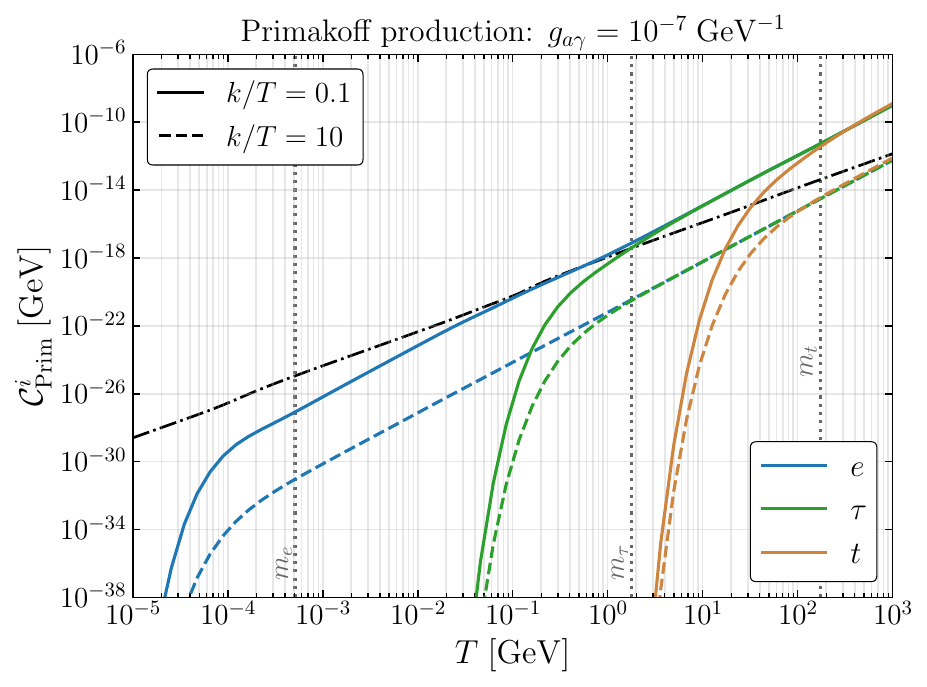}
    \includegraphics[width=0.5\linewidth]{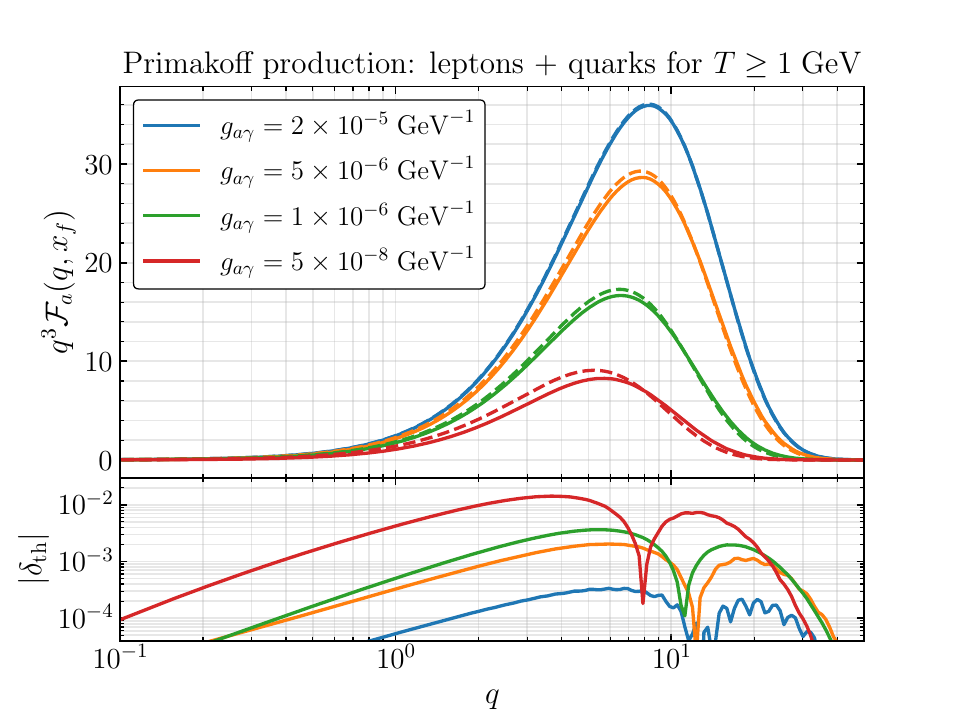}
    \caption{\textit{Left}: collision term for Primakoff production of ALPs through scattering off electrons (blue), taus (green), and top quarks (brown), as a function of the temperature of the primordial plasma. The curves correspond to two fixed values of the momentum, $k/T = 0.1$ (solid) and $k/T = 10$ (dashed). The ALP-photon coupling is set to $g_{a\gamma} = 10^{-7} \, {\rm GeV}^{-1}$. Vertical dotted lines mark the corresponding fermion masses. The dot-dashed line denotes the Hubble expansion rate.
    \textit{Right}: Exact ALP phase-space distributions produced via Primakoff interactions, including contributions from leptons and quarks for temperatures $T \geq 1\,\mathrm{GeV}$, for different values of the ALP–photon coupling $g_{a\gamma}$. Results are obtained by solving the momentum-dependent Boltzmann equation with the matrix element in eq.~\eqref{eq:MatrixPrimakoff}. As for the leptonic channels, we show for comparison the corresponding Bose-Einstein distributions (dashed lines) with temperatures rescaled to yield the same values of $\Delta N_{\rm eff}$.}
    \label{fig:Primakoff-collision-psd}
\end{figure}

As for the photon plasma mass, the computation of the total collision term for ALP production requires in principle summing the contributions from all charged species in the plasma.
We adopt a conservative approach in which only leptons are included for $ T < 1 \, {\rm GeV}$, while quark contributions are added for $T \ge 1 \, {\rm GeV}$:
\begin{equation}
    \label{eq:CPrim}
    \mathcal{C}_{\rm Prim}(T) = 
    \sum_{i \, \in \, \text{leptons}} \mathcal{C}_{\rm Prim}^{i}(T)
    + 
    \Theta(T - 1 \, \mathrm{GeV}) \sum_{j \, \in \, \text{quarks}} \mathcal{C}_{\rm Prim}^{j}(T) \, .
\end{equation}
The resulting collision term and the corresponding ALP phase-space distributions are shown in \cref{fig:Primakoff-collision-psd}. 
The left panel shows the contributions of various fermion species ($e$, $\tau$ and quark $t$) to the Primakoff collision term for two representative momenta, $k/T = 0.1$ and $k/T = 10$, and for the ALP-photon coupling $g_{a\gamma}=10^{-7}{\rm\,GeV^{-1}}$. Electrons dominate at temperatures $T<100 \, \mathrm{MeV}$, since they remain abundant and relativistic down to $T\sim m_e$. However, at large temperature the abundance of heavy tau and top particles dominates the rate. Moreover, soft ALP modes with $k/T = 0.1$ are enhanced with respect to hard ones because Primakoff scattering is dominated by small-angle and low-momentum transfer processes. 

Note that, since the ALP-photon interaction in eq.~\eqref{eq:ALP-photon} is non-renormalizable, the Primakoff collision term grows with temperature faster than the Hubble rate (black dot-dashed line), making the process UV-dominated.
As a result, the production of ALPs in the freeze-in regime is sensitive to the highest temperature from which the Boltzmann equation is evolved, $T_{\rm in}$. This is in principle connected to the reheating temperature of the Universe, since $T_{\rm in} \le T_{\rm reh}$. The latter is bounded from above by the non-observation of primordial tensor modes in the CMB, implying $T_{\rm reh} < 1.6 \times 10^{16} \, {\rm GeV}$~\cite{Planck:2018jri}.
At the same time, cosmological analyses of scenarios with very low reheating temperatures set the lower limit $T_{\rm reh} > 5.96 \, {\rm MeV}$~\cite{Barbieri:2025moq}.
The choice of $T_{\rm in}$ can also be interpreted as an assumption on the maximal temperature up to which our model remains valid (see e.g.~\cite{Carenza:2022ngg,Caloni:2024olo} for other examples of UV-dominated freeze-in scenarios), since at sufficiently large temperatures the EFT description adopted in eq.~\eqref{eq:lagrangian} is expected to break down. For these reasons, and in order to remain agnostic about the details of the reheating history and of the UV completion of the theory, we adopt $T_{\rm in}$ as our convention for the initial temperature.

As in the case of leptonic interactions, in the right panel of \cref{fig:Primakoff-collision-psd} we show the exact ALP phase-space distributions obtained by solving the Boltzmann equation for Primakoff production, together with the corresponding Bose-Einstein distributions with temperatures rescaled to reproduce the same values of $\Delta N_{\rm eff}$.

\subsection{Contribution to $\Delta N_{\rm eff}$}
\label{sec:DNeff}
Once we have obtained a solution for the ALP phase-space distribution, we compute the corresponding contribution to the effective number of relativistic species, $\Delta N_{\rm eff}$. This serves two main purposes. First, as a consistency check, allowing us to compare our results with previous computations available in the literature. Second, because $\Delta N_{\rm eff}$ plays a central role in our Markov chain Monte Carlo (MCMC) analysis, being the parameter that we vary in addition to the six $\Lambda$CDM parameters (see section~\ref{sec:MCMC} for details).
We compute the ALP contribution to $\Delta N_{\rm eff}$ at $T\ll 1$~MeV, i.e., the value relevant for CMB decoupling as
\begin{equation}
    \label{eq:DNeff}
    \Delta N_{\rm eff} = \frac{8}{7} \left( \frac{11}{4} \right)^{4/3} \frac{15}{2\pi^4} \left( \frac{g_{*s}(x_f)}{g_{*s}(x_{\rm in})} \right)^{4/3} \int_0^\infty {\rm d}q \, q^3 \mathcal{F}_a(q,x_f) \, ,
\end{equation}
where $x_{\rm in}$ and $x_f$ denote the values of $x$ evaluated at the initial and final temperatures, $T_{\rm in}$ and $T_f$, respectively.

\begin{figure}
    \centering
    \adjustbox{trim=3 0 15 0, clip}{\includegraphics[scale=0.53]{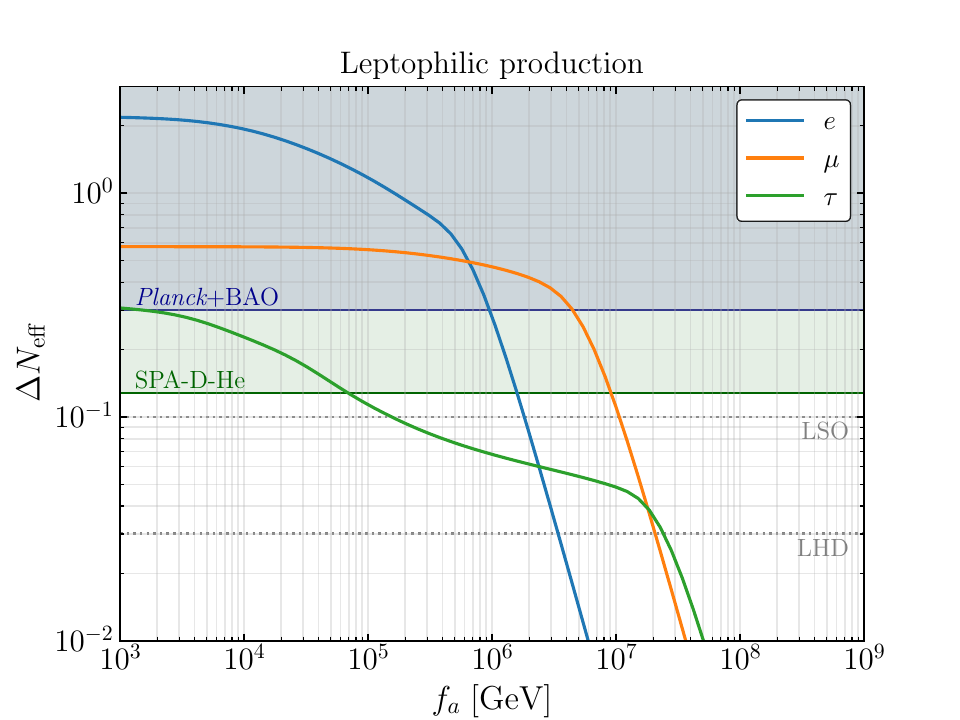}} 
    \hspace{-10pt}
    \adjustbox{trim=20 0 20 0, clip}{\includegraphics[scale=0.53]{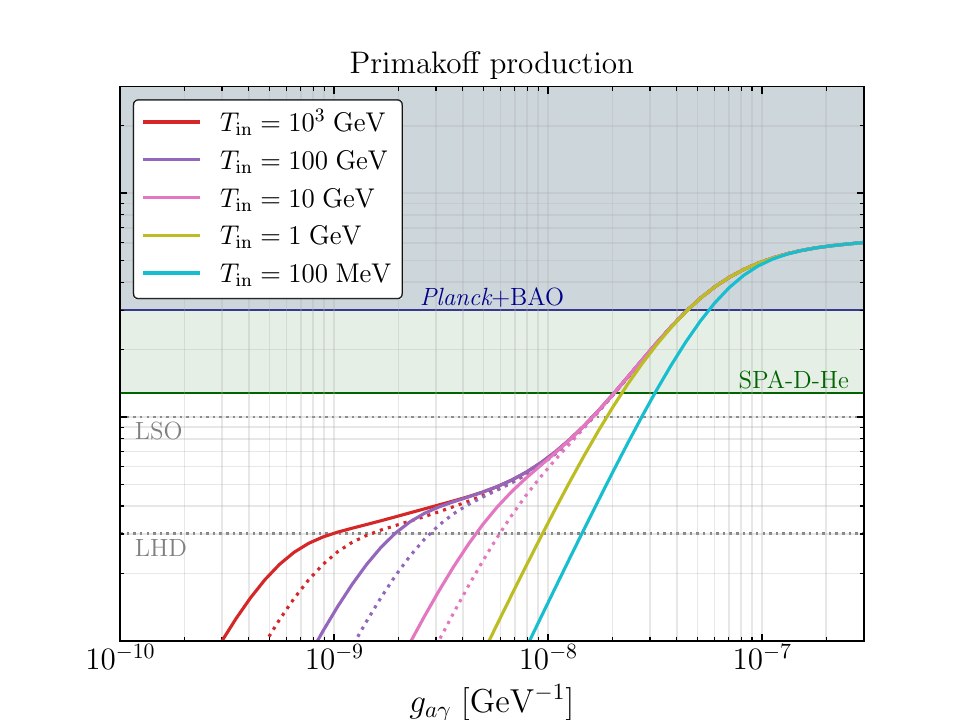}}
    \caption{\textit{Left:} Contribution to $\Delta N_{\rm eff}$ from leptophilic ALP production via pair annihilation and Compton-like processes, as a function of the ALP decay constant, $f_a$. The couplings are switched on one at a time. We set  $c_\ell=1$ without loss of generality; predictions for $\Delta N_{\rm eff}$ for $c_\ell\ne1$ can be obtained by replacing $f_a \to f_a/c_\ell$.
    \textit{Right:} Contribution to $\Delta N_{\rm eff}$ from ALP production via the Primakoff process, as a function of the ALP-photon coupling, $g_{a\gamma}$. This is shown for different values of the initial temperature, $T_{\rm in}$. The dotted curves represent the cases where only Primakoff production from leptons is included, while the solid curves also add the quark contributions for $T \ge 1 \, {\rm GeV}$, see eq.~\eqref{eq:CPrim}.
    In both panels we show the constraint on $\Delta N_{\rm eff}$ from {\it Planck}+BAO~\cite{Planck:2018vyg} as well as the bound obtained in our analysis from the combination of {\it Planck}, ACT and SPT data complemented with BBN measurements of the helium and deuterium abundances (SPA-D-He). The dotted horizontal lines indicate the forecast sensitivities from {\it LiteBIRD}+SO (LSO) and {\it LiteBIRD}+CMB-HD (LHD) derived in this work. All limits are reported at 95\% CL.} 
    \label{fig:DNeff_fermion_primakoff}
\end{figure}

\Cref{fig:DNeff_fermion_primakoff} shows the contribution to $\Delta N_{\rm eff}$ from the different ALP production channels considered in this work.
In the left panel, we display the contribution from leptophilic ALP production discussed in section~\ref{sec:ALP-lepton}, as a function of the ALP decay constant $f_a$. The ALP–lepton couplings are switched on one at a time, and we fix $c_\ell = 1$ without loss of generality. Results 
for $c_\ell\ne 1$ can be obtained by the replacement $f_a \to f_a/c_\ell$.
In the right panel, we show the contribution from Primakoff production as a function of the ALP–photon coupling $g_{a\gamma}$, for several choices of the initial temperature $T_{\rm in}$, ranging from $100 \, {\rm MeV}$ to $10^3 \, {\rm GeV}$. The dotted curves correspond to Primakoff production from leptons only, while the solid curves also include the quark contributions for $T \ge 1 \, {\rm GeV}$. As expected, the inclusion of quark interactions at high temperatures has a negligible impact on current bounds on $\Delta N_{\rm eff}$, but becomes important in view of the improved sensitivity of next-generation CMB experiments.
Current CMB constraints on $\Delta N_{\rm eff}$ already discriminate between different choices of the initial temperature $T_{\rm in}$ for $T_{\rm in} \lesssim 1\,\mathrm{GeV}$. Conversely, for larger initial temperatures the Primakoff contribution saturates and present data become insensitive to this dependence, which can instead be probed only by future ground-based CMB experiments.

\section{Constraints from current cosmological data}
\label{sec:constraints}

\subsection{Methodology and datasets}
\label{sec:MCMC}

We perform a MCMC analysis to obtain joint constraints on the ALP couplings and the cosmological parameters of the $\Lambda$CDM model, consistently incorporating the phase-space framework developed in the previous sections. This constitutes one of the main novelties of our work. The analysis is performed using \texttt{Cobaya}~\cite{Torrado:2020dgo} interfaced with the Boltzmann solver \texttt{CLASS}~\cite{Blas:2011rf}.\footnote{We adopt the same accuracy settings for \texttt{CLASS} as in the ACT DR6 analysis~\cite{AtacamaCosmologyTelescope:2025blo}, tuned to ensure sufficient precision at the small angular scales probed by the ACT and SPT data.} In our baseline setup, we extend the $\Lambda$CDM model by a single additional parameter, the ALP contribution to the effective number of relativistic species, $\Delta N_{\rm eff}$. In the case of ALP-electron interactions, we perform an additional run in which the sampling is carried out directly in terms of $f_a$. This allows us to test how the choice of the sampling parameter, and the associated prior, impacts the final constraints. 
A detailed comparison is presented in section~\ref{sec:KL}. In all cases, the ALP mass is fixed to a negligibly small value, $m_a = 10^{-3} \, {\rm eV}$, so that the ALP remains effectively massless at the epochs relevant for cosmological observations. We defer the analysis of scenarios with heavier ALPs to future work.

When sampling over $\Delta N_{\rm eff}$, this quantity is restricted to be strictly positive, as we only consider scenarios in which ALPs contribute extra dark radiation in addition to the SM neutrino background. Following the discussion in the previous sections, for each interaction channel we pre-compute the ALP PSD on a grid of values of $\Delta N_{\rm eff}$ (or, equivalently, of the corresponding coupling $g_{a\gamma}$ or $f_a$). For MCMC samples of $\Delta N_{\rm eff}$ that do not coincide with grid points, the ALP PSD is obtained by interpolation. By accounting for the full non-thermal PSD, this procedure captures possible spectral distortions beyond a simple thermal shape. The interpolated PSD is then passed to \texttt{CLASS}, which computes the CMB anisotropy spectra and matter power spectra consistently with the correct ALP distribution.

\begin{figure}
    \centering
    \adjustbox{trim=3 0 0 0, clip}{\includegraphics[scale=0.52]{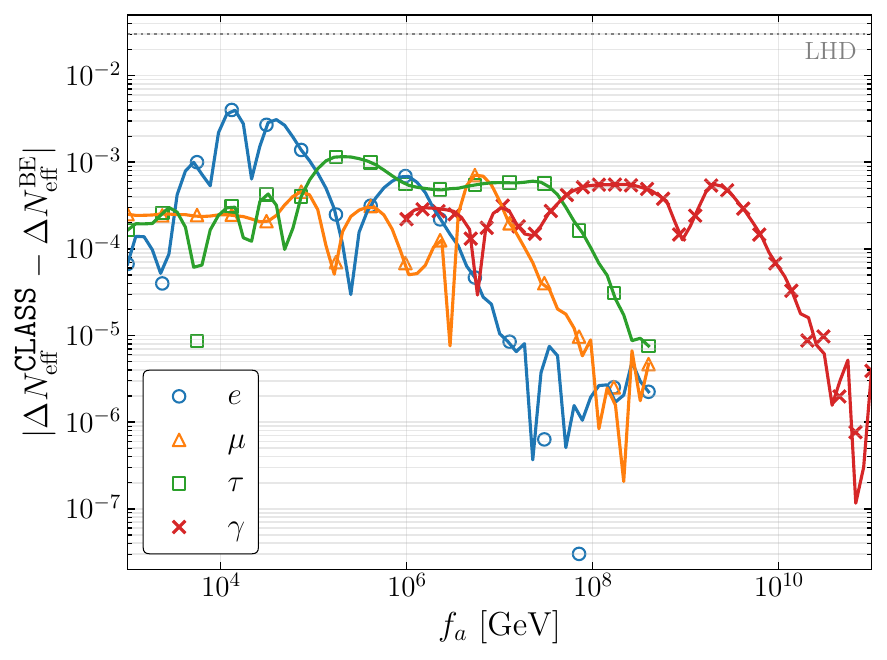}} 
    \hspace{-5pt}
    \adjustbox{trim=3 0 0 0, clip}{\includegraphics[scale=0.52]{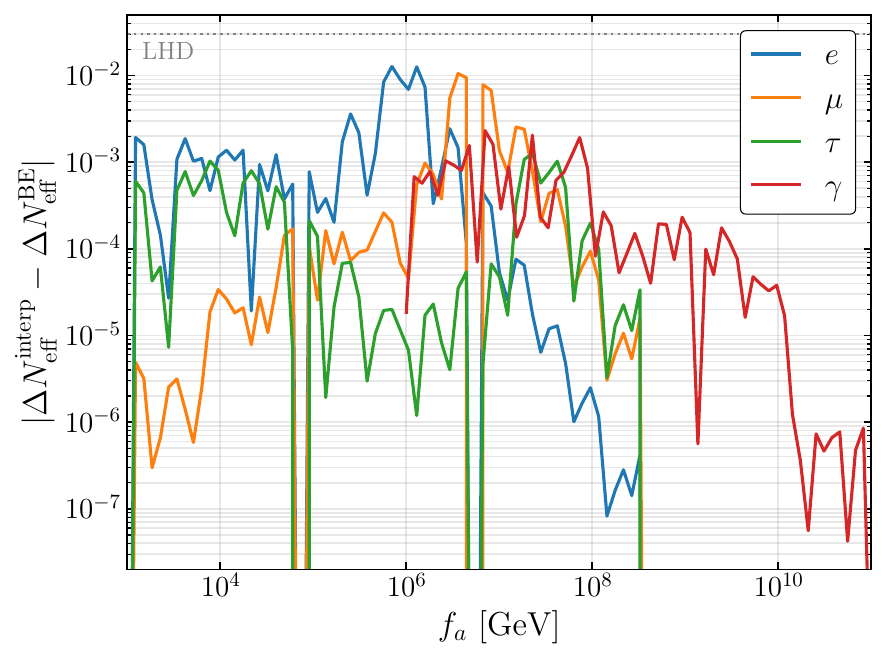}}
    \caption{\textit{Left:} Absolute difference between $\Delta N_{\rm eff}$ computed directly from the ALP PSD using eq.~\eqref{eq:DNeff} and the value obtained by \texttt{CLASS} when the ALP PSD is given through our interpolation procedure, shown as a function of $f_a$. For comparison with the Primakoff channel, we set $f_a = g_{a\gamma}^{-1}$ when plotting the photon coupling. Markers indicate the values used to construct the nodes of the interpolation grid employed in the analysis. \textit{Right:} The accuracy of the interpolation procedure is assessed with the difference between the exact PSD and the interpolated value at fixed coupling. In both panels, the numerical mismatch is well below the forecasted sensitivity of futuristic CMB LHD-like configuration (see \cref{tab:data_combinations}). This shows the robustness of our PSD grid and the stability of the interpolation used in the MCMC analysis. Results are given for ALP couplings to electrons ($e$), muons ($\mu$), taus ($\tau$), and photons ($\gamma$).} 
    \label{fig:Neff_accuracy}
\end{figure}

In \cref{fig:Neff_accuracy}, we assess the numerical accuracy of this procedure by showing the difference between the values of $\Delta N_{\rm eff}$ computed directly using eq.~\eqref{eq:DNeff} and those obtained with \texttt{CLASS} through our interpolation scheme. This discrepancy arises from two independent sources.
First (left panel), numerical uncertainties are introduced in \texttt{CLASS} because the integral over the PSD is evaluated on a grid of comoving momenta that differs from the one adopted for the pre-computation of the PSD within our PSD solver.\footnote{In other words, a discrepancy between the values of $\Delta N_{\rm eff}$ computed directly using eq.~\eqref{eq:DNeff} and those obtained with \texttt{CLASS} is expected even when the PSD is evaluated exactly at the interpolation grid points. We choose the momentum resolution used by \texttt{CLASS} such that this error remains below the sensitivity of future experiments. In particular, we employ five momentum bins selected according to the Gauss-Laguerre algorithm.} Since the PSDs pre-computed with the PSD solver are evaluated on a much finer momentum grid, we regard the corresponding values of $\Delta N_\mathrm{eff}$ as the reference ones.
Second (right panel), an additional numerical error arises from the interpolation of the PSD over a discrete grid of couplings, reflecting the finite resolution of the interpolation grid.
In the worst-case scenario, the total numerical uncertainty, given by the sum of the contributions shown in the two panels, is well below the forecasted sensitivity of a futuristic CMB LHD-like configuration~\cite{Sehgal:2019ewc, CMB-HD:2022bsz} (see \cref{tab:data_combinations}). This demonstrates the robustness and numerical stability of our approach, and that residual numerical uncertainties do not affect our cosmological constraints.

\begin{table}[!t]
    \centering
    \resizebox{0.65\textwidth}{!}{%
    \begin{tabular}{cc}
    \hline
    \hline
    \noalign{\vskip 1mm}
    \textbf{Data combination} & \textbf{Label} \\
    \noalign{\vskip 1mm}
    \hline
    \noalign{\vskip 1mm}
    \textit{Planck}+ACT+SPT & SPA  \\[1mm]
    \textit{Planck}+ACT+SPT+BBN (deuterium and helium) & SPA-D-He \\[1mm]
    \textit{Planck}+ACT+SPT+DESI & SPA-B \\[1mm]
    \textit{Planck}+ACT+SPT+BBN+DESI & SPA-D-He-B \\[1mm]
    {\it LiteBIRD}+Simons Observatory & LSO \\[1mm]
    {\it LiteBIRD}+CMB-HD & LHD \\[1mm]
    \hline
    \hline
    \end{tabular}
    }
    \caption{Summary of the data combinations used in this work and their corresponding labels. The last two lines correspond to combinations of forthcoming data or future experiments, see section~\ref{sec:forecasts} for details.}
    \label{tab:data_combinations}
\end{table}

We employ current state-of-the-art cosmological datasets, including:
\begin{itemize}
    \item {\it Planck}+ACT+SPT: for {\it Planck}, we use temperature data from the \texttt{Commander} likelihood~\cite{Planck:2018vyg,Planck:2019nip} and $E$-mode polarization data from \texttt{SRoll2}~\cite{Pagano:2019tci,Delouis:2019bub} at large angular scales ($2 \le \ell \le 29$), complemented at high multipoles by the \texttt{Plik\_lite} likelihood, which is the compressed high-$\ell$ \emph{Planck} temperature and polarization likelihood with foreground analytically marginalized~\cite{Planck:2018vyg}. These are combined with the \texttt{ACT-lite}\footnote{\url{https://github.com/ACTCollaboration/DR6-ACT-lite}} likelihood for ACT DR6~\cite{AtacamaCosmologyTelescope:2025blo} and the \texttt{SPT-lite}\footnote{\url{https://github.com/SouthPoleTelescope/spt_candl_data}} likelihood for SPT-3G D1~\cite{SPT-3G:2025bzu, Balkenhol:2024sbv}. Following the prescriptions described in refs.~\cite{AtacamaCosmologyTelescope:2025blo,SPT-3G:2025bzu}, we cut {\it Planck} data to multipoles $\ell_{\rm max} = 1000$ in temperature and $\ell_{\rm max} = 600$ in polarization, and assume no correlations between SPT and the other datasets. Furthermore, we include the joint {\it Planck}+ACT DR6 lensing likelihood\footnote{\url{https://github.com/ACTCollaboration/act_dr6_lenslike}}~\cite{ACT:2023kun,ACT:2023dou,ACT:2023ubw,Carron:2022eyg} and the MUSE lensing likelihood\footnote{\url{https://github.com/qujia7/spt_act_likelihood}} for SPT-3G D1~\cite{SPT-3G:2024atg}.
    \item DESI: measurements of BAO distance from recent results of DESI DR2~\cite{DESI:2025zpo,DESI:2025zgx}. These include data at multiple redshifts, obtained from the clustering of bright galaxies (BGS), luminous red galaxies (LRGs), emission line galaxies (ELGs), quasars (QSOs), and the Ly$\alpha$ forest. 
    \item BBN: measurements of the D/H ratio, inferred from observations of deuterium Lyman-$\alpha$ absorption lines in high-redshift, low-metallicity quasar absorption systems, as well as measurements of the primordial helium mass fraction, $Y_p$, derived from recombination emission lines of He and H in the most metal-poor, low-redshift extragalactic HII (ionized) regions. The values of both abundances are taken from the combined estimates of the most recent and precise determinations reported by the Particle Data Group 2025~\cite{ParticleDataGroup:2024cfk} and are incorporated through a Gaussian likelihood used to perform importance sampling on the MCMC chains obtained from CMB and BAO analyses.
\end{itemize}
These datasets are used in the combinations listed in \cref{tab:data_combinations}, together with the corresponding labels that we will use throughout the paper.

Throughout our analysis, we assume a spatially flat Universe with adiabatic initial conditions. Neutrinos are modeled as one massive and two massless species, with the sum of neutrino masses fixed to the minimal value allowed by flavor oscillation experiments in the normal ordering scenario, $\sum m_\nu = 0.06 \, {\rm eV}$~\cite{deSalas:2020pgw,Esteban:2020cvm,Capozzi:2021fjo}. The neutrino contribution to the effective number of relativistic species is fixed to $N_{\rm eff} = 3.044$~\cite{Mangano:2001iu, Bennett:2019ewm, Bennett:2020zkv, Akita:2020szl, Froustey:2020mcq, Cielo:2023bqp, Drewes:2024wbw}, and any contribution in excess to this value is attributed to ALPs. The parameter set varied in our baseline analysis is therefore $\{ \omega_b,\, \omega_c,\, 100 \, \theta_s, \, \tau_{\rm reio}, \, \log(10^{10}A_s), \, n_s, \, \Delta N_{\rm eff} \}$, where $\omega_b \equiv \Omega_b h^2$ and $\omega_c \equiv \Omega_c h^2$ represent the physical density of baryons and cold dark matter, respectively; $\theta_s$ is the angular size of the acoustic scale at recombination; $\tau_{\rm reio}$ is the reionization optical depth; $A_s$ is the amplitude of the primordial power spectrum at the pivot scale $k_* = 0.05 \, {\rm Mpc}^{-1}$; and $n_s$ is the scalar spectral index. The prior distributions adopted for the sampled parameters are summarized in \cref{tab:priors}. Convergence of the MCMC chains is assessed through the Gelman–Rubin statistic, requiring $R-1 < 0.01$~\cite{Gelman:1992zz} 
The posterior distributions are then obtained and analyzed using the \texttt{GetDist} package~\cite{Lewis:2019xzd}.

\begin{table}[t!]
\centering

\begin{tikzpicture}[remember picture,baseline=(tab.south)]
  \node (tab) {
	\begin{tabular}{ c c } 
		\hline
        \hline
        \noalign{\vskip 1mm}
		\textbf{Parameter} & \textbf{Prior} \\
        \noalign{\vskip 1mm}
		\hline
        \noalign{\vskip 1mm}
		$\omega_\mathrm{b}$ & $[0.005, 0.1]$ \\[1mm]
        $\omega_\mathrm{c}$ & $[0.001, 0.99]$ \\[1mm]
		$100 \, \theta_\mathrm{s}$ & $[0.5, 10]$ \\[1mm]
		$\tau_{\rm reio}$ & $[0.01, 0.8]$ \\[1mm]
		$\log(10^{10} A_\mathrm{s})$ & $[1.61, 3.91]$ \\[1mm]
		$n_\mathrm{s}$ & $[0.8, 1.2]$ \\[1mm] 
        \noalign{\vskip 1mm} 
        $\Delta N_{\rm eff}$ & $[0, ({\color{python_c0} 2}, \, {\color{python_c1} 0.55}, \, {\color{python_c2} 0.3}, \, {\color{python_c3} 0.77})]$ \\[1mm]
        \noalign{\vskip 1mm}
        \hdashline 
        \noalign{\vskip 1mm}
        $\log_{10} f_a$ & $[3, 8.5]$ \\[1mm]
        \hline
        \hline
	\end{tabular}
  };
\end{tikzpicture}%

\begin{tikzpicture}[remember picture,overlay]
  \node[anchor=south west, xshift=1em, yshift=1.06cm] at (tab.south east) {%
    \parbox{4cm}{\raggedright\footnotesize
      (${\color{python_c0}e}$, ${\color{python_c1}\mu}$, ${\color{python_c2}\tau}$, ${\color{python_c3}\gamma}$)%
    }%
  };
\end{tikzpicture}

\begin{tikzpicture}[remember picture,overlay]
  \node[anchor=south west, xshift=1em] at (tab.south east) {%
    \parbox{4cm}{\raggedright\footnotesize
      Only used in comparison run for $\Lambda\text{CDM}+e$ model.%
    }%
  };
\end{tikzpicture}
\vspace{-0.5cm}
\caption{Uniform priors adopted for the cosmological parameters in the MCMC analysis. The last line refers to the additional comparison run for the ALP-$e$ coupling in which $f_a$ is varied instead of $\Delta N_{\rm eff}$. The upper bounds of the $\Delta N_{\rm eff}$ prior ranges are chosen to match the maximal values shown in \cref{fig:DNeff_fermion_primakoff} and correspond to the largest values for which the ALP phase-space distributions have been precomputed.}
\label{tab:priors}
\end{table}

\subsection{Results and discussion}
The constraints derived in this work are summarized in \cref{tab:DNeff_constraints}, which reports the 95\% credible limits (CL) on the ALP decay constant $f_a$ (for ALP-lepton interactions, with $c_\ell = 1$) and on the ALP-photon coupling $g_{a\gamma}$ (for Primakoff production), assuming that only one coupling is active at a time.
The table includes results for all the dataset combinations considered in our analysis (see section~\ref{sec:MCMC}). We also report the forecast sensitivities for the \textit{LiteBIRD}+Simons Observatory and \textit{LiteBIRD}+CMB-HD configurations, which are discussed in section~\ref{sec:noise_forecasts}.

\begin{table}[!t]
    \centering
    \renewcommand{\arraystretch}{2} 
    \resizebox{\textwidth}{!}{%
    \begin{tabular}{l c c c c : c c} 
    \hline\hline
    \textbf{Model} 
      & 
      & \textbf{SPA}
      & \textbf{SPA-D-He}
      & \textbf{SPA-D-He-B} \hspace{0.9mm}
      & \textbf{LSO} 
      & \textbf{LHD} \\[1mm]
    \hline

    $\Lambda{\rm CDM}+e$
        & $f_a\ [{\rm GeV}]$
        & $> 1.51 \times 10^6$ & $> 1.63 \times 10^6$ 
        & $> 1.37 \times 10^6$ & $> 1.90 \times 10^6$ & $3.56 \times 10^6$ \\[1mm]

    \hline 

   $\Lambda{\rm CDM}+\mu$
        & $f_a\ [{\rm GeV}]$
        & $> 8.73 \times 10^6$ & $> 9.41 \times 10^6$ 
        & $> 7.50 \times 10^6$ & $> 1.05 \times 10^7$ & $2.17 \times 10^7$ \\[1mm]

    \hline

    $\Lambda{\rm CDM}+\tau$
        & $f_a\ [{\rm GeV}]$
        & $> 6.10 \times 10^4$ & $> 8.06 \times 10^4$ 
        & $> 2.85 \times 10^4$ & $> 1.34 \times 10^5$ & $2.45 \times 10^7$\\[1mm]

    \hline

    $\Lambda{\rm CDM}+\gamma$
        & $g_{a\gamma}\ [{\rm GeV}^{-1}]$
        & $< 2.18 \times 10^{-8}$ & $< 1.98 \times 10^{-8}$ 
        & $< 2.73 \times 10^{-8}$ & $< 1.66 \times 10^{-8}$ & $< 1.01 \times 10^{-9}$ \\[1mm]

    \hline\hline
    \end{tabular}
    }
    \caption{95\% credible limits on $f_a$ or $g_{a\gamma}$ for each model and dataset combination.}
    \label{tab:DNeff_constraints}
\end{table}

Compared to previous constraints obtained using {\it Planck} data alone~\cite{DEramo:2018vss,Green:2021hjh,DEramo:2020gpr}, the inclusion of ACT and SPT measurements has the largest impact on the ALP-$\tau$ coupling.
This is expected, since the improved sensitivity to $\Delta N_{\rm eff}$ provided by ground-based experiments allows to constrain light species produced at higher temperatures. Since ALP production from $\tau$ leptons occurs at the highest temperatures among the leptonic channels considered, the corresponding limit on $f_a$ benefits the most from the inclusion of ACT and SPT data.
A similar trend is observed when adding BBN information from helium and deuterium abundances. In this case, the limit on the ALP-$\tau$ coupling improves by approximately 32\%, while the constraints on the couplings to electrons, muons, and photons strengthen by about 8–9\%.

We also note that adding BAO measurements from DESI relaxes the limits on $\Delta N_{\rm eff}$, and therefore on $f_a$ and $g_{a\gamma}$. 
This trend is consistent with other recent analyses that combine CMB data with DESI BAO measurements to constrain $\Delta N_{\rm eff}$ (see e.g.\ refs.~\cite{AtacamaCosmologyTelescope:2025nti,SPT-3G:2025bzu,Elbers:2025vlz}). 
It can be understood from the DESI preference for lower values of $\Omega_m$ and higher values of $H_0$, which in turn allows for larger values of $\Delta N_{\rm eff}$.
For the specific case of light ALPs, a similar behaviour was also found in ref.~\cite{Caloni:2022uya} using earlier BAO measurements from BOSS DR12~\cite{BOSS:2016wmc}, 6dFGS~\cite{Beutler:2011hx} and SDSS-MGS~\cite{Ross:2014qpa}.

\begin{figure}[!t]
    \centering
    \includegraphics[width=0.6\linewidth]{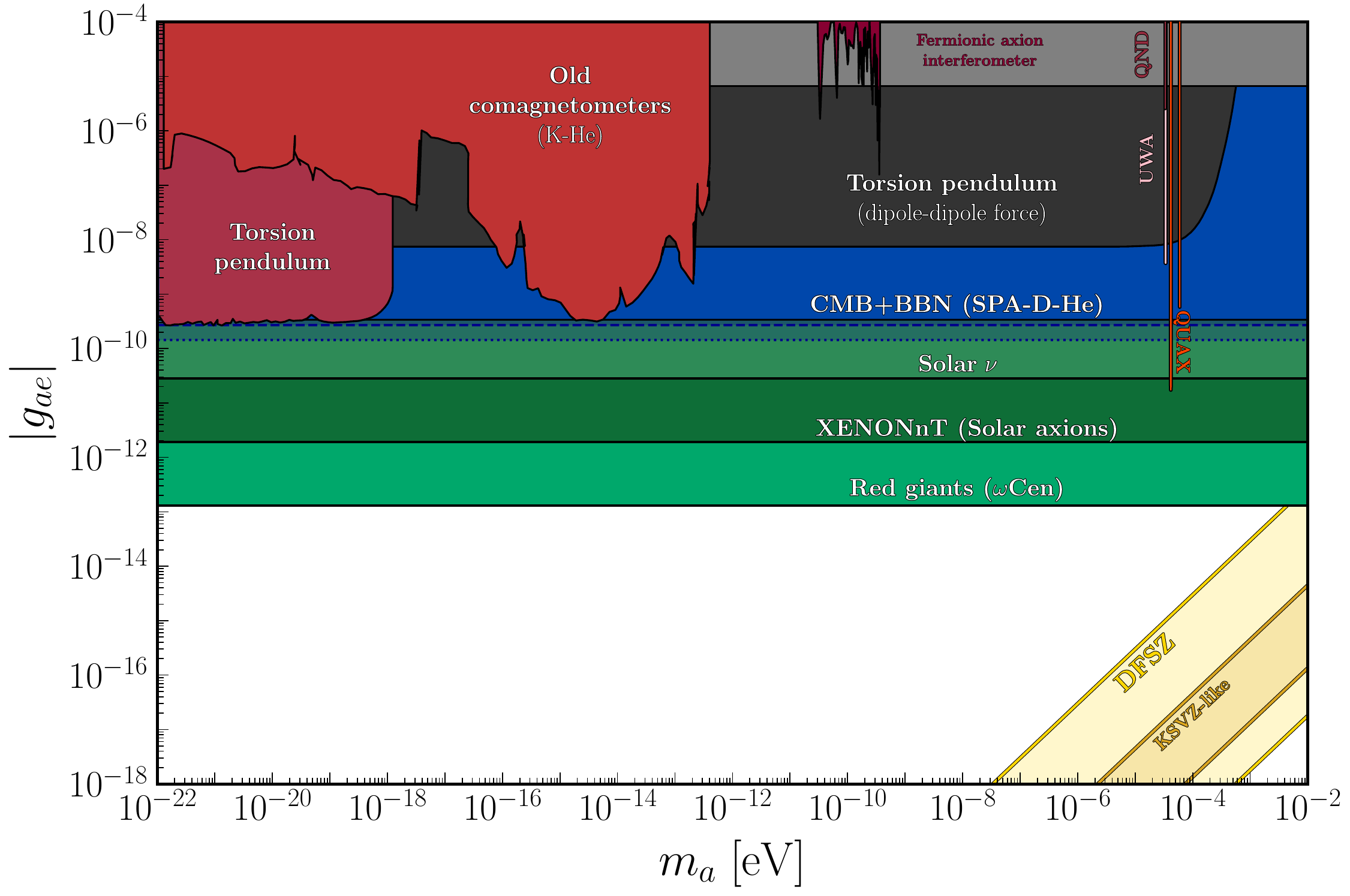} \\[15pt]
    \includegraphics[width=0.49\linewidth]{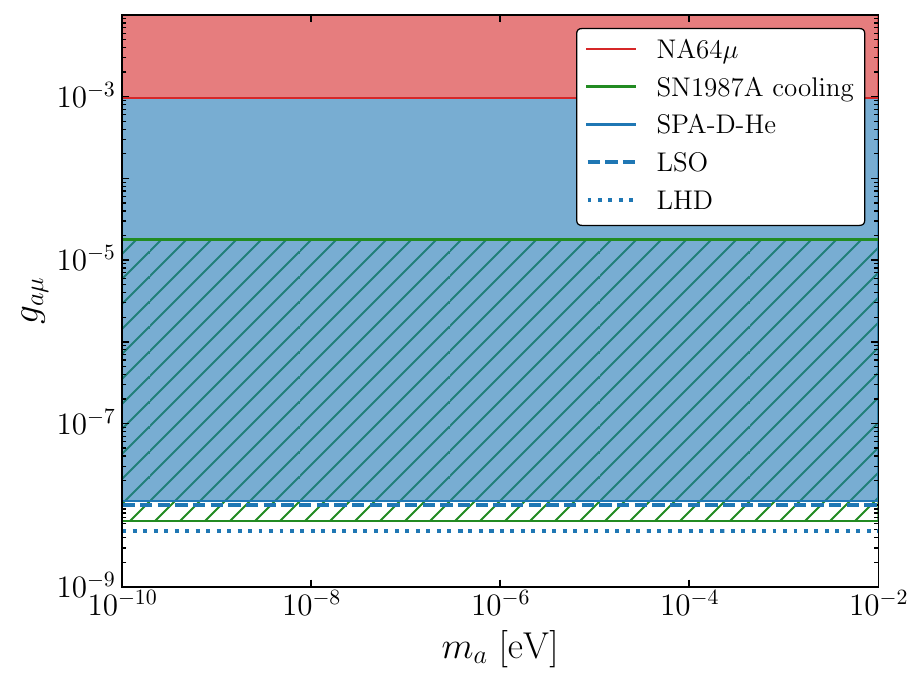}
    \includegraphics[width=0.49\linewidth]{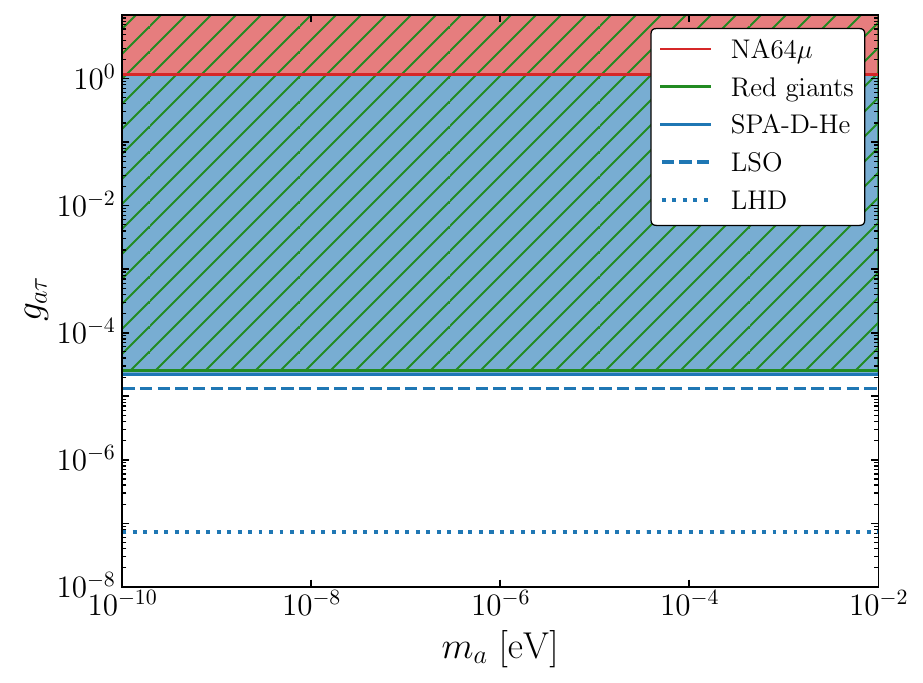}
    \caption{{\it Top}: Cosmological constraints on ALP-$e$ coupling $|g_{ae}|$. The solid blue contour shows the result obtained from our MCMC analysis using the SPA-D-He dataset. The dashed and dotted contours give the forecasts for the LSO and LHD configurations. For comparison, we include existing bounds from laboratory, astrophysical, and stellar probes. The figure is produced with the tools available with {\it AxionLimits}~\cite{AxionLimits}.
    {\it Bottom}: On the left, constraints on the ALP-$\mu$ coupling for light ALPs. In blue we show the bounds derived in this work from SPA-D-He (solid line) and our forecasts for LSO (dashed line) and LHD (dotted line). The red region corresponds to the parameter space excluded by the NA64$\mu$ experiment at CERN~\cite{NA64:2019auh}, as reported in ref.~\cite{Eberhart:2025lyu} (see also ref.~\cite{Li:2025yzb}). 
    The green region shows the limits from SN1987A cooling. These are taken from ref.~\cite{Ferreira:2025qui} and exploit the SFHo-18.8 model of ref.~\cite{Bollig:2020xdr} (see also ref.~\cite{Croon:2020lrf,Caputo:2021rux}). 
    On the right, constraints on the ALP-$\tau$ coupling for light ALPs. As in the left panel, the blue region denotes the bounds and forecasts derived in this work, while the red region shows the limit from NA64$\mu$~\cite{Eberhart:2025lyu}. The green region is excluded by the red giants limit on the loop-induced coupling to electrons~\cite{DEramo:2018vss,DEramo:2024jhn,Feng:1997tn}. Note that this limit is dependent on the UV details of the model, see ref.~\cite{DEramo:2018vss}.}
    \label{fig:constraints_axion_lepton}
\end{figure}

\begin{figure}[!t]
    \centering
    \includegraphics[width=0.6\linewidth]{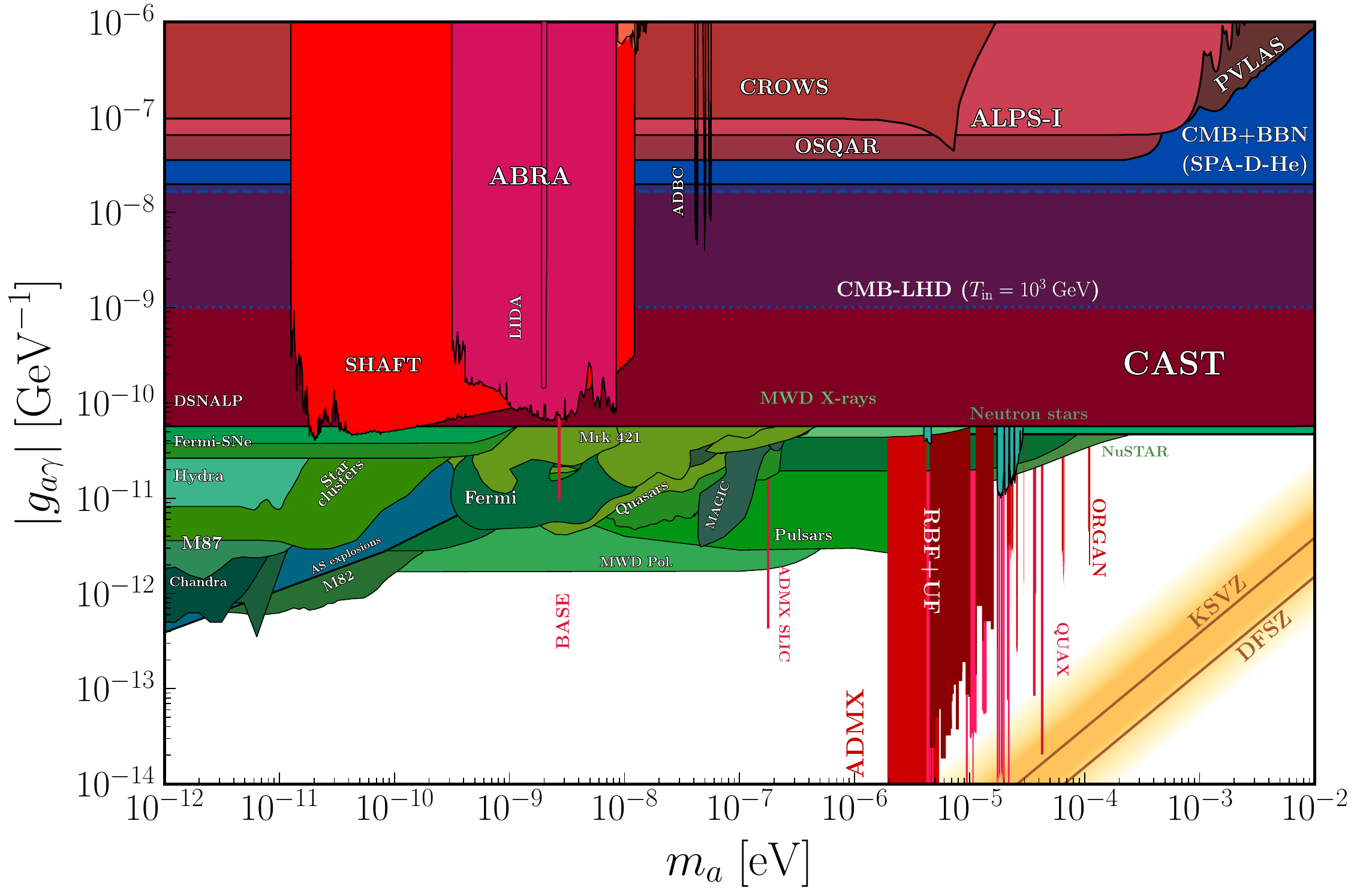}
    \caption{Cosmological constraints on ALP-photon coupling $g_{a\gamma}$. The solid blue contour shows the result obtained from our MCMC analysis using the SPA-D-He dataset. The dashed and dotted contours give the forecasts for the LSO and LHD configurations. For comparison, we include existing bounds from laboratory, astrophysical, and stellar probes. The figure is produced with the tools available with {\it AxionLimits}~\cite{AxionLimits}.}
    \label{fig:constraints_axion_photon}
\end{figure}

To facilitate a direct comparison with laboratory and astrophysical constraints, our results can equivalently be recast in terms of the couplings $g_{a\ell} = m_\ell c_\ell/f_a$. Figures~\ref{fig:constraints_axion_lepton}–\ref{fig:constraints_axion_photon} show the corresponding constraints alongside existing bounds, for ALP couplings to leptons and photons. While cosmological limits on the ALP couplings to electrons and photons are weaker than laboratory and astrophysical bounds, the constraints on the ALP-$\mu$ and ALP-$\tau$ couplings are already competitive with existing limits from other probes. This motivates a dedicated forecast study for future CMB experiments, which will be the subject of section~\ref{sec:forecasts}.
 
Beyond deriving state-of-the-art bounds on ALP couplings using the most recent cosmological data, a central aspect of our analysis is the use of the exact ALP phase-space distribution function.
To evaluate the effect on the inferred bounds, compared to the common approach of assuming a thermal (Bose-Einstein) distribution, we perform a test MCMC run in which the ALP distribution is instead assumed to be thermal.
We find that the difference between the bounds obtained with the exact PSD and those derived under the thermal assumption decreases as the mass of the lepton involved increases. This happens because electrons remain relativistic for a longer period, so that ALP production occurs at lower temperatures, where thermalization is more efficient. This leads to larger spectral distortions for the electron channel. Instead, for the $\mu$ and $\tau$ channels, the production is concentrated at higher temperatures, where distributions are closer to thermal.
A more detailed discussion of the impact of adopting the exact PSD, as opposed to a thermal one, is presented in section~\ref{sec:spectral_shape}.

\newpage

Finally, we stress that our main results are obtained by sampling on the effective number of relativistic species, $\Delta N_{\rm eff}$. As mentioned in section~\ref{sec:MCMC}, we also performed an alternative run for the electron channel in which $\log_{10} f_a$ is sampled directly. A detailed discussion of the impact of this prior choice is deferred to section~\ref{sec:KL}.

\subsection{Prior sensitivity analysis}
\label{sec:KL}
We derive constraints on the model parameters in a Bayesian framework. From Bayes' theorem~\cite{Bayes:1764vd}, the posterior probability $\calP\left(\bth|\bD\right)$ for the parameter vector, $\bth$, given the observed data, $\bD$, is written as
\begin{equation}
    \mathcal{P} \round{ \bth | \bD} = \frac{\mathcal{P} \round{\bD | \bth} \calP \round{\bth}}{\calP \round{\bD}} \; . 
\end{equation}
Credible one- and two- dimensional regions for subset of the parameters are derived from this posterior after marginalization.

In the context of parameter estimation, the one mainly relevant for this work, the evidence $\mathcal{Z} = \calP\left(\bD\right)$ appearing at the denominator is just a normalization constant, since it does not depend on $\bth$. The two main ingredients entering the analysis are thus the likelihood $\calL\left(\bth\right)\equiv \calP\left(\bD|\bth\right)$, i.e., the probability of the observed data as a function of the parameters, and the prior $\Pi\left(\bth\right)\equiv \calP\left(\bth\right)$, i.e.,  the a priori (before we observe the data) probability of the parameters. The choice of both the likelihood and the prior presents some degree of arbitrariness. The likelihood is based on the statistical modeling of the measurement process, which often comes with assumptions, e.g. about the properties of instrumental noise or systematics, or with approximations to make the model tractable in practice. We have described the likelihoods used in our analysis in \cref{sec:MCMC} and we refer the reader to the references in that section for details about how they are constructed. The prior should instead reflect our knowledge about the parameters before performing the experiment. This can be informed by previous observations, a process that mitigates the subjectivity of the prior, particularly as the accuracy of those observations increases. However, in the case of searches for new physics, previous information  is typically not available or very limited, due to the very nature on the problem, and several reasonable choices for the prior exist. In this situation, it is important to assess the robustness of the analysis with respect to different prior choices. This process is known, in the context of Bayesian data analysis, as \emph{prior sensitivity analysis}, and is the subject of this section.

We consider separable priors, i.e., priors that can be written as the product of the priors of the individual parameters. In other words, we take the parameters as a priori independent. Furthermore, given that current cosmological data are very informative on the six $\Lambda$CDM parameters, we do not explore different choices for the corresponding priors, and simply assume wide uniform priors on $\{\omega_b,\,\omega_c,\,\theta_s,\,\tau,\,\log(10^{10}A_s),\,n_s\}$. The focus of our sensitivity analysis will thus be the prior on the ALP decay constant $f_a$.

We consider two priors: a uniform prior on $\log_{10} f_a$, and a uniform prior of $\Delta N_\mathrm{eff}$ (see \cref{tab:priors} for the ranges). This choice of the uniform prior on $\log_{10} f_a$ is motivated by the fact that this prior assigns equal weight to each decade of the coupling parameter space. It might thus appear appropriate when testing BSM models with new, weakly constrained, couplings. However, this approach brings along some potential issues. Since models with very large values of $f_a$ (very weak couplings) yield no observable cosmological effects, the prior volume may become dominated by models that are essentially indistinguishable from $\Lambda$CDM, depending on the chosen range. A uniform prior on $\Delta N_\mathrm{eff}$ is instead chosen because this is the quantity most directly constrained by the data; specifically, observations are sensitive to $f_a$ primarily through its contribution to $\Delta N_\mathrm{eff}$. From a heuristic perspective, this makes it a motivated candidate for a prior that limits volume effects and maximizes the influence of the observations on our posterior inferences. We will expand further on this point at the end of the section.

    \begin{figure}
        \centering
        \includegraphics[scale=0.49]{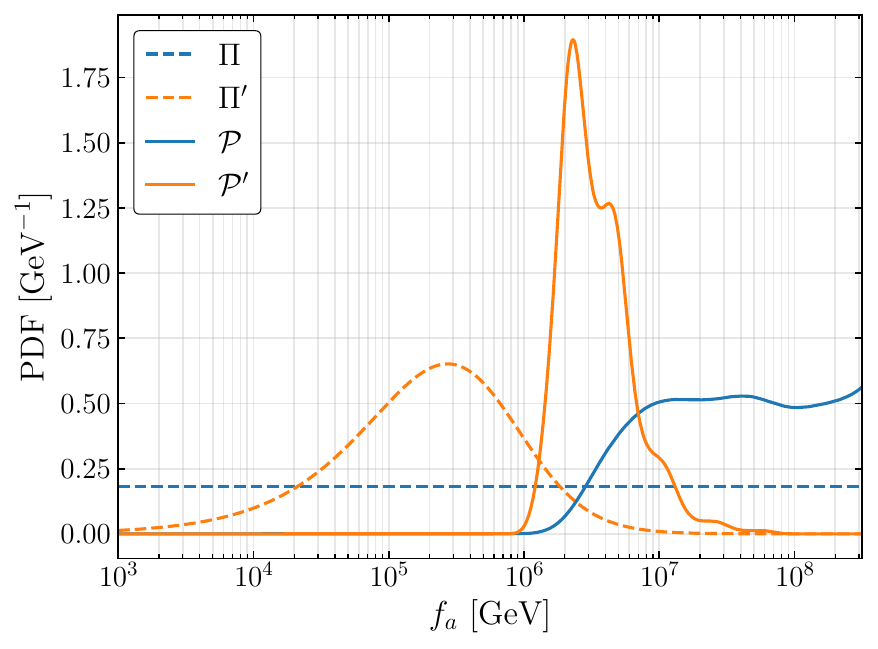}
        \includegraphics[scale=0.49]{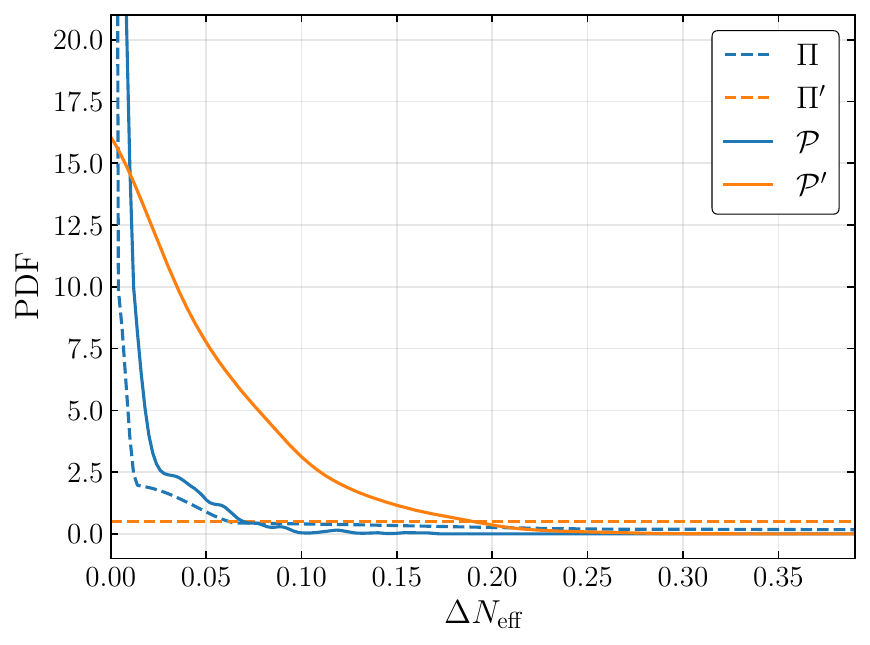}
        \caption{\textit{Left:} Prior and posterior distributions for $f_a$ corresponding to two different prior choices. The blue dashed and solid lines are the flat prior on $\log_{10} f_a$ and the corresponding posterior, respectively. In the same way, the orange dashed and solid lines are instead the non-uniform $\log_{10} f_a$ prior, induced by a flat $\Delta N_\text{eff}$ prior, and the corresponding posterior. All the PDFs are normalized and posteriors were obtained from runs with the SPA dataset. \textit{Right:} Same as if left panel but with all distributions expressed in terms of $\Delta N_\mathrm{eff}$.}
    \label{fig:kl_div}
    \end{figure}

    We visually compare the two priors in \cref{fig:kl_div} (dashed curves), in the case of ALPs coupling to electrons. To allow for the comparison, the priors have to be expressed in terms of the same variable.
    This is chosen to be $\log_{10} f_a$ in the left panel of the figure. We use $\Pi$ and $\Pi'$ to denote the flat prior on $\log_{10}f_a$ and $\Delta N_\mathrm{eff}$, respectively. We thus have:
    \begin{equation}   \label{eq:pr_logfa}
        \Pi(\log_{10}f_a) \propto 1 \, , \qquad \Pi'(\log_{10}f_a) \propto \left|\frac{\mathrm{d} \dneff}{\mathrm {d}\log_{10}f_a} \right| \, ,
    \end{equation}
    where in the identity on the right we have used the law of transformation of probabilities.
    When expressed in terms of $\log_{10}f_a$, the flat prior on $\Delta N_\mathrm{eff}$ assigns larger weight to the decade in $f_a$ between $ 10^{5}\,\GeV$ and $10^{6}\,\GeV$, and progressively smaller weight to values outside this range. The reason is that extreme values of $\log_{10} f_a$ map to very small regions around $\Delta N_\mathrm{eff}=0$ or $\Delta N_\mathrm{eff}=2.2$, for the higher and lower end of the prior range, respectively. Since the flat prior is weighting intervals in $\dneff$ only based on their length, this results in the shape shown in the figure.

This comparison might suggest that the uniform $\Delta N_\mathrm{eff}$ prior is somehow ``more informative'' because it assigns varying density across $\log_{10} f_a$. However, this is merely an artifact of the choice of coordinates. The situation is reversed in the right panel of \cref{fig:kl_div}, where we show both priors\footnote{With a slight abuse of notation, we are using here, for the pdf's of $\dneff$,  the same symbols used in Eqs.~\ref{eq:pr_logfa} for those of $\logfa$. The argument should make clear which one we are referring to.} expressed in terms of $\dneff$
\begin{equation}
        \Pi(\dneff) \propto \left|\frac{\mathrm {d}\log_{10}f_a} {\mathrm{d} \Delta N_\mathrm{eff}}\right| \, , \qquad \Pi'(\dneff) \propto 1 \, .
\end{equation}
In this representation, the uniform $\log_{10} f_a$ prior appears more informative, as it heavily weights values of $\Delta N_\mathrm{eff}$ near $0$. This illustrates that there is nothing inherently ``uninformative'' about a flat prior; since nonlinear transformations do not preserve the shape of a density, a distribution that is flat in one parameterization might well be peaked or skewed in another.

To move beyond visual comparisons, we adopt the Kullback-Leibler (KL) divergence~\cite{Kullback:1951zyt} as a metric for the information gain. This allows us to quantify the transition from prior to posterior and provides a robust measure of how the volume of the parameter space is compressed by the likelihood. The KL divergence, or relative entropy, of a distribution $P$ for the random variable $x$ from another distribution $Q$ for the same variable is defined as 
\begin{equation} \label{eq:kl_def_text}
        D_\mathrm{KL} \round{P  \parallel Q } = 
            \int {\rm d} x \, P \round{x} \log \sround{\frac{P \round{x}}{Q \round{x}}} \; .
\end{equation}
The KL divergence is an asymmetric measure of the difference between the two distributions, and is invariant under reparameterization. In a Bayesian context, the KL divergence of the posterior from the prior, $D_\mathrm{KL} \round{\calP  \parallel \Pi}$, quantifies the information gained after the data have been observed: more informative priors generally yield a smaller KL divergence. We compute this quantity\footnote{Note that here we are computing the KL divergence \emph{given the particular, observed realization of the data.}} for the two choices of prior discussed above, in the case of ALPs coupling to electrons. This yields:
\begin{align}
        D_\mathrm{KL} &= 0.94 \, \text{nat} \quad (\text{uniform $\log_{10} f_a$ prior}) \, , \\
        D_\mathrm{KL} &= 2.72 \, \text{nat} \quad (\text{uniform $\dneff$ prior}) \,  ,
\end{align}
where we have used the natural basis for the logarithm in eq.~\eqref{eq:kl_def_text}, so that the information gain is measured in ``nats''. These results clearly show that the uniform $\dneff$ prior is the less informative of the two, confirming the intuition that using the parameter most directly affecting the observations tends to minimize volume effects. 
    
Further insight can be gained re-examining \cref{fig:kl_div}, focusing on the shift from each prior (dashed) to the corresponding posterior (solid). In the left panel, the likelihood redistributes the probability mass initially assigned to the region $f_a \lesssim 10^6\,\GeV$ toward higher values of the decay constant. While this shift occurs for both priors, the flat $\Delta N_\mathrm{eff}$ prior initially assigned significantly more weight to this region. Consequently, the likelihood must move a larger fraction of the total probability mass, resulting in a higher information gain. Due to the reparameterization invariance of relative entropy, the same conclusion is reached by inspecting the right panel.

Based on the information-theoretic results presented in this section, we adopt the flat $\Delta N_{\mathrm{eff}}$ prior as our baseline. Unless stated otherwise, all results throughout this work have been derived using this prior.


\section{Forecasts for future CMB surveys}
\label{sec:forecasts}

In this section, we assess the expected sensitivity of forthcoming CMB surveys to ALP couplings with leptons and photons.
As described below, we model the instrumental noise of each experiment using the publicly available specifications or noise curves provided by the corresponding collaboration.

\subsection{Noise modeling and simulated data}
\label{sec:noise_forecasts}
The instrumental sensitivity of each experiment is characterized through its angular power-spectrum noise.
When noise curves are not publicly available, we compute them from the reported instrumental specifications, assuming white noise with a Gaussian beam (see e.g.\ refs.~\cite{Perotto:2006rj,Wu:2014hta,Gerbino:2019okg}):
\begin{equation}
    \label{eq:noise}
    N_\ell^{XX} = \sigma_X^2 \exp\left[ \frac{\ell(\ell+1) \theta_{\rm FWHM}^2}{8 \log 2} \right] \, ,  
\end{equation}
with $X \in \{ T,E \}$. Here, $\sigma_X$ denotes the sensitivity in units of $\mu$K-arcmin (with $\sigma_{E} = \sqrt{2} \sigma_T$) and $\theta_{\rm FWHM}$ is the full-width half-maximum beam size in radians. We assume that temperature and polarization noises are uncorrelated, so that $N_\ell^{TE} = 0$. 

Simulated data are modeled as the sum of fiducial CMB spectra and the experimental noise contributions:
\begin{equation}
    \widehat{C}_\ell^{\,XX} = C_\ell^{XX, \, \mathrm{fid}} + N_\ell^{XX} \,,
    \label{eq:mock_data}
\end{equation}
where the fiducial spectra are computed assuming a $\Lambda$CDM model with $\Delta N_{\rm eff}=0$ and cosmological parameters fixed to the mean values from the CMB-SPA analysis (see \cref{tab:comparison}).
We explore the following combinations of future CMB datasets:
\begin{itemize}
    \item {\it LiteBIRD}+SO: we combine the noise information from the large-aperture telescope (LAT) of the Simons Observatory (SO) and the {\it LiteBIRD} satellite. For SO LAT, we use the publicly available noise curves provided by the collaboration at the goal noise level~\cite{SimonsObservatory:2018koc}.\footnote{\url{https://github.com/simonsobs/so_noise_models}} For {\it LiteBIRD}, we coadd the noise power spectra using eq.~\eqref{eq:noise}, with sensitivities and beam widths of the individual frequency channels from ref.~\cite{LiteBIRD:2022cnt}, and assuming that the rotating half-wave plate will effectively mitigate the $1/f$ noise at large angular scales~\cite{LiteBIRD:2022cnt,Micheli:2024hfe}.
    We combine these datasets as follows.
    At large angular scales ($2 \le \ell \le 50$), we use the {\it LiteBIRD} noise curves for temperature and $E$-mode polarization, over a sky fraction $f_{\rm sky}=0.7$.
    For $51 \le \ell \le 3000$, we use the SO noise curves over the common sky fraction observed by both experiments ($f_{\rm sky}=0.4$), whereas the additional {\it LiteBIRD}-only sky coverage ($f_{\rm sky}=0.3$) is included up to $\ell=1000$.\footnote{The SO noise curves account for extra variance coming from removal of extragalactic foregrounds that dominate the sky signal at small angular scales. We safely neglect this contribution in the \textit{LiteBIRD} noise curves given the \textit{LiteBIRD} resolution and the conservative choice for the maximum multipole retained in the likelihood analysis}
    We also include the SO lensing reconstruction with noise provided by the collaboration~\cite{SimonsObservatory:2018koc}.
    In the following, we refer to this configuration as LSO.
    \item {\it LiteBIRD}+CMB-HD: to assess the ultimate sensitivity achievable with CMB observations, we combine {\it LiteBIRD} with the futuristic ground-based CMB-HD survey~\cite{CMB-HD:2022bsz, MacInnis:2023vif, MacInnis:2024znd}, designed to achieve ultra-deep sensitivity and unprecedented angular resolution. We include \textit{LiteBIRD} on large angular scales ($2 \le \ell \le 50$) over a sky fraction of $f_{\rm sky} = 0.7$. On smaller angular scales ($51 \le \ell \le 5000$), we adopt the publicly available noise curves for CMB-HD~\cite{MacInnis:2024znd},\footnote{\url{https://github.com/CMB-HD/hdMockData.git}} considering the sky fraction common to both experiments ($f_{\rm sky} = 0.6$). The noise curves already include the effect of residual foreground contamination, as described in ref.~\cite{MacInnis:2024znd}. The additional sky coverage observed exclusively by \textit{LiteBIRD} ($f_{\rm sky} = 0.1$) is included up to $\ell = 1000$, following the same approach adopted for the LSO case. As for LSO configuration, we include the CMB-HD lensing reconstruction with noise provided by the collaboration.
\end{itemize}
We perform the MCMC analysis with \texttt{Cobaya}, following the methodology outlined in section~\ref{sec:MCMC}. We adopt the inverse Wishart likelihood~\cite{Gerbino:2019okg} as implemented in ref.~\cite{Rashkovetskyi:2021rwg}.\footnote{\url{https://github.com/misharash/cobaya_mock_cmb}}
    
\subsection{Results and discussion}
The forecast results are summarized in \cref{tab:DNeff_constraints}, which reports the projected constraints on the ALP couplings (last two columns). The complete set of parameter constraints and the corresponding triangle plots are presented in appendix~\ref{app:triangle_plots}.
For the LSO configuration, we find a projected sensitivity to extra radiation at the level of $\Delta N_{\rm eff} \simeq 0.1$ (95\% CL), consistent with previous forecasts for the combination of SO with \textit{Planck}~\cite{SimonsObservatory:2018koc}. This is not surprising, as this sensitivity is mainly driven by the high-resolution measurements of temperature and polarization at small angular scales provided by SO.
The improvement in the bounds on the ALP couplings is particularly pronounced for the $\tau$ channel, where the bound on $f_a$ strengthens by a factor of about 1.7 with respect to SPA-D-He. 
This follows from the same physical mechanism responsible for the improvement observed when including ACT and SPT data over {\it Planck} alone: ALP production in the $\tau$ channel occurs at higher temperatures compared to the other leptonic channels, and therefore benefits more from the increased sensitivity to $\Delta N_{\rm eff}$.
For the $e$ and $\mu$ channels, as well as for Primakoff production, the improvement is more modest and lies around 15\%.

For the LHD configuration, we find a projected sensitivity of $\Delta N_{\rm eff} \simeq 0.03$ (95\% CL), in agreement with the analyses of refs.~\cite{CMB-HD:2022bsz,MacInnis:2023vif}. This leads to a dramatic improvement for the $\tau$ channel, with the limit on $f_a$ tightening by more than two orders of magnitude. For the $e$ and $\mu$ channels, the bounds on $f_a$ strengthens by factors of 2.2 and 2.3, respectively.
A substantial improvement is also observed for the Primakoff case, where the bound on $g_{a\gamma}$ tightens by more than one order of magnitude with respect to the limits obtained with current data. Note that the projected sensitivity for this channel is strongly dependent on the choice of the initial temperature $T_{\rm in}$, since the Primakoff production is probed in the UV freeze-in regime (see \cref{fig:DNeff_fermion_primakoff}).
The bound reported in \cref{tab:DNeff_constraints} is obtained for $T_{\rm in} = 10^3 \, {\rm GeV}$, while for larger values of $T_{\rm in}$ the corresponding constraint on $g_{a\gamma}$ would become even stronger, see ref.~\cite{Caloni:2024olo}. A systematic exploration of this dependence goes beyond the scope of this paper and is left for future work.

The bounds on ALP couplings to leptons can equivalently be recast in terms of the effective couplings $g_{a\ell} = m_\ell c_\ell / f_a$. The corresponding forecast constraints on both leptonic and photon couplings are shown in figures~\ref{fig:constraints_axion_lepton} and~\ref{fig:constraints_axion_photon}, together with existing laboratory, astrophysical and cosmological bounds. As for current data, future constraints are weaker than existing astrophysical bounds for the ALP-photon coupling and in the electron channel, and competitive with SN constraints in the muonic channel. The largest improvement in the tauonic channel would result in much stringent (up to more than two orders of magnitude for LHD) constraints on the ALP-$\tau$ coupling than those from SN. 

Our phase-space treatment establishes a robust framework for accurately interpreting future CMB measurements. In the following we proceed to a detailed comparison with simplified treatments that assume purely thermal spectra.

\subsection{Accuracy of the thermal approximation} 
    \label{sec:spectral_shape}
    We note that the constraints on $\dneff$ reported above are remarkably stable, for a given dataset, across the different production channels. This is because observations are directly sensitive to $N_\mathrm{eff}$ at leading order, and secondarily to the exact shape of the distribution function. Since the actual distribution functions are fairly close to thermal, this introduces small differences in the constraints on $\dneff$. On the other hand, the relation between $f_a$ and $\dneff$ strongly depends on the production channel, so the corresponding limits on $f_a$ can differ by orders of magnitude. A promising strategy to obtain constraints on $f_a$ without the need to compute the exact distribution function and to modify Boltzmann codes would then be to obtain constraints on $\dneff$ assuming a Bose-Einstein distribution for ALPs, and the to express those results in terms of $f_a$ by using the relevant $\dneff$ vs $f_a$ relation. In this section we assess the error introduced by approximating a non-thermal distribution function, computed as detailed in the previous section, with a thermal distribution function with the same $\Delta N_\mathrm{eff}$.

    We focus on $95\%$ credible intervals, and compute the empirical Monte Carlo uncertainty on the upper edge of the interval. To this purpose, we use the subsampling boostrap technique, following the implementation\footnote{This is based on taking many contiguous and overlapping subsamples of a Markov chain, and computing the statistic of interest (the 95\% upper limit in our case) on each subsample. The variability across all the subsamples is used to estimate the asymptotic variance governing the Monte Carlo fluctuations.} of the \texttt{mcmcse} code~\cite{mcmcse}. 
    
    For each dataset, we then consider the set of differences
    \begin{equation}
        \xi_i \equiv \frac{\dneff^{95,\mathrm{th}} - \dneff^{95,i}}{\dneff^{95,i}} \, ,
    \end{equation}
    where $i=\{e,\,\mu,\,\tau,\,\gamma\}$, and $\dneff^{95,i}$ and $\dneff^{95,\mathrm{th}}$ are the 95\% upper limits for production channel $i$ and for a Bose-Einstein distribution, respectively, as estimated from our chains. To each $\xi_i$ we attach an uncertainty $\sigma^i_\text{MC}$ given by propagating the Monte Carlo standard errors on $\dneff^{95,i}$ and $\dneff^{95,\mathrm{th}}$, estimated with the subsampling bootstrap method. We then use the $\xi_i \pm 3\sigma^i_\text{MC}$ interval to bound the systematic error introduced by the thermal approximation. This bound is conservative and inherently tied to the precision of our MCMC estimates; it could, in principle, be further tightened by increasing the number of independent samples in our chains.
    
    We report our results in \cref{tab:mc_error}, one for each dataset. In each table, we list the values of $\xi_i$, $\sigma^i_\text{MC}$, along with the $\pm 3\sigma$ interval. The numbers listed in the table can be used to assess whether, given a desired level of accuracy on $\dneff^{95}$, it is safe to approximate the distribution function with a Bose-Einstein. This is particularly relevant for future data. To give a concrete example based on our Monte Carlo uncertainties in the LSO chains, and focusing on the ALP–$\tau$ coupling, the exact phase-space distribution can be safely replaced by its thermal approximation if one is willing to tolerate a relative error of at most $\sim0.1$ (at 99\% CL); otherwise, the full distribution must be retained. 
    Finally, we emphasize that any such comparison must account for the Monte Carlo error of the specific chains being analyzed. The thermal approximation is justified if: (i) the systematic shifts $\xi_i$ are smaller than the required accuracy, and (ii) the chains are sufficiently converged such that their associated Monte Carlo error is significantly smaller than the residuals quantified here.

\begin{table}[!t]
    \centering

    \begin{minipage}{0.48\textwidth}
        \centering
        \textbf{\footnotesize SPA-D-He}\\[0.6mm]
        \renewcommand{\arraystretch}{1.3}
        \resizebox{\textwidth}{!}{%
        \begin{tabular}{cccc}
        \hline\hline
        \noalign{\vskip 1mm}
        \textbf{Model} 
        & $\xi_i$ 
        & $\sigma_\mathrm{MC}^i$ 
        & $[\xi_i-3\sigma_\mathrm{MC}^i,\,\xi_i+3\sigma_\mathrm{MC}^i]$ \\
        \noalign{\vskip 1mm}
        \hline
        \noalign{\vskip 1mm}
        $e$      & $-0.026$ & 0.043 & $[-0.103, \,0.154]$ \\
        $\mu$    & 0.019 & 0.049 & $[-0.127, \,0.165]$ \\
        $\tau$   & 0.053 & 0.047 & $[-0.087, \,0.192]$ \\
        $\gamma$ & 0.032 & 0.043 & $[-0.097, \,0.161]$ \\
        \hline\hline
        \end{tabular}
        }
    \end{minipage}
    \hfill
    \begin{minipage}{0.48\textwidth}
        \centering
        \textbf{\footnotesize SPA-D-He-B}\\[0.6mm]
        \renewcommand{\arraystretch}{1.3}
        \resizebox{\textwidth}{!}{%
        \begin{tabular}{cccc}
        \hline\hline
        \noalign{\vskip 1mm}
        \textbf{Model} 
        & $\xi_i$ 
        & $\sigma_\mathrm{MC}^i$ 
        & $[\xi_i-3\sigma_\mathrm{MC}^i,\,\xi_i+3\sigma_\mathrm{MC}^i]$ \\
        \noalign{\vskip 1mm}
        \hline
        \noalign{\vskip 1mm}
        $e$      &  0.004 & 0.025 & $[-0.070,\,0.078]$ \\
        $\mu$    & $-0.002$ & 0.027 & $[-0.083,\,0.080]$ \\
        $\tau$   &  0.000 & 0.027 & $[-0.080,\,0.080]$ \\
        $\gamma$ & $-0.021$ & 0.028 & $[-0.106,\,0.064]$ \\
        \hline\hline
        \end{tabular}
        }
    \end{minipage}

    \vspace{4mm}

    \begin{minipage}{0.48\textwidth}
        \centering
        \textbf{\footnotesize LSO}\\[0.6mm]
        \renewcommand{\arraystretch}{1.3}
        \resizebox{\textwidth}{!}{%
        \begin{tabular}{cccc}
        \hline\hline
        \noalign{\vskip 1mm}
        \textbf{Model} 
        & $\xi_i$ 
        & $\sigma_\mathrm{MC}^i$ 
        & $[\xi_i-3\sigma_\mathrm{MC}^i,\,\xi_i+3\sigma_\mathrm{MC}^i]$ \\
        \noalign{\vskip 1mm}
        \hline
        \noalign{\vskip 1mm}
        $e$      &  0.018 & 0.027 & $[-0.061,\,0.098]$ \\
        $\mu$    & $-0.025$ & 0.025 & $[-0.098,\,0.049]$ \\
        $\tau$   & $-0.025$ & 0.028 & $[-0.109,\,0.058]$ \\
        $\gamma$ & $-0.007$ & 0.030 & $[-0.096,\,0.082]$ \\
        \hline\hline
        \end{tabular}
        }
    \end{minipage}
    \hfill
    \begin{minipage}{0.48\textwidth}
        \centering
        \textbf{\footnotesize LHD}\\[0.6mm]
        \renewcommand{\arraystretch}{1.3}
        \resizebox{\textwidth}{!}{%
        \begin{tabular}{cccc}
        \hline\hline
        \noalign{\vskip 1mm}
        \textbf{Model} 
        & $\xi_i$ 
        & $\sigma_\mathrm{MC}^i$ 
        & $[\xi_i-3\sigma_\mathrm{MC}^i,\,\xi_i+3\sigma_\mathrm{MC}^i]$ \\
        \noalign{\vskip 1mm}
        \hline
        \noalign{\vskip 1mm}
        $e$      & $-0.022$ & 0.029 & $[-0.110,\,0.065]$ \\
        $\mu$    & $-0.021$ & 0.033 & $[-0.118,\,0.077]$ \\
        $\tau$   &  0.026 & 0.034 & $[-0.075,\,0.126]$ \\
        $\gamma$ & $-0.023$ & 0.027 & $[-0.103,\,0.057]$ \\
        \hline\hline
        \end{tabular}
        }
    \end{minipage}

    \caption{Relative deviations $\xi_i \equiv \left( \Delta N_{\rm eff}^{\rm th.} - \Delta N_{\rm eff}^i \right) / \Delta N_{\rm eff}^i $, corresponding Monte Carlo uncertainties $\sigma_\mathrm{MC}^i$, and $3\sigma$ intervals for all data combinations considered: SPA-D-He, SPA-B-D-He, LSO, and LHD.}
    \label{tab:mc_error}
\end{table}


\section{Conclusions}
\label{sec:conclusions}

In this work we have derived updated cosmological bounds on light ALPs coupled to leptons and photons, using state-of-the-art cosmological data. 
From a methodological perspective, the main novelty of our analysis lies in the implementation of the exact non-thermal ALP phase-space distribution, obtained by solving the momentum-dependent Boltzmann equation for each production channel. This distribution is then consistently propagated into cosmological observables using the Boltzmann solver \texttt{CLASS} and incorporated into a full MCMC analysis performed with \texttt{Cobaya} to derive constraints from cosmological data.
This work provides the first constraints on ALP couplings from the combination of \textit{Planck}~\cite{Planck:2018vyg}, ACT~\cite{AtacamaCosmologyTelescope:2025nti} and SPT~\cite{SPT-3G:2025bzu} data from CMB measurements, further complemented with BBN determinations of the primordial deuterium and helium abundances and with BAO measurements from DESI DR2~\cite{DESI:2025zgx}.

For ALP coupled to leptons, we obtain the following 95\% CL on the ALP decay constant from the CMB+BBN combination: $f_a > 1.63 \times 10^6 \, {\rm GeV}$, $9.41 \times 10^6 \, {\rm GeV}$ and $8.06 \times 10^4 \, {\rm GeV}$ for couplings to electrons, muons and taus, respectively. For the ALP-photon coupling, we find $g_{a\gamma} < 1.98 \times 10^{-8} \, {\rm GeV}^{-1}$.
When BAO measurements from DESI are included, these constraints are mildly relaxed, reflecting the DESI preference for lower values of $\Omega_m$ and higher values of $H_0$.
Compared to astrophysical and laboratory limits, the cosmological bounds derived in this work are particularly competitive for the ALP couplings to $\mu$ and $\tau$.

We further extend our analysis by presenting forecasts for next-generation CMB surveys. We consider two experimental configurations.
The first combines the space-based mission {\it LiteBIRD}~\cite{LiteBIRD:2022cnt} with the Simons Observatory~\cite{SimonsObservatory:2018koc}, representing a realistic benchmark for forthcoming CMB observations. The second pairs {\it LiteBIRD} with the futuristic CMB-HD experiment~\cite{CMB-HD:2022bsz}, and is intended as a proxy to assess the ultimate sensitivity to ALP couplings achievable with CMB observations.
For the {\it LiteBIRD}+SO configuration, the projected bound on $\Delta N_\mathrm{eff}$ at 95\% CL is $\simeq 0.01$, corresponding to an improvement in the couplings of a factor $\simeq 1.7$ for the $\tau$ channel and by 12-17\% for the $e$, $\mu$, and photon cases, relative to the current CMB+BBN constraints.
Instead, for the {\it LiteBIRD}+CMB-HD configuration we find $\Delta N_\mathrm{eff} \simeq 0.03$ (95\% CL). The improvement is dramatic for the $\tau$ coupling, with the limit on $f_a$ tightening by more than two orders of magnitude, while for the $e$ and $\mu$ channels the bounds strengthen by factors larger than two. In the Primakoff case, the projected constraint on $g_{a\gamma}$ improves by more than one order of magnitude. This result is obtained for $T_{\rm in}=10^3\,\mathrm{GeV}$, while larger initial temperatures would lead to significantly stronger bounds, due to the UV-dominated nature of the Primakoff production.

We performed a prior sensitivity analysis showing that sampling over $\Delta N_\mathrm{eff}$, rather than the more intuitive parameter $\log_{10} f_a$, enables a more efficient exploration of the parameter space and results in more robust and reliable constraints. We further demonstrated that this prior choice is significantly less informative than the alternative, ensuring that the constraints are driven predominantly by the data rather than by prior assumptions.
Moreover, we performed a direct comparison between results obtained using the exact non-thermal ALP phase-space distribution and those derived under the assumption of a purely thermal spectrum. We found that non-thermal spectral distortions can induce shifts in $\Delta N_{\rm eff}$, with the magnitude of the effect depending on the specific production channel. The relevance of these differences for current and next-generation CMB experiments depends critically on the level of accuracy required in the analysis of $\Delta N_{\rm eff}$. To this end, we provided a practical prescription to rapidly assess the impact based on the numerical results of our study.

Our framework is fundamental in providing a robust treatment of ALP cosmology in view of forthcoming high-precision CMB measurements, ensuring that theoretical uncertainties associated with the ALP momentum distribution do not bias constraints and remain under control.

\vfill
\noindent{\bf Note added}
\\
While this paper was being finalized, a related analysis appeared  in ref.~\cite{Badziak:2025mkt}, which also derives cosmological bounds on ALP-lepton couplings using the exact ALP phase-space distribution.
Our analysis, however, differs in several important aspects.
First, ref.~\cite{Badziak:2025mkt} focuses on {\it Planck}-only data for the CMB, whereas our analysis incorporates the most recent data from ACT and SPT, as well as BBN determinations of the helium and deuterium abundances.
The enhanced sensitivity to $\Delta N_{\rm eff}$ provided by these datasets is particularly relevant for deriving substantially stronger constraints on ALP-$\tau$ couplings for light ALPs.
Second, our work also presents updated bounds on ALP-photon interactions based on a full phase-space treatment of Primakoff production, a channel not explored in ref.~\cite{Badziak:2025mkt}.
We also provide forecasts for the reach of upcoming CMB surveys.
Finally, we perform a dedicated analysis of the impact of different prior choices on the inferred cosmological bounds, showing that sampling directly in $\Delta N_{\rm eff}$, compared to $\log_{10}f_a$, corresponds to a less informative prior and results in more robust bounds on ALP couplings.

\acknowledgments
We thank Ricardo Zambujal Ferreira, Serena Giardiello and Ali Rida Khalife for useful discussions. 
L.C.\ acknowledges the financial support provided through national funds by FCT -- Funda\c{c}\~ao para a Ci\^encia e Tecnologia, I.P., with DOI identifiers 10.54499/2023.11681.PEX, 10.54499/UID/04564/2025 and by the project 2024.00249.CERN funded by measure RE-C06-i06.m02 -- ``Reinforcement of funding for International Partnerships in Science, Technology and Innovation'' of the Recovery and Resilience Plan -- RRP, within the framework of the financing contract signed between the Recover Portugal Mission Structure (EMRP) and the Foundation for Science and Technology I.P.\ (FCT), as an intermediate beneficiary. N.B.\ is supported by the Spanish grants PID2023-147306NB-I00 and CEX2023-001292-S (MCIU/AEI/10.13039/501100011033). N.B., M.G. and M.L.\ are funded by the European Union (ERC, RELiCS, project number 101116027). Views and opinions expressed are however those of the authors only and do not necessarily reflect those of the European Union or the European Research Council Executive Agency. Neither the European Union nor the granting authority can be held responsible for them.
L.V.\ acknowledges support by Istituto Nazionale di Fisica Nucleare (INFN) through the Commissione Scientifica Nazionale 4 (CSN4) Iniziativa Specifica ``Quantum Universe'' (QGSKY). This publication is based upon work from the COST Action ``COSMIC WISPers'' (CA21106), supported by COST (European Cooperation in Science and Technology). This is not an official SO paper.

\appendix

\section{General \texorpdfstring{$2 \rightarrow 2$}{2to2} collision integral}
\label{app:collision-integral}

The most general collision integral for binary scatterings can be written as 
\begin{align}
    &\mathbb{C} \sround{f_1, f_2, f_3} = \notag \\
    & \quad \, = \frac{1}{2 E_k} \int \, {\rm d}\Pi_1 {\rm d}\Pi_2 {\rm d}\Pi_3 \round[4]{2 \pi} \delta^{\round{4}} \round{P_1 + P_2 - P_3 - K} \abs{\mathcal{M}_{12\rightarrow3X} \round{s,t,u} }^2 f_1 f_2 \round{1 \mp f_3} \; .
\end{align}
This has dimensions of inverse time and gives a momentum-dependent production rate density, which reduces to the usual total production rate per unit volume when integrating over momentum.
The phase space measure is 
\begin{equation} \label{eq:phase_spase_measure}
    {\rm d}\Pi_i \equiv \frac{g_i}{\round[3]{2\pi}} \frac{{\rm d}^3 \mathbf{p}_i}{2E_{p_i}} \; .
\end{equation}
The first step of our reduction procedure is the integration of the $3-$momentum $p_3$, i.e. the momentum of the outgoing particle of which we are not studying the evolution. Thanks to the properties of the Dirac's delta function, one can easily prove that 
\begin{equation}
    \int \frac{{\rm d}^3 \mathbf{p}_3}{2E_{p_3}} \, \delta^{\round{4}} \round{P_1 + P_2 - P_3 - K} = \delta \round{\sround{ E_{p_1} + E_{p_2} - E_{k} }^2 - E^2_{p_1+p_2-k} } \Theta \round{ E_{p_1} + E_{p_2} - E_{k} } \; .
    \end{equation}
Our collision integral then becomes 
\begin{align}
    \mathbb{C} \sround{f_1, f_2, f_3} &= \frac{1}{2 E_k} \frac{g_1g_2g_3}{\round[5]{2\pi}} \int \frac{{\rm d}^3 p_1}{2E_1} \frac{{\rm d}^3 p_2}{2E_2} \, \abs{\mathcal{M}_{12\rightarrow3X} \round{s,t,u} }^2 f_1 f_2 \round{1 \mp f_3} \notag \\
    & \quad \, \times \delta \round{\sround{ E_{p_1} + E_{p_2} - E_{k} }^2 - E^2_{p_1+p_2-k} } \Theta \round{ E_{p_1} + E_{p_2} - E_{k} } \; . \label{eq:coll_step2}
\end{align}
For the remaining $3-$momenta exploit the spherical coordinates parameterization in a frame with the $z$ axis aligned with our momentum $k$. Thanks to the symmetry of the system we are also allowed to rotate our frame to have one of our momenta, let's say $\mathbf{p}_1$ laying on the $xz$ plain. We thus end up with the following parameterization 
\begin{align}
    \mathbf{k} &=  k \round{ 0, 0, 1 } \; , \notag \\
    \mathbf{p}_1 &=  p_1 \round{ \sin\theta_1, 0, \cos \theta_1 } \; , \notag \\
    \mathbf{p}_2 &=  p_2 \round{ \sin\theta_2 \cos \phi_2, \sin \theta_2 \sin\phi_2, \cos \theta_2 } \; ;
\end{align}
where $p_1, p_2 \in [ 0, +\infty )$, $\theta_1, \theta_2 \in [ 0, \pi )$ and $\phi \in [0, 2\pi)$. Integration measures can be written now as 
\begin{equation}
    {\rm d}^3 \mathbf{p}_1 = p_1^2 {\rm d}p_1 {\rm d}c_1 {\rm d}\Phi \; , \qquad {\rm d}^3 \mathbf{p}_2 = p_2^2 {\rm d}p_2 {\rm d}c_2 {\rm d}\phi \; .
    \end{equation}
    In particular, we notice that 
    \begin{equation}
        \frac{\mathbf{p}_1 \cdot \mathbf{k}}{p_1 k} \equiv \cos \theta_1 \equiv c_1 \; ,  \quad \frac{\mathbf{p}_2 \cdot \mathbf{k}}{p_2 k} \equiv \cos \theta_2 \equiv c_2 \; , \quad \frac{\mathbf{p}_1 \cdot \mathbf{p}_2}{p_1 p_2} = c_1 c_2 + s_1 s_2 \cos \round{\phi} \; .
    \end{equation}
    The integration over $\Phi$ is trivial and gives an overall $2\pi$ factor, leaving our collision integral to be 
    \begin{align}
        \mathbb{C} \sround{f_1, f_2, f_3} &= \frac{1}{2 E_k} \frac{g_1g_2g_3}{\round[4]{2\pi}} \int \frac{p_1^2 {\rm d} p_1}{2E_1} \frac{p_1^3 {\rm d} p_2}{2E_2} {\rm d}c_1 {\rm d}c_2 {\rm d}\phi \, \abs{\mathcal{M}_{12\rightarrow3X} \round{s,t,u} }^2 f_1 f_2 \round{1 \mp f_3} \; . \notag \\
        & \quad \, \times \delta \round{\sround{ E_{p_1} + E_{p_2} - E_{k} }^2 - E^2_{p_1+p_2-k} } \Theta \round{ E_{p_1} + E_{p_2} - E_{k} } \; .
        \label{eq:coll_step3}
    \end{align}
    
    It is possible to carry out another analytical integration, still leaving the integral completely general, and applicable to any $2\rightarrow 2$ scattering. For the $\phi$ integral, we use the well-known identity of the Dirac's delta function 
    \begin{equation}
        \int_0^{2\pi} {\rm d}\phi \, \delta \round{g\round{\phi}} F \round{\phi} = \int_0^{2\pi} {\rm d}\phi \sum_{i=1}^{g \text{ roots}} \abs{\frac{\partial g}{\partial \phi}}^{-1} F\round{\phi} \delta \round{\phi-\phi_i} = \sum_{i=1}^{g \text{ roots}} \abs{\frac{\partial g}{\partial \phi}}^{-1}_{\phi_i} F\round{\phi_i} \; . \label{eq:delta_integral}
    \end{equation}
    In our specific case we have to find the roots of the function 
    \begin{align}
        g\round{\phi} &\equiv \sround{ E_{p_1} + E_{p_2} - E_{k} }^2 - E^2_{p_1+p_2-k} \notag \\
        & = \Delta^2 + 2\varepsilon - 2 p_1 c_1 p_2 c_2 + 2 k \round{p_1 c_1 + p_2 c_2} - 2 p_1 p_2 s_1 s_2 \cos \phi \; ,
    \end{align}
    where for convenience we defined 
    \begin{equation}
        \Delta \equiv m_1^2 + m_2^2 - m_3^2 + m_X^2 \; , \quad \text{and} \quad \varepsilon \equiv E_{p_1} E_{p_2} - E_k \round{E_{p_1} + E_{p_2}} \; .
    \end{equation}
    We thus have
    \begin{equation}
        \cos \bar{\phi} \equiv \frac{\Delta^2/2 + \varepsilon - p_1 c_1 p_2 c_2 + k \round{p_1 c_1 + p_2 c_2}}{p_1s_2p_2s_2} \; . \label{eq:cosphi_sol}
    \end{equation}
    Because $\cos\bar{\phi} = \cos \round{-\bar{\phi}}$, eq.~\eqref{eq:cosphi_sol} has one solution for $\phi_i$ in $[0,\pi]$ and one in $[\pi,2\pi]$. Then, eq.~\eqref{eq:delta_integral} can be simply rewritten as 
    \begin{equation}
        \int_0^{2\pi} {\rm d}\phi \, \delta \round{g\round{\phi}} F \round{\phi} = 2 \int_0^{\pi} {\rm d}\phi \, \abs{\frac{\partial g}{\partial \phi}}^{-1} F\round{\phi} \delta \round{\phi-\bar{\phi}} = 2 \, \abs{\frac{\partial g}{\partial \phi}}^{-1}_{\bar{\phi}} F\round{\bar{\phi}} \; . \label{eq:delta_integral_2}
    \end{equation}
    Note that $\cos \phi_i$ must lie between $-1$ and $+1$, which is equivalent to requiring $\cos^2\phi_i \le 1$. When this constraint is applied to eq.~\eqref{eq:cosphi_sol}, it restricts the allowed values of the outgoing variables $p_2$ and $c_2$ for any fixed set of incoming parameters $k$, $p_1$ and $c_1$ (recall that eq.~\eqref{eq:cosphi_sol} follows from energy–momentum conservation). To enforce this physical restriction in the subsequent integrals over ${\rm d}p_2$, ${\rm d}c_2$, etc., we insert a step function into eq.~\eqref{eq:delta_integral_2}
    \begin{align}
        \int_0^{2\pi} {\rm d}\phi \, \delta \round{g\round{\phi}} F \round{\phi} &= 2 \, \abs{\frac{\partial g}{\partial \phi}}^{-1}_{\bar{\phi}} \Theta \round{1 - \cos^2 \bar{\phi}} F\round{\bar{\phi}} = 2 \, \abs{\frac{\partial g}{\partial \phi}}^{-1}_{\bar{\phi}} \Theta \round{ \abs{\frac{\partial g}{\partial \phi}}^2_{\bar{\phi}} } F\round{\bar{\phi}} \notag \\
        &\equiv 2 \times \frac{\Theta \round{\mathcal{S} \round{p_1, p_2, k , c_1, c_2}}}{\sqrt{\mathcal{S} \round{p_1, p_2, k , c_1, c_2}}} \; ,
    \end{align}
    where we defined 
    \begin{equation}
        \mathcal{S} \round{p_1, p_2, k , c_1, c_2} \equiv \abs{\frac{\partial g}{\partial \phi}}^2_{\bar{\phi}} = 4 p_1^2 p_2^2 s_1^2 s_2^2 - \sround{ \Delta^2 + 2\varepsilon - 2 p_1 c_1 p_2 c_2 + 2 k \round{p_1 c_1 + p_2 c_2} }^2 \; .
    \end{equation}
    We can thus write our reduced formula for the collision integral as 
    \begin{align}
        \mathbb{C} \sround{f_1, f_2, f_3} &= \frac{1}{E_k} \frac{g_1g_2g_3}{\round[4]{2\pi}} \int \frac{p_1^2 {\rm d} p_1}{2E_1} \frac{p_2^2 {\rm d} p_2}{2E_2} {\rm d}c_1 {\rm d}c_2 \, \abs{\mathcal{M}_{12\rightarrow3X} \round{s,t,u} }^2 f_1 f_2 \round{1 \mp f_3} \notag \\
        & \quad \, \times \frac{\Theta \round{\mathcal{S} \round{p_1, p_2, k , c_1, c_2}}}{\sqrt{\mathcal{S} \round{p_1, p_2, k , c_1, c_2}}} \Theta \round{ E_{p_1} + E_{p_2} - E_{k} } \; .
        \label{eq:coll_step3_red}
    \end{align}
    In the general case, the Mandelstam's variables read
    \begin{align}
        s &= \round{P_1 + P_2}^2 = m_1^2 + m_2^2 + 2 E_{p_1} E_{p_2} - 2 p_1 p_2 \round{ c_1 c_2 + s_1 s_2 \cos \bar{\phi} } \; , \\
        t &= \round{P_2 - K}^2 = m_2^2 + m_4^4 - 2 E_{p_2} E_{k} + 2 p_2 k c_2 \; , \\ 
        u &= \round{P_1 - K}^2 = m_1^2 + m_4^4 - 2 E_{p_1} E_{k} + 2 p_1 k c_1 \; .
    \end{align}

\section{Full triangle plots and constraints}
\label{app:triangle_plots}
\begin{figure}[tbh]
    \centering
    \includegraphics[width=0.95\linewidth]{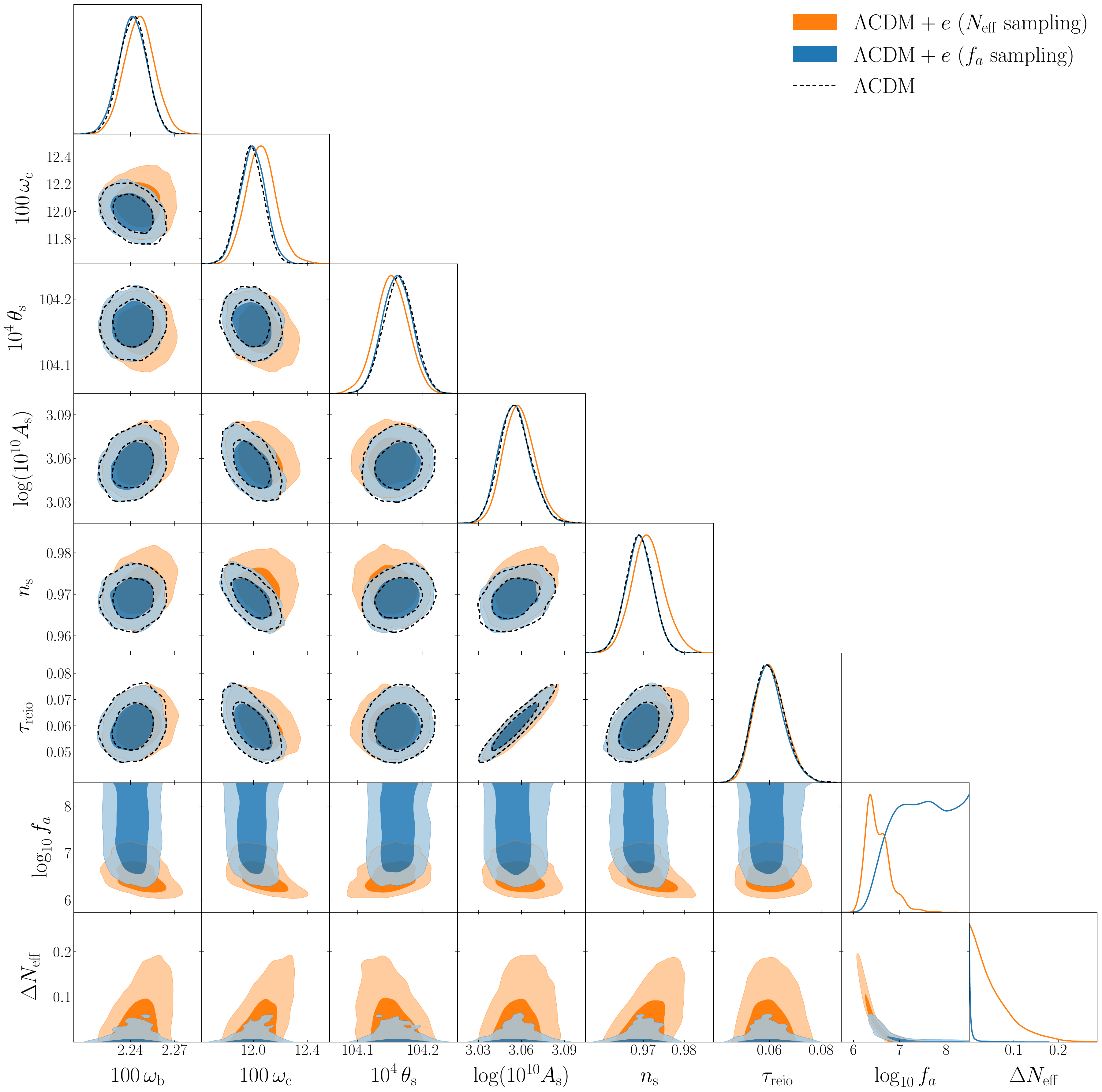}
    \caption{Triangle plot including one–dimensional posteriors and two-dimensional $68\%$ and $95\%$ credible regions for the cosmological parameters of the $\Lambda\text{CDM}+e$ model, obtained from runs with the SPA dataset. In blue (orange) we show results obtained with the $\Delta N_\text{eff}$ ($\log_{10} f_a$) sampling strategy. Although both $\Delta N_{\rm eff}$ and $f_a$ are shown, they are related by a one-to-one mapping and do not represent independent degrees of freedom. Results for the standard $\Lambda\text{CDM}$ model are shown as a benchmark reference. Note in particular the impact of the different sampling strategies on the posteriors of the various parameters. Numerical results with error bars are shown in \cref{tab:comparison}.}
    \label{fig:DNeff_vs_fa_triangle}
\end{figure}

In this appendix we complement the information presented in the main text by providing the full set of posterior distributions and numerical constraints for all cosmological parameters sampled in our MCMC analysis. \Cref{fig:DNeff_vs_fa_triangle} compares the posterior distributions obtained by sampling either $\Delta N_{\rm eff}$ or $\log_{10} f_a$, showing how different prior choices propagate to the constraints inferred for the cosmological parameters, despite the one-to-one mapping between the two quantities. The corresponding numerical constraints that quantify the differences introduced by the two sampling choices are reported in \cref{tab:comparison}.

\begin{table}[t]
    \renewcommand{\arraystretch}{1.5}
    \footnotesize
    \centering
    \begin{tabular}{ l c c c }
        \hline
        \hline
            & {\boldmath $\Lambda\textbf{CDM}+e$}
            & {\boldmath $\Lambda\textbf{CDM}+e$}
            & {\boldmath $\Lambda\textbf{CDM}$} \\[-5pt]
            & ($\Delta N_\mathrm{eff}$ sampling)
            & ($\log_{10} f_a$ sampling)
            &  \\
        \hline
            {\boldmath $100 \, \omega_\mathrm{b}$} & 
            $2.246\pm 0.010$ & 
            $2.2415\pm 0.0096$ & 
            $2.2420\pm 0.0093$ \\

            {\boldmath $100 \, \omega_\text{c}$} & 
            $12.06\pm 0.11$ & 
            $11.996\pm 0.096$ & 
            $11.984\pm 0.093$ \\

            {\boldmath$10^4 \, \theta_\mathrm{s}$} & 
            $104.152\pm 0.025$ & 
            $104.161\pm 0.023$ & 
            $104.163\pm 0.023$ \\

            {\boldmath$\log(10^{10} A_\mathrm{s})$} & 
            $3.0587^{+0.0098}_{-0.012}$ & 
            $3.0556^{+0.0095}_{-0.012}$ & 
            $3.056^{+0.010}_{-0.012}$ \\

            {\boldmath$n_\mathrm{s}$} & 
            $0.9712^{+0.0036}_{-0.0041}$ & 
            $0.9691\pm 0.0034$ & 
            $0.9691\pm 0.0033$ \\

            {\boldmath$\tau_\mathrm{reio}$} & 
            $0.0603^{+0.0052}_{-0.0065}$ & 
            $0.0597^{+0.0051}_{-0.0067}$ & 
            $0.0600^{+0.0055}_{-0.0065}$ \\
        \hdashline
            {\boldmath$\Delta N_\mathrm{eff}$} & 
            $< 0.148$ & 
            $< 0.0320$ &
             - \\
            {\boldmath$f_a$} &
            $> 1.51 \times 10^6$ &
            $> 3.52 \times 10^6$ &
             - \\
        \hline
        \hline
    \end{tabular}
    \caption{Bayesian credible intervals for the cosmological parameters of the $\Lambda\text{CDM}+e$ model, obtained from runs with the SPA dataset. The corresponding posterior distributions are shown in \cref{fig:DNeff_vs_fa_triangle}. When sampling on $\Delta N_\text{eff}$ ($f_a$), limits on $f_a$ ($\Delta N_\text{eff}$) have been derived through the one-to-one mapping shown in \cref{fig:DNeff_fermion_primakoff}. As of common use in the literature, bounds on the six parameters of $\Lambda\text{CDM}$ and on $\Delta N_\text{eff}$ ($f_a$) are quoted as $68\%$ and $95\%$ credible intervals, respectively. Results for the standard $\Lambda\text{CDM}$ model shown as a benchmark reference.}
    \label{tab:comparison}
\end{table}

\begin{figure}
    \centering
    \includegraphics[width=0.95\linewidth]{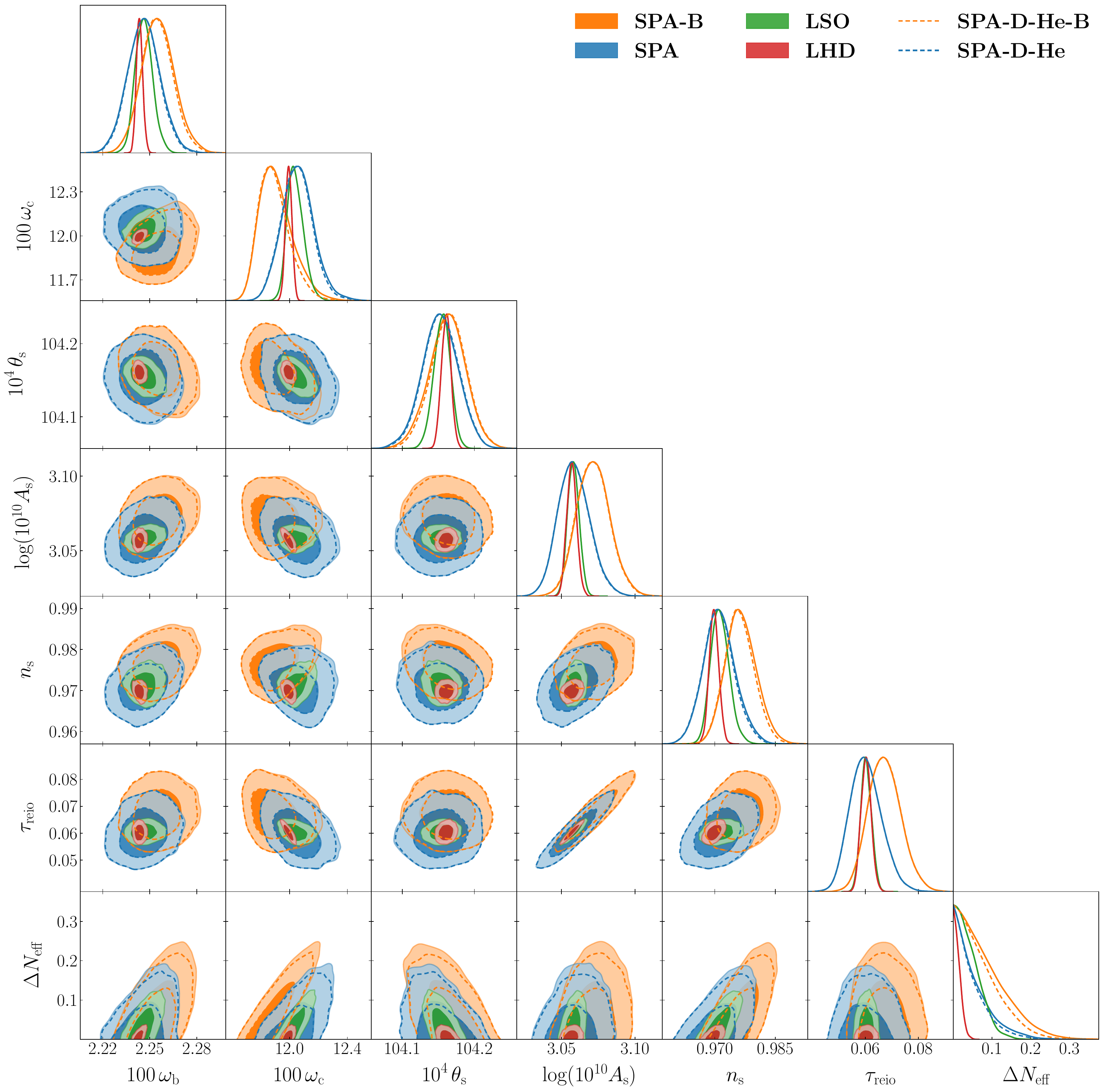}
    \caption{Triangle plot including one–dimensional posteriors and two-dimensional $68\%$ and $95\%$ credible regions for the seven cosmological parameters of the $\Lambda\text{CDM}+e$ model, obtained from MCMC runs on all the data sets considered in this work. Numerical results with error bars are shown in \cref{tab:bounds_electron}.}
    \label{fig:electron_datasets}
\end{figure}

\Cref{fig:electron_datasets} illustrates the posterior distribution and two-dimensional credible regions obtained for the $\Lambda\text{CDM}+e$ model, combining MCMC runs for all data sets considered in this work. The figure legend and the corresponding numerical constraints are reported in \cref{tab:bounds_electron}. Also shown are tables for the other production channels besides the electronic one. \Cref{tab:bounds_muon} summarizes the marginalized constraints for the $\Lambda$CDM$+\mu$ model obtained from all dataset combinations considered in this work. Compared to the electron channel, ALP production through muons occurs at higher temperatures, resulting in a slightly enhanced sensitivity to $\Delta N_{\rm eff}$ when forecasted small-scale CMB data are included. This trend is consistent with the progressive tightening of the bounds when moving from current datasets to the LSO and LHD configurations. \Cref{tab:bounds_tau} reports the full set of marginalized constraints for the $\Lambda$CDM$+\tau$ model. As discussed in the main text, the sensitivity to ALP production in the $\tau$ channel improves significantly with increasing experimental sensitivity, since ALP production from the $\tau$ channel occurs at higher temperatures than other leptonic channels and is therefore more sensitive to small deviations in $\Delta N_{\rm eff}$. This behavior is clearly visible when moving from current datasets to the LSO and LHD configurations, where the bounds on both $\Delta N_{\rm eff}$ and $f_a$ tighten considerably. For all channels, the standard cosmological parameters are consistent across all datasets, indicating that the improved constraints mainly derive from the increased sensitivity to extra radiation.

\begin{table}[t]
    \renewcommand{\arraystretch}{1.5}
    \centering
    {\boldmath$\Lambda\textbf{CDM}+e$}
    \vskip 1mm 
    \resizebox{\textwidth}{!}{
    \begin{tabular}{ l c c c c c c}
        \hline
        \hline
            & \textbf{SPA}
            & \textbf{SPA-D-He}
            & \textbf{SPA-B}
            & \textbf{SPA-D-He-B}
            & \textbf{LSO} 
            & \textbf{LHD} \\
        \hline
            {\boldmath $100 \, \omega_\mathrm{b}$} & 
            $2.246 \pm 0.010$ & 
            $2.246 \pm 0.010$ &
            $2.255 \pm 0.011$ & 
            $2.255 \pm 0.010$ & 
            $2.2464^{+0.0051}_{-0.0058}$ & 
            $2.2435^{+0.0019}_{-0.0021}$\\

            {\boldmath $100 \, \omega_\text{c}$} & 
            $12.06 \pm 0.11$ & 
            $12.05 \pm 0.11$ & 
            $11.909^{+0.083}_{-0.14}$ & 
            $11.894^{+0.080}_{-0.12}$ & 
            $12.042^{+0.047}_{-0.063}$ & 
            $11.994 \pm 0.023$\\

            {\boldmath$10^4 \, \theta_\mathrm{s}$} & 
            $104.152 \pm 0.025$ & 
            $104.153 \pm 0.025$ & 
            $104.162 \pm 0.026$ & 
            $104.164 \pm 0.025$ & 
            $104.155\pm 0.012$ & 
            $104.1609\pm 0.0066$\\

            {\boldmath$\log(10^{10} A_\mathrm{s})$} & 
            $3.0587^{+0.0099}_{-0.012}$ & 
            $3.0585^{+0.0099}_{-0.011}$ & 
            $3.072^{+0.011}_{-0.012}$ & 
            $3.072\pm 0.011$ & 
            $3.0581\pm 0.0039$ & 
            $3.0570\pm 0.0034$\\

            {\boldmath$n_\mathrm{s}$} & 
            $0.9712^{+0.0037}_{-0.0041}$ & 
            $0.9710 \pm 0.0039$ & 
            $0.9765^{+0.0035}_{-0.0040}$ & 
            $0.9761^{+0.0034}_{-0.0038}$ & 
            $0.9713^{+0.0020}_{-0.0024}$ & 
            $0.9698\pm 0.0012$\\

            {\boldmath$\tau_\mathrm{reio}$} & 
            $0.0603^{+0.0053}_{-0.0065}$ & 
            $0.0603^{+0.0053}_{-0.0065}$ & 
            $0.0673^{+0.0057}_{-0.0065}$ & 
            $0.0673^{+0.0057}_{-0.0065}$ & 
            $0.0603\pm 0.0020$ & 
            $0.0603\pm 0.0019$\\
        \hdashline
            {\boldmath$\Delta N_\mathrm{eff}$} & 
            $< 0.148$ & 
            $< 0.131$ & 
            $< 0.196$ & 
            $< 0.173$ & 
            $< 0.100$ & 
            $< 0.0302$ \\

            {\boldmath$f_a$} &
            $> 1.51 \times 10^6$ &
            $> 1.63 \times 10^6$ &
            $> 1.26 \times 10^6$ &
            $> 1.37 \times 10^6$ &
            $> 1.90 \times 10^6$ &
            $> 3.56 \times 10^6$  \\
        \hline
        \hline
    \end{tabular}
    }
    \caption{Bayesian credible intervals for the seven parameters of the $\Lambda\text{CDM}+\Delta N_\text{eff}$ model, with $\Delta N_\text{eff}$ referring to the electron production channel, obtained from MCMC runs on all the data sets considered in this work. The corresponding posterior distributions are shown in \cref{fig:electron_datasets}. As of common use in the literature, bounds on the six parameters of $\Lambda\text{CDM}$ and on $\Delta N_\text{eff}$ are quoted as $68\%$ and $95\%$ credible intervals, respectively. Limits on $f_a$ have been derived through the one-to-one mapping shown in \cref{fig:DNeff_fermion_primakoff}.}
    \label{tab:bounds_electron}
\end{table}

\begin{table}[t]
    \renewcommand{\arraystretch}{1.5}
    \centering
    {\boldmath$\Lambda\textbf{CDM}+\mu$}
    \vskip 1mm 
    \resizebox{\textwidth}{!}{
    \begin{tabular}{ l c c c c c c}
        \hline
        \hline
            & \textbf{SPA}
            & \textbf{SPA-D-He}
            & \textbf{SPA-B}
            & \textbf{SPA-D-He-B}
            & \textbf{LSO} 
            & \textbf{LHD} \\
        \hline
            {\boldmath $100 \, \omega_\mathrm{b}$} & 
            $2.246 \pm 0.010$ & 
            $2.246 \pm 0.010$ & 
            $2.255 \pm 0.011$ & 
            $2.254 \pm 0.010$ & 
            $2.2463\pm 0.0056$ & 
            $2.2435\pm 0.0020$\\

            {\boldmath $100 \, \omega_\text{c}$} & 
            $12.05^{+0.10}_{-0.11}$ & 
            $12.05 \pm 0.11$ & 
            $11.913^{+0.084}_{-0.14}$ & 
            $11.897^{+0.081}_{-0.12}$ & 
            $12.042^{+0.046}_{-0.064}$ & 
            $11.995 \pm 0.023$\\

            {\boldmath$10^4 \, \theta_\mathrm{s}$} & 
            $104.153 \pm 0.024$ & 
            $104.154 \pm 0.024$ & 
            $104.162 \pm 0.025$ & 
            $104.164 \pm 0.025$ & 
            $104.155\pm 0.011$ & 
            $104.1607\pm 0.0066$\\

            {\boldmath$\log(10^{10} A_\mathrm{s})$} & 
            $3.059^{+0.010}_{-0.012}$ & 
            $3.059^{+0.010}_{-0.012}$ & 
            $3.071 \pm 0.011$ & 
            $3.071 \pm 0.011$ & 
            $3.0579 \pm 0.0039$ & 
            $3.0568 \pm 0.0034$\\

            {\boldmath$n_\mathrm{s}$} & 
            $0.9713^{+0.0036}_{-0.0042}$ & 
            $0.9710^{+0.0035}_{-0.0040}$ & 
            $0.9766^{+0.0033}_{-0.0041}$ & 
            $0.9762^{+0.0032}_{-0.0039}$ & 
            $0.9713^{+0.0020}_{-0.0025}$ & 
            $0.9698\pm 0.0012$\\

            {\boldmath$\tau_\mathrm{reio}$} & 
            $0.0604^{+0.0054}_{-0.0064}$ & 
            $0.0604^{+0.0054}_{-0.0063}$ & 
            $0.0670^{+0.0057}_{-0.0065}$ & 
            $0.0670^{+0.0057}_{-0.0065}$ & 
            $0.0603\pm 0.0020$ & 
            $0.0602\pm 0.0019$\\

        \hdashline
            {\boldmath$\Delta N_\mathrm{eff}$} & 
            $< 0.140$ & 
            $< 0.125$ & 
            $< 0.199$ & 
            $< 0.174$ & 
            $< 0.105$ & 
            $< 0.0301$\\

            {\boldmath$f_a$} &
            $> 8.73 \times 10^6$ &
            $> 9.41 \times 10^6$ &
            $> 6.79 \times 10^6$ &
            $> 7.50 \times 10^6$ &
            $> 1.05 \times 10^7$ &
            $> 2.17 \times 10^7$ \\
        \hline
        \hline
    \end{tabular}
    }
    \caption{Bayesian credible intervals for the seven parameters of the $\Lambda\text{CDM}+\Delta N_\text{eff}$ model, with $\Delta N_\text{eff}$ referring to the muon production channel, obtained from MCMC runs on all the data sets considered in this work. As of common use in the literature, bounds on the six parameters of $\Lambda\text{CDM}$ and on $\Delta N_\text{eff}$ are quoted as $68\%$ and $95\%$ credible intervals, respectively. Limits on $f_a$ have been derived through the one-to-one mapping shown in \cref{fig:DNeff_fermion_primakoff}.}
    \label{tab:bounds_muon}
\end{table}

\begin{table}[t]
    \renewcommand{\arraystretch}{1.5}
    \centering
    {\boldmath$\Lambda\textbf{CDM}+\tau$}
    \vskip 1mm 
    \resizebox{\textwidth}{!}{
    \begin{tabular}{ l c c c c c c}
        \hline
        \hline
            & \textbf{SPA}
            & \textbf{SPA-D-He}
            & \textbf{SPA-B}
            & \textbf{SPA-D-He-B}
            & \textbf{LSO} 
            & \textbf{LHD} \\
        \hline
            {\boldmath $100 \, \omega_\mathrm{b}$} & 
            $2.247 \pm 0.010        $ & 
            $2.2467 \pm 0.0098      $ & 
            $2.255 \pm 0.010        $ & 
            $2.254 \pm 0.010        $ & 
            $2.2466\pm 0.0056$ & 
            $2.2436\pm 0.0020$\\

            {\boldmath $100 \, \omega_\text{c}$} & 
            $12.05^{+0.10}_{-0.11}$ & 
            $12.04 \pm 0.11          $ & 
            $11.913^{+0.088}_{-0.13}$ & 
            $11.897^{+0.083}_{-0.12}$ & 
            $12.042^{+0.046}_{-0.064}$ & 
            $11.995\pm 0.023$\\

            {\boldmath$10^4 \, \theta_\mathrm{s}$} & 
            $104.154\pm 0.025$ & 
            $104.155\pm 0.024$ & 
            $104.163\pm 0.025$ & 
            $104.165\pm 0.024$ & 
            $104.156\pm 0.011$ & 
            $104.1607\pm 0.0065$\\

            {\boldmath$\log(10^{10} A_\mathrm{s})$} & 
            $3.059^{+0.010}_{-0.012}$ & 
            $3.059^{+0.010}_{-0.012}$ & 
            $3.071 \pm 0.011$ & 
            $3.071 \pm 0.011$ & 
            $3.0581\pm 0.0039$ & 
            $3.0569\pm 0.0034$\\

            {\boldmath$n_\mathrm{s}$} & 
            $0.9714^{+0.0035}_{-0.0041}$ & 
            $0.9712^{+0.0035}_{-0.0040}$ & 
            $0.9766\pm 0.0039          $ & 
            $0.9762\pm 0.0038          $ & 
            $0.9714^{+0.0020}_{-0.0025}$ & 
            $0.9698\pm 0.0012$\\

            {\boldmath$\tau_\mathrm{reio}$} & 
            $0.0606^{+0.0054}_{-0.0065}$ & 
            $0.0606^{+0.0054}_{-0.0065}$ & 
            $0.0672^{+0.0056}_{-0.0064}$ & 
            $0.0673^{+0.0056}_{-0.0064}$ & 
            $0.0604\pm 0.0020$ & 
            $0.0602\pm 0.0020$\\

        \hdashline
            {\boldmath$\Delta N_\mathrm{eff}$} & 
            $< 0.134$ & 
            $< 0.121$ & 
            $< 0.198$ & 
            $< 0.174$ & 
            $< 0.105$ & 
            $< 0.0288$\\

            {\boldmath$f_a$} &
            $> 6.10 \times 10^4$ &
            $> 8.06 \times 10^4$ &
            $> 1.80 \times 10^4$ &
            $> 2.85 \times 10^4$ &
            $> 1.34 \times 10^5$ &
            $> 2.45 \times 10^7$\\
            
        \hline
        \hline
    \end{tabular}
    }
    \caption{Same as \cref{tab:bounds_muon} for the tau production channel.}
    \label{tab:bounds_tau}
\end{table}

Finally, \cref{tab:bounds_primakoff} presents the constraints obtained for the Primakoff production channel, in terms of the ALP-photon coupling $g_{a\gamma}$. As in the leptonic cases, the bounds on the $\Lambda$CDM parameters are unaffected by the inclusion of ALPs, and future observations drive the improved sensitivity to $\Delta N_{\rm eff}$. Constraints on $g_{a\gamma}$ are particularly sensitive to the initial temperature $T_{\rm in}$ assumed, because of the dependence of the Primakoff production from the UV freeze-in details. Here, we fix $T_{\rm in}$ to the representative choice $T_{\rm in} = 10^3\,{\rm GeV}$. The strong improvement observed for the LHD configuration benefits from the potential of next-generation CMB experiments to probe extremely feeble ALP-photon couplings.

\begin{table}[t]
    \renewcommand{\arraystretch}{1.5}
    \centering
    {\boldmath$\Lambda\textbf{CDM}+\gamma$}
    \vskip 1mm 
    \resizebox{\textwidth}{!}{
    \begin{tabular}{ l c c c c c c}
        \hline
        \hline
            & \textbf{SPA}
            & \textbf{SPA-D-He}
            & \textbf{SPA-B}
            & \textbf{SPA-D-He-B}
            & \textbf{LSO} 
            & \textbf{LHD} \\
        \hline
            {\boldmath $100 \, \omega_\mathrm{b}$} & 
            $2.2461^{+0.0093}_{-0.011}$ & 
            $2.2459^{+0.0091}_{-0.010}$ & 
            $2.255\pm 0.010$ & 
            $2.255\pm 0.010$ & 
            $2.2465^{+0.0050}_{-0.0060}$ & 
            $2.2436^{+0.0018}_{-0.0021}$\\

            {\boldmath $100 \, \omega_\text{c}$} & 
            $12.05^{+0.10}_{-0.12}$ & 
            $12.04\pm 0.11          $ & 
            $11.912^{+0.083}_{-0.14}$ & 
            $11.895^{+0.079}_{-0.12}$ & 
            $12.041^{+0.044}_{-0.064}$ & 
            $11.995\pm 0.023$\\

            {\boldmath$10^4 \, \theta_\mathrm{s}$} & 
            $104.154\pm 0.024$ & 
            $104.154\pm 0.024$ & 
            $104.162\pm 0.026$ & 
            $104.164\pm 0.025$ & 
            $104.155\pm 0.012$ & 
            $104.1606\pm 0.0065$\\

            {\boldmath$\log(10^{10} A_\mathrm{s})$} & 
            $3.059^{+0.010}_{-0.011}$ & 
            $3.058^{+0.010}_{-0.011}$ & 
            $3.072^{+0.011}_{-0.012}$ & 
            $3.071^{+0.011}_{-0.012}$ & 
            $3.0581\pm 0.0039$ & 
            $3.0569\pm 0.0035$\\

            {\boldmath$n_\mathrm{s}$} & 
            $0.9712\pm 0.0039$ & 
            $0.9710\pm 0.0038$ & 
            $0.9767^{+0.0035}_{-0.0041}$ & 
            $0.9763^{+0.0034}_{-0.0038}$ & 
            $0.9712^{+0.0020}_{-0.0026}$ & 
            $0.9698\pm 0.0012$\\

            {\boldmath$\tau_\mathrm{reio}$} & 
            $0.0604^{+0.0054}_{-0.0062}$ & 
            $0.0604^{+0.0054}_{-0.0062}$ & 
            $0.0674^{+0.0057}_{-0.0067}$ & 
            $0.0674^{+0.0057}_{-0.0067}$ & 
            $0.0603\pm 0.0021$ & 
            $0.0602^{+0.0019}_{-0.0021}$\\

        \hdashline
            {\boldmath$\Delta N_\mathrm{eff}$} & 
            $< 0.138$ & 
            $< 0.124$ & 
            $< 0.204$ & 
            $< 0.178$ & 
            $< 0.103$ & 
            $< 0.0302$\\

            {\boldmath$g_{a\gamma}$} &
            $< 2.18 \times 10^{-8}$ &
            $< 1.98 \times 10^{-8}$ &
            $< 3.08 \times 10^{-8}$ &
            $< 2.73 \times 10^{-8}$ &
            $< 1.66 \times 10^{-8}$ &
            $< 1.01 \times 10^{-9}$ \\
        \hline
        \hline
    \end{tabular}
    }
    \caption{Same as \cref{tab:bounds_muon} for the Primakoff production channel.}
    \label{tab:bounds_primakoff}
\end{table}

\newpage
\bibliographystyle{JHEP}
\bibliography{bibliography.bib}

@article{Conlon:2006tq,
    author = "Conlon, Joseph P.",
    title = "{The QCD axion and moduli stabilisation}",
    eprint = "hep-th/0602233",
    archivePrefix = "arXiv",
    reportNumber = "DAMTP-2006-17",
    doi = "10.1088/1126-6708/2006/05/078",
    journal = "JHEP",
    volume = "05",
    pages = "078",
    year = "2006"
}

@article{Grimm:2007hs,
    author = "Grimm, Thomas W.",
    title = "{Axion inflation in type II string theory}",
    eprint = "0710.3883",
    archivePrefix = "arXiv",
    primaryClass = "hep-th",
    reportNumber = "BONN-TH-2007-11, MAD-TH-07-11",
    doi = "10.1103/PhysRevD.77.126007",
    journal = "Phys. Rev. D",
    volume = "77",
    pages = "126007",
    year = "2008"
}

@article{Melchiorri:2007cd,
    author = "Melchiorri, Alessandro and Mena, Olga and Slosar, Anze",
    title = "{An improved cosmological bound on the thermal axion mass}",
    eprint = "0705.2695",
    archivePrefix = "arXiv",
    primaryClass = "astro-ph",
    doi = "10.1103/PhysRevD.76.041303",
    journal = "Phys. Rev. D",
    volume = "76",
    pages = "041303",
    year = "2007"
}

@article{Battye:1994au,
    author = "Battye, R. A. and Shellard, E. P. S.",
    title = "{Axion string constraints}",
    eprint = "astro-ph/9403018",
    archivePrefix = "arXiv",
    reportNumber = "DAMTP-R-94-8",
    doi = "10.1103/PhysRevLett.73.2954",
    journal = "Phys. Rev. Lett.",
    volume = "73",
    pages = "2954--2957",
    year = "1994",
    note = "[Erratum: Phys.Rev.Lett. 76, 2203--2204 (1996)]"
}

@article{Davis:1986xc,
    author = "Davis, Richard Lynn",
    title = "{Cosmic Axions from Cosmic Strings}",
    reportNumber = "SLAC-PUB-3895",
    doi = "10.1016/0370-2693(86)90300-X",
    journal = "Phys. Lett. B",
    volume = "180",
    pages = "225--230",
    year = "1986"
}

@article{Dine:1982ah,
    author = "Dine, Michael and Fischler, Willy",
    editor = "Srednicki, M. A.",
    title = "{The Not So Harmless Axion}",
    reportNumber = "UPR-0201T",
    doi = "10.1016/0370-2693(83)90639-1",
    journal = "Phys. Lett. B",
    volume = "120",
    pages = "137--141",
    year = "1983"
}

@article{Abbott:1982af,
    author = "Abbott, L. F. and Sikivie, P.",
    editor = "Srednicki, M. A.",
    title = "{A Cosmological Bound on the Invisible Axion}",
    reportNumber = "PRINT-82-0695 (BRANDEIS)",
    doi = "10.1016/0370-2693(83)90638-X",
    journal = "Phys. Lett. B",
    volume = "120",
    pages = "133--136",
    year = "1983"
}

@article{Preskill:1982cy,
    author = "Preskill, John and Wise, Mark B. and Wilczek, Frank",
    editor = "Srednicki, M. A.",
    title = "{Cosmology of the Invisible Axion}",
    reportNumber = "HUTP-82-A048, NSF-ITP-82-103",
    doi = "10.1016/0370-2693(83)90637-8",
    journal = "Phys. Lett. B",
    volume = "120",
    pages = "127--132",
    year = "1983"
}

@article{Kim:2015yna,
    author = "Kim, Jihn E. and Marsh, David J. E.",
    title = "{An ultralight pseudoscalar boson}",
    eprint = "1510.01701",
    archivePrefix = "arXiv",
    primaryClass = "hep-ph",
    reportNumber = "KCL-PH-TH-2015-44",
    doi = "10.1103/PhysRevD.93.025027",
    journal = "Phys. Rev. D",
    volume = "93",
    number = "2",
    pages = "025027",
    year = "2016"
}

@article{Alonso-Alvarez:2019ssa,
    author = "Alonso-{\'A}lvarez, Gonzalo and Gupta, Rick S. and Jaeckel, Joerg and Spannowsky, Michael",
    title = "{On the Wondrous Stability of ALP Dark Matter}",
    eprint = "1911.07885",
    archivePrefix = "arXiv",
    primaryClass = "hep-ph",
    reportNumber = "IPPP/19/84",
    doi = "10.1088/1475-7516/2020/03/052",
    journal = "JCAP",
    volume = "03",
    pages = "052",
    year = "2020"
}

@article{Bauer:2017ris,
    author = "Bauer, Martin and Neubert, Matthias and Thamm, Andrea",
    title = "{Collider Probes of Axion-Like Particles}",
    eprint = "1708.00443",
    archivePrefix = "arXiv",
    primaryClass = "hep-ph",
    reportNumber = "MITP-17-047",
    doi = "10.1007/JHEP12(2017)044",
    journal = "JHEP",
    volume = "12",
    pages = "044",
    year = "2017"
}

@article{Graf:2010tv,
    author = "Graf, Peter and Steffen, Frank Daniel",
    title = "{Thermal axion production in the primordial quark-gluon plasma}",
    eprint = "1008.4528",
    archivePrefix = "arXiv",
    primaryClass = "hep-ph",
    reportNumber = "MPP-2010-20",
    doi = "10.1103/PhysRevD.83.075011",
    journal = "Phys. Rev. D",
    volume = "83",
    pages = "075011",
    year = "2011"
}

@article{Masso:2002np,
    author = "Masso, Eduard and Rota, Francesc and Zsembinszki, Gabriel",
    title = "{On axion thermalization in the early universe}",
    eprint = "hep-ph/0203221",
    archivePrefix = "arXiv",
    reportNumber = "UAB-FT-522",
    doi = "10.1103/PhysRevD.66.023004",
    journal = "Phys. Rev. D",
    volume = "66",
    pages = "023004",
    year = "2002"
}

@article{Chang:1993gm,
    author = "Chang, Sanghyeon and Choi, Kiwoon",
    title = "{Hadronic axion window and the big bang nucleosynthesis}",
    eprint = "hep-ph/9306216",
    archivePrefix = "arXiv",
    reportNumber = "SNUTP-93-11",
    doi = "10.1016/0370-2693(93)90656-3",
    journal = "Phys. Lett. B",
    volume = "316",
    pages = "51--56",
    year = "1993"
}

@article{Turner:1986tb,
    author = "Turner, Michael S.",
    title = "{Thermal Production of Not SO Invisible Axions in the Early Universe}",
    reportNumber = "FERMILAB-PUB-86-150-A",
    doi = "10.1103/PhysRevLett.59.2489",
    journal = "Phys. Rev. Lett.",
    volume = "59",
    pages = "2489",
    year = "1987",
    note = "[Erratum: Phys.Rev.Lett. 60, 1101 (1988)]"
}

@article{Witten:1984dg,
    author = "Witten, Edward",
    title = "{Some Properties of O(32) Superstrings}",
    reportNumber = "Print-84-0838 (PRINCETON)",
    doi = "10.1016/0370-2693(84)90422-2",
    journal = "Phys. Lett. B",
    volume = "149",
    pages = "351--356",
    year = "1984"
}

@article{Hannestad:2005df,
    author = "Hannestad, Steen and Mirizzi, Alessandro and Raffelt, Georg",
    title = "{New cosmological mass limit on thermal relic axions}",
    eprint = "hep-ph/0504059",
    archivePrefix = "arXiv",
    doi = "10.1088/1475-7516/2005/07/002",
    journal = "JCAP",
    volume = "07",
    pages = "002",
    year = "2005"
}

@article{Weinberg:1977ma,
    author = "Weinberg, Steven",
    title = "{A New Light Boson?}",
    reportNumber = "HUTP-77/A074",
    doi = "10.1103/PhysRevLett.40.223",
    journal = "Phys. Rev. Lett.",
    volume = "40",
    pages = "223--226",
    year = "1978"
}

@article{Wilczek:1977pj,
    author = "Wilczek, Frank",
    title = "{Problem of Strong  $P$  and  $T$  Invariance in the Presence of Instantons}",
    reportNumber = "Print-77-0939 (COLUMBIA)",
    doi = "10.1103/PhysRevLett.40.279",
    journal = "Phys. Rev. Lett.",
    volume = "40",
    pages = "279--282",
    year = "1978"
}

@article{Peccei:1977hh,
    author = "Peccei, R. D. and Quinn, Helen R.",
    title = "{CP Conservation in the Presence of Instantons}",
    reportNumber = "ITP-568-STANFORD",
    doi = "10.1103/PhysRevLett.38.1440",
    journal = "Phys. Rev. Lett.",
    volume = "38",
    pages = "1440--1443",
    year = "1977"
}

@article{Iocco:2008va,
    author = "Iocco, Fabio and Mangano, Gianpiero and Miele, Gennaro and Pisanti, Ofelia and Serpico, Pasquale D.",
    title = "{Primordial Nucleosynthesis: from precision cosmology to fundamental physics}",
    eprint = "0809.0631",
    archivePrefix = "arXiv",
    primaryClass = "astro-ph",
    reportNumber = "DSF-20-2008, FERMILAB-PUB-08-216-A, IFIC-08-37",
    doi = "10.1016/j.physrep.2009.02.002",
    journal = "Phys. Rept.",
    volume = "472",
    pages = "1--76",
    year = "2009"
}

@article{Pisanti:2007hk,
    author = "Pisanti, O. and Cirillo, A. and Esposito, S. and Iocco, F. and Mangano, G. and Miele, G. and Serpico, P. D.",
    title = "{PArthENoPE: Public Algorithm Evaluating the Nucleosynthesis of Primordial Elements}",
    eprint = "0705.0290",
    archivePrefix = "arXiv",
    primaryClass = "astro-ph",
    reportNumber = "DSF-13-07, FERMILAB-PUB-07-079-A, SLAC-PUB-12488",
    doi = "10.1016/j.cpc.2008.02.015",
    journal = "Comput. Phys. Commun.",
    volume = "178",
    pages = "956--971",
    year = "2008"
}

@article{LiteBIRD:2022cnt,
    author = "Allys, E. and others",
    collaboration = "LiteBIRD",
    title = "{Probing Cosmic Inflation with the LiteBIRD Cosmic Microwave Background Polarization Survey}",
    eprint = "2202.02773",
    archivePrefix = "arXiv",
    primaryClass = "astro-ph.IM",
    doi = "10.1093/ptep/ptac150",
    journal = "PTEP",
    volume = "2023",
    number = "4",
    pages = "042F01",
    year = "2023"
}

@article{Yeh:2020mgl,
    author = "Yeh, Tsung-Han and Olive, Keith A. and Fields, Brian D.",
    title = "{The impact of new $d(p,\gamma)$3 rates on Big Bang Nucleosynthesis}",
    eprint = "2011.13874",
    archivePrefix = "arXiv",
    primaryClass = "astro-ph.CO",
    reportNumber = "UMN--TH--4004/20, FTPI--MINN--20/35",
    doi = "10.1088/1475-7516/2021/03/046",
    journal = "JCAP",
    volume = "03",
    pages = "046",
    year = "2021"
}

@article{Yeh:2022heq,
    author = "Yeh, Tsung-Han and Shelton, Jessie and Olive, Keith A. and Fields, Brian D.",
    title = "{Probing physics beyond the standard model: limits from BBN and the CMB independently and combined}",
    eprint = "2207.13133",
    archivePrefix = "arXiv",
    primaryClass = "astro-ph.CO",
    reportNumber = "UMN-TH-4125/22, FTPI-MINN-22/16",
    doi = "10.1088/1475-7516/2022/10/046",
    journal = "JCAP",
    volume = "10",
    pages = "046",
    year = "2022"
}

@article{Pospelov:2010hj,
    author = "Pospelov, Maxim and Pradler, Josef",
    title = "{Big Bang Nucleosynthesis as a Probe of New Physics}",
    eprint = "1011.1054",
    archivePrefix = "arXiv",
    primaryClass = "hep-ph",
    doi = "10.1146/annurev.nucl.012809.104521",
    journal = "Ann. Rev. Nucl. Part. Sci.",
    volume = "60",
    pages = "539--568",
    year = "2010"
}

@article{Pitrou:2018cgg,
    author = "Pitrou, Cyril and Coc, Alain and Uzan, Jean-Philippe and Vangioni, Elisabeth",
    title = "{Precision big bang nucleosynthesis with improved Helium-4 predictions}",
    eprint = "1801.08023",
    archivePrefix = "arXiv",
    primaryClass = "astro-ph.CO",
    doi = "10.1016/j.physrep.2018.04.005",
    journal = "Phys. Rept.",
    volume = "754",
    pages = "1--66",
    year = "2018"
}

@article{Steigman:2007xt,
    author = "Steigman, Gary",
    title = "{Primordial Nucleosynthesis in the Precision Cosmology Era}",
    eprint = "0712.1100",
    archivePrefix = "arXiv",
    primaryClass = "astro-ph",
    doi = "10.1146/annurev.nucl.56.080805.140437",
    journal = "Ann. Rev. Nucl. Part. Sci.",
    volume = "57",
    pages = "463--491",
    year = "2007"
}

@article{DEramo:2020gpr,
    author = "D'Eramo, Francesco and Lenoci, Alessandro",
    title = "{Lower mass bounds on FIMP dark matter produced via freeze-in}",
    eprint = "2012.01446",
    archivePrefix = "arXiv",
    primaryClass = "hep-ph",
    reportNumber = "DESY 20-219, DESY-20-219",
    doi = "10.1088/1475-7516/2021/10/045",
    journal = "JCAP",
    volume = "10",
    pages = "045",
    year = "2021"
}

@article{DEramo:2023nzt,
    author = "D'Eramo, Francesco and Hajkarim, Fazlollah and Lenoci, Alessandro",
    title = "{Dark radiation from the primordial thermal bath in momentum space}",
    eprint = "2311.04974",
    archivePrefix = "arXiv",
    primaryClass = "hep-ph",
    reportNumber = "DESY-23-177",
    doi = "10.1088/1475-7516/2024/03/009",
    journal = "JCAP",
    volume = "03",
    pages = "009",
    year = "2024"
}

@article{Arvanitaki:2009fg,
    author = "Arvanitaki, Asimina and Dimopoulos, Savas and Dubovsky, Sergei and Kaloper, Nemanja and March-Russell, John",
    title = "{String Axiverse}",
    eprint = "0905.4720",
    archivePrefix = "arXiv",
    primaryClass = "hep-th",
    doi = "10.1103/PhysRevD.81.123530",
    journal = "Phys. Rev. D",
    volume = "81",
    pages = "123530",
    year = "2010"
}

@article{Svrcek:2006yi,
    author = "Svrcek, Peter and Witten, Edward",
    title = "{Axions In String Theory}",
    eprint = "hep-th/0605206",
    archivePrefix = "arXiv",
    reportNumber = "SLAC-PUB-11894",
    doi = "10.1088/1126-6708/2006/06/051",
    journal = "JHEP",
    volume = "06",
    pages = "051",
    year = "2006"
}

@article{Consiglio:2017pot,
	Archiveprefix = {arXiv},
	Author = {Consiglio, R. and de Salas, P. F. and Mangano, G. and Miele, G. and Pastor, S. and Pisanti, O.},
	Doi = {10.1016/j.cpc.2018.06.022},
	Eprint = {1712.04378},
	Journal = {Comput. Phys. Commun.},
	Pages = {237--242},
	Primaryclass = {astro-ph.CO},
	Title = {{PArthENoPE reloaded}},
	Volume = {233},
	Year = {2018}}

@article{Saikawa:2018rcs,
	author = "Saikawa, Ken'ichi and Shirai, Satoshi",
	title = "{Primordial gravitational waves, precisely: The role of thermodynamics in the Standard Model}",
	eprint = "1803.01038",
	archivePrefix = "arXiv",
	primaryClass = "hep-ph",
	reportNumber = "IPMU18-0037, MPP-2018-19",
	doi = "10.1088/1475-7516/2018/05/035",
	journal = "JCAP",
	volume = "05",
	pages = "035",
	year = "2018"
}

@article{DiLuzio:2021vjd,
	author = "Di Luzio, Luca and Martinelli, Guido and Piazza, Gioacchino",
	title = "{Breakdown of chiral perturbation theory for the axion hot dark matter bound}",
	eprint = "2101.10330",
	archivePrefix = "arXiv",
	primaryClass = "hep-ph",
	reportNumber = "DESY-21-012, DESY 21-012",
	doi = "10.1103/PhysRevLett.126.241801",
	journal = "Phys. Rev. Lett.",
	volume = "126",
	number = "24",
	pages = "241801",
	year = "2021"
}

@article{Bolz:2000fu,
	author = "Bolz, M. and Brandenburg, A. and Buchmuller, W.",
	title = "{Thermal production of gravitinos}",
	eprint = "hep-ph/0012052",
	archivePrefix = "arXiv",
	reportNumber = "DESY-00-167",
	doi = "10.1016/S0550-3213(01)00132-8",
	journal = "Nucl. Phys. B",
	volume = "606",
	pages = "518--544",
	year = "2001",
	note = "[Erratum: Nucl.Phys.B 790, 336--337 (2008)]"
}

@article{Archidiacono:2013cha,
    author = "Archidiacono, Maria and Hannestad, Steen and Mirizzi, Alessandro and Raffelt, Georg and Wong, Yvonne Y. Y.",
    title = "{Axion hot dark matter bounds after Planck}",
    eprint = "1307.0615",
    archivePrefix = "arXiv",
    primaryClass = "astro-ph.CO",
    reportNumber = "MPP-2013-113",
    doi = "10.1088/1475-7516/2013/10/020",
    journal = "JCAP",
    volume = "10",
    pages = "020",
    year = "2013"
}

@article{Cadamuro:2011fd,
	author = "Cadamuro, Davide and Redondo, Javier",
	title = "{Cosmological bounds on pseudo Nambu-Goldstone bosons}",
	eprint = "1110.2895",
	archivePrefix = "arXiv",
	primaryClass = "hep-ph",
	reportNumber = "MPP-2011-116",
	doi = "10.1088/1475-7516/2012/02/032",
	journal = "JCAP",
	volume = "02",
	pages = "032",
	year = "2012"
}

@article{Cadamuro:2010cz,
	author = "Cadamuro, Davide and Hannestad, Steen and Raffelt, Georg and Redondo, Javier",
	title = "{Cosmological bounds on sub-MeV mass axions}",
	eprint = "1011.3694",
	archivePrefix = "arXiv",
	primaryClass = "hep-ph",
	reportNumber = "MPP-2010-148",
	doi = "10.1088/1475-7516/2011/02/003",
	journal = "JCAP",
	volume = "02",
	pages = "003",
	year = "2011"
}

@article{Carenza:2021ebx,
	author = "Carenza, Pierluca and Lattanzi, Massimiliano and Mirizzi, Alessandro and Forastieri, Francesco",
	title = "{Thermal axions with multi-eV masses are possible in low-reheating scenarios}",
	eprint = "2104.03982",
	archivePrefix = "arXiv",
	primaryClass = "astro-ph.CO",
	doi = "10.1088/1475-7516/2021/07/031",
	journal = "JCAP",
	volume = "07",
	pages = "031",
	year = "2021"
}

@article{Salvio:2013iaa,
	author = "Salvio, Alberto and Strumia, Alessandro and Xue, Wei",
	title = "{Thermal axion production}",
	eprint = "1310.6982",
	archivePrefix = "arXiv",
	primaryClass = "hep-ph",
	reportNumber = "FTUAM-13-29, IFT-UAM-CSIC-13-113",
	doi = "10.1088/1475-7516/2014/01/011",
	journal = "JCAP",
	volume = "01",
	pages = "011",
	year = "2014"
}

@article{Lewis:2019xzd,
	author         = "Lewis, Antony",
	title          = "{GetDist: a Python package for analysing Monte Carlo
	samples}",
	year           = "2019",
	eprint         = "1910.13970",
	archivePrefix  = "arXiv",
	primaryClass   = "astro-ph.IM",
	SLACcitation   = "%%CITATION = ARXIV:1910.13970;%%",
	url            = "https://getdist.readthedocs.io"
}

@article{Mangano:2001iu,
	author = "Mangano, G. and Miele, G. and Pastor, S. and Peloso, M.",
	title = "{A Precision calculation of the effective number of cosmological neutrinos}",
	eprint = "astro-ph/0111408",
	archivePrefix = "arXiv",
	reportNumber = "DSF-37-2001, MPI-PHT-2001-51",
	doi = "10.1016/S0370-2693(02)01622-2",
	journal = "Phys. Lett. B",
	volume = "534",
	pages = "8--16",
	year = "2002"
}

@article{Bennett:2019ewm,
	author = "Bennett, Jack J. and Buldgen, Gilles and Drewes, Marco and Wong, Yvonne Y. Y.",
	title = "{Towards a precision calculation of the effective number of neutrinos $N_{\rm eff}$ in the Standard Model I: the QED equation of state}",
	eprint = "1911.04504",
	archivePrefix = "arXiv",
	primaryClass = "hep-ph",
	doi = "10.1088/1475-7516/2020/03/003",
	journal = "JCAP",
	volume = "03",
	pages = "003",
	year = "2020",
	note = "[Addendum: JCAP 03, A01 (2021)]"
}

@article{Bennett:2020zkv,
	author = "Bennett, Jack J. and Buldgen, Gilles and De Salas, Pablo F. and Drewes, Marco and Gariazzo, Stefano and Pastor, Sergio and Wong, Yvonne Y. Y.",
	title = "{Towards a precision calculation of $N_{\rm eff}$ in the Standard Model II: Neutrino decoupling in the presence of flavour oscillations and finite-temperature QED}",
	eprint = "2012.02726",
	archivePrefix = "arXiv",
	primaryClass = "hep-ph",
	reportNumber = "CPPC-2020-10",
	doi = "10.1088/1475-7516/2021/04/073",
	journal = "JCAP",
	volume = "04",
	pages = "073",
	year = "2021"
}

@article{Akita:2020szl,
	author = "Akita, Kensuke and Yamaguchi, Masahide",
	title = "{A precision calculation of relic neutrino decoupling}",
	eprint = "2005.07047",
	archivePrefix = "arXiv",
	primaryClass = "hep-ph",
	doi = "10.1088/1475-7516/2020/08/012",
	journal = "JCAP",
	volume = "08",
	pages = "012",
	year = "2020"
}

@article{Froustey:2020mcq,
	author = "Froustey, Julien and Pitrou, Cyril and Volpe, Maria Cristina",
	title = "{Neutrino decoupling including flavour oscillations and primordial nucleosynthesis}",
	eprint = "2008.01074",
	archivePrefix = "arXiv",
	primaryClass = "hep-ph",
	doi = "10.1088/1475-7516/2020/12/015",
	journal = "JCAP",
	volume = "12",
	pages = "015",
	year = "2020"
}

@article{Planck:2018vyg,
	author = "Aghanim, N. and others",
	collaboration = "Planck",
	title = "{Planck 2018 results. VI. Cosmological parameters}",
	eprint = "1807.06209",
	archivePrefix = "arXiv",
	primaryClass = "astro-ph.CO",
	doi = "10.1051/0004-6361/201833910",
	journal = "Astron. Astrophys.",
	volume = "641",
	pages = "A6",
	year = "2020",
	note = "[Erratum: Astron.Astrophys. 652, C4 (2021)]"
}

@article{Beutler:2011hx,
	author = "Beutler, Florian and Blake, Chris and Colless, Matthew and Jones, D. Heath and Staveley-Smith, Lister and Campbell, Lachlan and Parker, Quentin and Saunders, Will and Watson, Fred",
	title = "{The 6dF Galaxy Survey: Baryon Acoustic Oscillations and the Local Hubble Constant}",
	eprint = "1106.3366",
	archivePrefix = "arXiv",
	primaryClass = "astro-ph.CO",
	doi = "10.1111/j.1365-2966.2011.19250.x",
	journal = "Mon. Not. Roy. Astron. Soc.",
	volume = "416",
	pages = "3017--3032",
	year = "2011"
}

@article{Ross:2014qpa,
	author = "Ross, Ashley J. and Samushia, Lado and Howlett, Cullan and Percival, Will J. and Burden, Angela and Manera, Marc",
	title = "{The clustering of the SDSS DR7 main Galaxy sample \textendash{} I. A 4 per cent distance measure at $z = 0.15$}",
	eprint = "1409.3242",
	archivePrefix = "arXiv",
	primaryClass = "astro-ph.CO",
	doi = "10.1093/mnras/stv154",
	journal = "Mon. Not. Roy. Astron. Soc.",
	volume = "449",
	number = "1",
	pages = "835--847",
	year = "2015"
}

@article{BOSS:2016wmc,
	author = "Alam, Shadab and others",
	collaboration = "BOSS",
	title = "{The clustering of galaxies in the completed SDSS-III Baryon Oscillation Spectroscopic Survey: cosmological analysis of the DR12 galaxy sample}",
	eprint = "1607.03155",
	archivePrefix = "arXiv",
	primaryClass = "astro-ph.CO",
	doi = "10.1093/mnras/stx721",
	journal = "Mon. Not. Roy. Astron. Soc.",
	volume = "470",
	number = "3",
	pages = "2617--2652",
	year = "2017"
}

@article{DiLuzio:2020wdo,
	author = "Di Luzio, Luca and Giannotti, Maurizio and Nardi, Enrico and Visinelli, Luca",
	title = "{The landscape of QCD axion models}",
	eprint = "2003.01100",
	archivePrefix = "arXiv",
	primaryClass = "hep-ph",
	reportNumber = "DESY 20-036, DESY-20-036",
	doi = "10.1016/j.physrep.2020.06.002",
	journal = "Phys. Rept.",
	volume = "870",
	pages = "1--117",
	year = "2020"
}

@article{Marsh:2015xka,
	author = "Marsh, David J. E.",
	title = "{Axion Cosmology}",
	eprint = "1510.07633",
	archivePrefix = "arXiv",
	primaryClass = "astro-ph.CO",
	reportNumber = "KCL-PH-TH-2015-50",
	doi = "10.1016/j.physrep.2016.06.005",
	journal = "Phys. Rept.",
	volume = "643",
	pages = "1--79",
	year = "2016"
}

@article{Baumann:2016wac,
	author = "Baumann, Daniel and Green, Daniel and Wallisch, Benjamin",
	title = "{New Target for Cosmic Axion Searches}",
	eprint = "1604.08614",
	archivePrefix = "arXiv",
	primaryClass = "astro-ph.CO",
	doi = "10.1103/PhysRevLett.117.171301",
	journal = "Phys. Rev. Lett.",
	volume = "117",
	number = "17",
	pages = "171301",
	year = "2016"
}

@article{Bashinsky:2003tk,
	author = "Bashinsky, Sergei and Seljak, Uros",
	title = "{Neutrino perturbations in CMB anisotropy and matter clustering}",
	eprint = "astro-ph/0310198",
	archivePrefix = "arXiv",
	doi = "10.1103/PhysRevD.69.083002",
	journal = "Phys. Rev. D",
	volume = "69",
	pages = "083002",
	year = "2004"
}

@article{Hou:2011ec,
	author = "Hou, Zhen and Keisler, Ryan and Knox, Lloyd and Millea, Marius and Reichardt, Christian",
	title = "{How Massless Neutrinos Affect the Cosmic Microwave Background Damping Tail}",
	eprint = "1104.2333",
	archivePrefix = "arXiv",
	primaryClass = "astro-ph.CO",
	doi = "10.1103/PhysRevD.87.083008",
	journal = "Phys. Rev. D",
	volume = "87",
	pages = "083008",
	year = "2013"
}

@article{Ferreira:2020bpb,
	author = "Ferreira, Ricardo Z. and Notari, Alessio and Rompineve, Fabrizio",
	title = "{Dine-Fischler-Srednicki-Zhitnitsky axion in the CMB}",
	eprint = "2012.06566",
	archivePrefix = "arXiv",
	primaryClass = "hep-ph",
	doi = "10.1103/PhysRevD.103.063524",
	journal = "Phys. Rev. D",
	volume = "103",
	number = "6",
	pages = "063524",
	year = "2021"
}

@article{Caloni:2022uya,
    author = "Caloni, Luca and Gerbino, Martina and Lattanzi, Massimiliano and Visinelli, Luca",
    title = "{Novel cosmological bounds on thermally-produced axion-like particles}",
    eprint = "2205.01637",
    archivePrefix = "arXiv",
    primaryClass = "astro-ph.CO",
    doi = "10.1088/1475-7516/2022/09/021",
    journal = "JCAP",
    volume = "09",
    pages = "021",
    year = "2022"
}

@article{CMB-HD:2022bsz,
    author = "Aiola, Simone and others",
    collaboration = "CMB-HD",
    title = "{Snowmass2021 CMB-HD White Paper}",
    eprint = "2203.05728",
    archivePrefix = "arXiv",
    primaryClass = "astro-ph.CO",
    reportNumber = "FERMILAB-PUB-22-344-PPD",
    month = "3",
    year = "2022"
}

@article{Blennow:2012de,
    author = "Blennow, Mattias and Fernandez-Martinez, Enrique and Mena, Olga and Redondo, Javier and Serra, Paolo",
    title = "{Asymmetric Dark Matter and Dark Radiation}",
    eprint = "1203.5803",
    archivePrefix = "arXiv",
    primaryClass = "hep-ph",
    reportNumber = "CERN-PH-TH-2012-070, MPP-2012-56",
    doi = "10.1088/1475-7516/2012/07/022",
    journal = "JCAP",
    volume = "07",
    pages = "022",
    year = "2012"
}

@article{Gondolo:1990dk,
    author = "Gondolo, Paolo and Gelmini, Graciela",
    title = "{Cosmic abundances of stable particles: Improved analysis}",
    reportNumber = "UCLA-90-TEP-68",
    doi = "10.1016/0550-3213(91)90438-4",
    journal = "Nucl. Phys. B",
    volume = "360",
    pages = "145--179",
    year = "1991"
}

@article{Cheng:2025cmb,
    author = "Cheng, Hanyu and Yin, Ziwen and Di Valentino, Eleonora and Marsh, David J. E. and Visinelli, Luca",
    title = "{Constraining exotic high-$z$ reionization histories with Gaussian processes and the Cosmic Microwave Background}",
    eprint = "2506.19096",
    archivePrefix = "arXiv",
    primaryClass = "astro-ph.CO",
    reportNumber = "CA21106; CA21136",
    month = "6",
    year = "2025"
}

@article{Yin:2025amn,
    author = "Yin, Ziwen and Cheng, Hanyu and Di Valentino, Eleonora and Gendler, Naomi and Marsh, David J. E. and Visinelli, Luca",
    title = "{Constraining the axiverse with reionization}",
    eprint = "2507.03535",
    archivePrefix = "arXiv",
    primaryClass = "hep-ph",
    reportNumber = "CA21106; CA21136",
    month = "7",
    year = "2025"
}

@article{Cielo:2023bqp,
    author = "Cielo, Mattia and Escudero, Miguel and Mangano, Gianpiero and Pisanti, Ofelia",
    title = "{Neff in the Standard Model at NLO is 3.043}",
    eprint = "2306.05460",
    archivePrefix = "arXiv",
    primaryClass = "hep-ph",
    reportNumber = "CERN-TH-2023-103",
    doi = "10.1103/PhysRevD.108.L121301",
    journal = "Phys. Rev. D",
    volume = "108",
    number = "12",
    pages = "L121301",
    year = "2023"
}

@article{Drewes:2024wbw,
    author = "Drewes, Marco and Georis, Yannis and Klasen, Michael and Wiggering, Luca Paolo and Wong, Yvonne Y. Y.",
    title = "{Towards a precision calculation of N $_{eff}$ in the Standard Model. Part III. Improved estimate of NLO contributions to the collision integral}",
    eprint = "2402.18481",
    archivePrefix = "arXiv",
    primaryClass = "hep-ph",
    reportNumber = "CPPC-2024-01, MS-TP-24-06",
    doi = "10.1088/1475-7516/2024/06/032",
    journal = "JCAP",
    volume = "06",
    pages = "032",
    year = "2024"
}

@article{Hall:2009bx,
    author = "Hall, Lawrence J. and Jedamzik, Karsten and March-Russell, John and West, Stephen M.",
    title = "{Freeze-In Production of FIMP Dark Matter}",
    eprint = "0911.1120",
    archivePrefix = "arXiv",
    primaryClass = "hep-ph",
    reportNumber = "OUTP-09-18-P, UCB-PTH-09-32",
    doi = "10.1007/JHEP03(2010)080",
    journal = "JHEP",
    volume = "03",
    pages = "080",
    year = "2010"
}

@article{SimonsObservatory:2018koc,
    author = "Ade, Peter and others",
    collaboration = "Simons Observatory",
    title = "{The Simons Observatory: Science goals and forecasts}",
    eprint = "1808.07445",
    archivePrefix = "arXiv",
    primaryClass = "astro-ph.CO",
    doi = "10.1088/1475-7516/2019/02/056",
    journal = "JCAP",
    volume = "02",
    pages = "056",
    year = "2019"
}

@article{DEramo:2022nvb,
    author = "D'Eramo, Francesco and Di Valentino, Eleonora and Giar\`e, William and Hajkarim, Fazlollah and Melchiorri, Alessandro and Mena, Olga and Renzi, Fabrizio and Yun, Seokhoon",
    title = "{Cosmological bound on the QCD axion mass, redux}",
    eprint = "2205.07849",
    archivePrefix = "arXiv",
    primaryClass = "astro-ph.CO",
    doi = "10.1088/1475-7516/2022/09/022",
    journal = "JCAP",
    volume = "09",
    pages = "022",
    year = "2022"
}

@article{Carenza:2022ngg,
    author = "Carenza, Pierluca and Lucente, Giuseppe and Gerbino, Martina and Giannotti, Maurizio and Lattanzi, Massimiliano",
    title = "{Strong cosmological constraints on the neutrino magnetic moment}",
    eprint = "2211.10432",
    archivePrefix = "arXiv",
    primaryClass = "hep-ph",
    doi = "10.1103/PhysRevD.110.023510",
    journal = "Phys. Rev. D",
    volume = "110",
    number = "2",
    pages = "023510",
    year = "2024"
}

@article{DEramo:2018vss,
    author = "D'Eramo, Francesco and Ferreira, Ricardo Z. and Notari, Alessio and Bernal, Jos\'e Luis",
    title = "{Hot Axions and the $H_0$ tension}",
    eprint = "1808.07430",
    archivePrefix = "arXiv",
    primaryClass = "hep-ph",
    doi = "10.1088/1475-7516/2018/11/014",
    journal = "JCAP",
    volume = "11",
    pages = "014",
    year = "2018"
}

@article{Jain:2024dtw,
    author = "Jain, Mudit and Maggi, Angelo and Ai, Wen-Yuan and Marsh, David J. E.",
    title = "{New insights into axion freeze-in}",
    eprint = "2406.01678",
    archivePrefix = "arXiv",
    primaryClass = "hep-ph",
    reportNumber = "KCL-PH-TH/2024-31",
    doi = "10.1007/JHEP11(2024)166",
    journal = "JHEP",
    volume = "11",
    pages = "166",
    year = "2024"
}

@article{Ferreira:2018vjj,
    author = "Ferreira, Ricardo Z. and Notari, Alessio",
    title = "{Observable Windows for the QCD Axion Through the Number of Relativistic Species}",
    eprint = "1801.06090",
    archivePrefix = "arXiv",
    primaryClass = "hep-ph",
    doi = "10.1103/PhysRevLett.120.191301",
    journal = "Phys. Rev. Lett.",
    volume = "120",
    number = "19",
    pages = "191301",
    year = "2018"
}

@article{DEramo:2024jhn,
    author = "D'Eramo, Francesco and Lenoci, Alessandro",
    title = "{Back to the phase space: Thermal axion dark radiation via couplings to standard model fermions}",
    eprint = "2410.21253",
    archivePrefix = "arXiv",
    primaryClass = "hep-ph",
    doi = "10.1103/PhysRevD.110.116028",
    journal = "Phys. Rev. D",
    volume = "110",
    number = "11",
    pages = "116028",
    year = "2024"
}

@article{Cima:2025zmc,
    author = "Cima, Federico and D'Eramo, Francesco",
    title = "{Probing Non-Minimal Dark Sectors via the 21 cm Line at Cosmic Dawn}",
    eprint = "2507.10664",
    archivePrefix = "arXiv",
    primaryClass = "hep-ph",
    month = "7",
    year = "2025"
}

@article{Caloni:2024olo,
    author = "Caloni, Luca and Stengel, Patrick and Lattanzi, Massimiliano and Gerbino, Martina",
    title = "{Constraining UV freeze-in of light relics with current and next-generation CMB observations}",
    eprint = "2405.09449",
    archivePrefix = "arXiv",
    primaryClass = "astro-ph.CO",
    reportNumber = "CA21106",
    doi = "10.1088/1475-7516/2024/10/106",
    journal = "JCAP",
    volume = "10",
    pages = "106",
    year = "2024"
}

@article{Langhoff:2022bij,
    author = "Langhoff, Kevin and Outmezguine, Nadav Joseph and Rodd, Nicholas L.",
    title = "{Irreducible Axion Background}",
    eprint = "2209.06216",
    archivePrefix = "arXiv",
    primaryClass = "hep-ph",
    reportNumber = "CERN-TH-2022-148",
    doi = "10.1103/PhysRevLett.129.241101",
    journal = "Phys. Rev. Lett.",
    volume = "129",
    number = "24",
    pages = "241101",
    year = "2022"
}

@article{Torrado:2020dgo,
    author = "Torrado, Jesus and Lewis, Antony",
    title = "{Cobaya: Code for Bayesian Analysis of hierarchical physical models}",
    eprint = "2005.05290",
    archivePrefix = "arXiv",
    primaryClass = "astro-ph.IM",
    reportNumber = "TTK-20-15",
    doi = "10.1088/1475-7516/2021/05/057",
    journal = "JCAP",
    volume = "05",
    pages = "057",
    year = "2021"
}

@article{Blas:2011rf,
    author = "Blas, Diego and Lesgourgues, Julien and Tram, Thomas",
    title = "{The Cosmic Linear Anisotropy Solving System (CLASS) II: Approximation schemes}",
    eprint = "1104.2933",
    archivePrefix = "arXiv",
    primaryClass = "astro-ph.CO",
    reportNumber = "CERN-PH-TH-2011-082, LAPTH-010-11",
    doi = "10.1088/1475-7516/2011/07/034",
    journal = "JCAP",
    volume = "07",
    pages = "034",
    year = "2011"
}

@article{Gelman:1992zz,
    author = "Gelman, Andrew and Rubin, Donald B.",
    title = "{Inference from Iterative Simulation Using Multiple Sequences}",
    doi = "10.1214/ss/1177011136",
    journal = "Statist. Sci.",
    volume = "7",
    pages = "457--472",
    year = "1992"
}

@article{Planck:2019nip,
    author = "Aghanim, N. and others",
    collaboration = "Planck",
    title = "{Planck 2018 results. V. CMB power spectra and likelihoods}",
    eprint = "1907.12875",
    archivePrefix = "arXiv",
    primaryClass = "astro-ph.CO",
    doi = "10.1051/0004-6361/201936386",
    journal = "Astron. Astrophys.",
    volume = "641",
    pages = "A5",
    year = "2020"
}

@article{Jaeckel:2010ni,
    author = "Jaeckel, Joerg and Ringwald, Andreas",
    title = "{The Low-Energy Frontier of Particle Physics}",
    eprint = "1002.0329",
    archivePrefix = "arXiv",
    primaryClass = "hep-ph",
    reportNumber = "CPT-10-18, DESY-10-016, IPPP-10-09",
    doi = "10.1146/annurev.nucl.012809.104433",
    journal = "Ann. Rev. Nucl. Part. Sci.",
    volume = "60",
    pages = "405--437",
    year = "2010"
}

@article{Sehgal:2019ewc,
    author = "Sehgal, Neelima and others",
    title = "{CMB-HD: An Ultra-Deep, High-Resolution Millimeter-Wave Survey Over Half the Sky}",
    eprint = "1906.10134",
    archivePrefix = "arXiv",
    primaryClass = "astro-ph.CO",
    journal = "Bull. Am. Astron. Soc.",
    volume = "51",
    number = "7",
    pages = "1--23",
    year = "2019"
}

@article{AtacamaCosmologyTelescope:2025blo,
    author = "Louis, Thibaut and others",
    collaboration = "Atacama Cosmology Telescope",
    title = "{The Atacama Cosmology Telescope: DR6 power spectra, likelihoods and {\ensuremath{\Lambda}}CDM parameters}",
    eprint = "2503.14452",
    archivePrefix = "arXiv",
    primaryClass = "astro-ph.CO",
    reportNumber = "FERMILAB-PUB-25-0071-PPD",
    doi = "10.1088/1475-7516/2025/11/062",
    journal = "JCAP",
    volume = "11",
    pages = "062",
    year = "2025"
}

@article{SPT-3G:2025bzu,
    author = "Camphuis, E. and others",
    collaboration = "SPT-3G",
    title = "{SPT-3G D1: CMB temperature and polarization power spectra and cosmology from 2019 and 2020 observations of the SPT-3G Main field}",
    eprint = "2506.20707",
    archivePrefix = "arXiv",
    primaryClass = "astro-ph.CO",
    reportNumber = "FERMILAB-PUB-25-0144-PPD",
    month = "6",
    year = "2025"
}

@article{SPT-3G:2024atg,
    author = "Ge, F. and others",
    collaboration = "SPT-3G",
    title = "{Cosmology from CMB lensing and delensed EE power spectra using 2019{\textendash}2020 SPT-3G polarization data}",
    eprint = "2411.06000",
    archivePrefix = "arXiv",
    primaryClass = "astro-ph.CO",
    reportNumber = "FERMILAB-PUB-24-0840-PPD",
    doi = "10.1103/PhysRevD.111.083534",
    journal = "Phys. Rev. D",
    volume = "111",
    number = "8",
    pages = "083534",
    year = "2025"
}

@article{ACT:2023kun,
    author = "Madhavacheril, Mathew S. and others",
    collaboration = "ACT",
    title = "{The Atacama Cosmology Telescope: DR6 Gravitational Lensing Map and Cosmological Parameters}",
    eprint = "2304.05203",
    archivePrefix = "arXiv",
    primaryClass = "astro-ph.CO",
    reportNumber = "FERMILAB-PUB-23-206-PPD",
    doi = "10.3847/1538-4357/acff5f",
    journal = "Astrophys. J.",
    volume = "962",
    number = "2",
    pages = "113",
    year = "2024"
}

@article{ACT:2023dou,
    author = "Qu, Frank J. and others",
    collaboration = "ACT",
    title = "{The Atacama Cosmology Telescope: A Measurement of the DR6 CMB Lensing Power Spectrum and Its Implications for Structure Growth}",
    eprint = "2304.05202",
    archivePrefix = "arXiv",
    primaryClass = "astro-ph.CO",
    reportNumber = "FERMILAB-PUB-23-237-PPD, FERMILAB-PUB-23-237-PPD",
    doi = "10.3847/1538-4357/acfe06",
    journal = "Astrophys. J.",
    volume = "962",
    number = "2",
    pages = "112",
    year = "2024"
}

@article{ACT:2023ubw,
    author = "MacCrann, Niall and others",
    collaboration = "ACT",
    title = "{The Atacama Cosmology Telescope: Mitigating the Impact of Extragalactic Foregrounds for the DR6 Cosmic Microwave Background Lensing Analysis}",
    eprint = "2304.05196",
    archivePrefix = "arXiv",
    primaryClass = "astro-ph.CO",
    reportNumber = "FERMILAB-PUB-23-236-PPD",
    doi = "10.3847/1538-4357/ad2610",
    journal = "Astrophys. J.",
    volume = "966",
    number = "1",
    pages = "138",
    year = "2024"
}

@article{AtacamaCosmologyTelescope:2025nti,
    author = "Calabrese, Erminia and others",
    collaboration = "Atacama Cosmology Telescope",
    title = "{The Atacama Cosmology Telescope: DR6 constraints on extended cosmological models}",
    eprint = "2503.14454",
    archivePrefix = "arXiv",
    primaryClass = "astro-ph.CO",
    reportNumber = "FERMILAB-PUB-25-0157-PPD",
    doi = "10.1088/1475-7516/2025/11/063",
    journal = "JCAP",
    volume = "11",
    pages = "063",
    year = "2025"
}

@article{Carron:2022eyg,
    author = "Carron, Julien and Mirmelstein, Mark and Lewis, Antony",
    title = "{CMB lensing from Planck PR4~maps}",
    eprint = "2206.07773",
    archivePrefix = "arXiv",
    primaryClass = "astro-ph.CO",
    doi = "10.1088/1475-7516/2022/09/039",
    journal = "JCAP",
    volume = "09",
    pages = "039",
    year = "2022"
}

@article{Balkenhol:2024sbv,
    author = "Balkenhol, L. and Trendafilova, C. and Benabed, K. and Galli, S.",
    title = "{candl: cosmic microwave background analysis with a differentiable likelihood}",
    eprint = "2401.13433",
    archivePrefix = "arXiv",
    primaryClass = "astro-ph.CO",
    doi = "10.1051/0004-6361/202449432",
    journal = "Astron. Astrophys.",
    volume = "686",
    pages = "A10",
    year = "2024"
}

@article{Perotto:2006rj,
    author = "Perotto, Laurence and Lesgourgues, Julien and Hannestad, Steen and Tu, Huitzu and Wong, Yvonne Y. Y.",
    title = "{Probing cosmological parameters with the CMB: Forecasts from full Monte Carlo simulations}",
    eprint = "astro-ph/0606227",
    archivePrefix = "arXiv",
    reportNumber = "MPP-2006-76, LAPTH-1148-06",
    doi = "10.1088/1475-7516/2006/10/013",
    journal = "JCAP",
    volume = "10",
    pages = "013",
    year = "2006"
}

@article{Wu:2014hta,
    author = "Wu, W. L. K. and Errard, J. and Dvorkin, C. and Kuo, C. L. and Lee, A. T. and McDonald, P. and Slosar, A. and Zahn, O.",
    title = "{A Guide to Designing Future Ground-based Cosmic Microwave Background Experiments}",
    eprint = "1402.4108",
    archivePrefix = "arXiv",
    primaryClass = "astro-ph.CO",
    doi = "10.1088/0004-637X/788/2/138",
    journal = "Astrophys. J.",
    volume = "788",
    pages = "138",
    year = "2014"
}

@article{Gerbino:2019okg,
    author = "Gerbino, Martina and Lattanzi, Massimiliano and Migliaccio, Marina and Pagano, Luca and Salvati, Laura and Colombo, Loris and Gruppuso, Alessandro and Natoli, Paolo and Polenta, Gianluca",
    title = "{Likelihood methods for CMB experiments}",
    eprint = "1909.09375",
    archivePrefix = "arXiv",
    primaryClass = "astro-ph.CO",
    doi = "10.3389/fphy.2020.00015",
    journal = "Front. in Phys.",
    volume = "8",
    pages = "15",
    year = "2020"
}

@article{MacInnis:2024znd,
    author = "MacInnis, Amanda and Sehgal, Neelima",
    title = "{CMB-HD as a probe of dark matter on sub-galactic scales}",
    eprint = "2405.12220",
    archivePrefix = "arXiv",
    primaryClass = "astro-ph.CO",
    doi = "10.1088/1475-7516/2025/02/048",
    journal = "JCAP",
    volume = "02",
    pages = "048",
    year = "2025"
}

@article{MacInnis:2023vif,
    author = "MacInnis, Amanda and Sehgal, Neelima and Rothermel, Miriam",
    title = "{Cosmological parameter forecasts for a CMB-HD survey}",
    eprint = "2309.03021",
    archivePrefix = "arXiv",
    primaryClass = "astro-ph.CO",
    doi = "10.1103/PhysRevD.109.063527",
    journal = "Phys. Rev. D",
    volume = "109",
    number = "6",
    pages = "063527",
    year = "2024"
}

@article{Lepage:1977sw,
    author = "Lepage, G. Peter",
    title = "{A New Algorithm for Adaptive Multidimensional Integration}",
    reportNumber = "SLAC-PUB-1839-REV, SLAC-PUB-1839",
    doi = "10.1016/0021-9991(78)90004-9",
    journal = "J. Comput. Phys.",
    volume = "27",
    pages = "192",
    year = "1978"
}

@article{Hahn:2004fe,
    author = "Hahn, T.",
    title = "{CUBA: A Library for multidimensional numerical integration}",
    eprint = "hep-ph/0404043",
    archivePrefix = "arXiv",
    reportNumber = "MPP-2004-40",
    doi = "10.1016/j.cpc.2005.01.010",
    journal = "Comput. Phys. Commun.",
    volume = "168",
    pages = "78--95",
    year = "2005"
}

@article{Pagano:2019tci,
    author = "Pagano, L. and Delouis, J. -M. and Mottet, S. and Puget, J. -L. and Vibert, L.",
    title = "{Reionization optical depth determination from Planck HFI data with ten percent accuracy}",
    eprint = "1908.09856",
    archivePrefix = "arXiv",
    primaryClass = "astro-ph.CO",
    doi = "10.1051/0004-6361/201936630",
    journal = "Astron. Astrophys.",
    volume = "635",
    pages = "A99",
    year = "2020"
}

@article{Delouis:2019bub,
    author = "Delouis, J. -M. and Pagano, L. and Mottet, S. and Puget, J. -L. and Vibert, L.",
    title = "{SRoll2: an improved mapmaking approach to reduce large-scale systematic effects in the Planck High Frequency Instrument legacy maps}",
    eprint = "1901.11386",
    archivePrefix = "arXiv",
    primaryClass = "astro-ph.CO",
    doi = "10.1051/0004-6361/201834882",
    journal = "Astron. Astrophys.",
    volume = "629",
    pages = "A38",
    year = "2019"
}

@article{DESI:2025zpo,
    author = "Abdul Karim, M. and others",
    collaboration = "DESI",
    title = "{DESI DR2 results. I. Baryon acoustic oscillations from the Lyman alpha forest}",
    eprint = "2503.14739",
    archivePrefix = "arXiv",
    primaryClass = "astro-ph.CO",
    reportNumber = "FERMILAB-PUB-25-0167-PPD",
    doi = "10.1103/2wwn-xjm5",
    journal = "Phys. Rev. D",
    volume = "112",
    number = "8",
    pages = "083514",
    year = "2025"
}

@article{DESI:2025zgx,
    author = "Abdul Karim, M. and others",
    collaboration = "DESI",
    title = "{DESI DR2 results. II. Measurements of baryon acoustic oscillations and cosmological constraints}",
    eprint = "2503.14738",
    archivePrefix = "arXiv",
    primaryClass = "astro-ph.CO",
    reportNumber = "FERMILAB-PUB-25-0169-PPD",
    doi = "10.1103/tr6y-kpc6",
    journal = "Phys. Rev. D",
    volume = "112",
    number = "8",
    pages = "083515",
    year = "2025"
}

@article{Carrington:1997sq,
    author = "Carrington, Magaret E. and Hou, De-fu and Thoma, Markus H.",
    title = "{Equilibrium and nonequilibrium hard thermal loop resummation in the real time formalism}",
    eprint = "hep-ph/9708363",
    archivePrefix = "arXiv",
    reportNumber = "UGI-97-12",
    doi = "10.1007/s100520050412",
    journal = "Eur. Phys. J. C",
    volume = "7",
    pages = "347--354",
    year = "1999"
}

@article{Eberhart:2025lyu,
    author = "Eberhart, Alexander and Fedele, Marco and Kahlhoefer, Felix and Ravensburg, Eike and Ziegler, Robert",
    title = "{Leptophilic ALPs in laboratory experiments}",
    eprint = "2504.05873",
    archivePrefix = "arXiv",
    primaryClass = "hep-ph",
    reportNumber = "TTP25-011, P3H-25-026, MITP-25-027",
    doi = "10.1007/JHEP12(2025)055",
    journal = "JHEP",
    volume = "12",
    pages = "055",
    year = "2025"
}

@article{Li:2025yzb,
    author = "Li, Haotian and Liu, Zuowei and Song, Ningqiang",
    title = "{Probing axion and flavored new physics with the NA64{\ensuremath{\mu}} experiment}",
    eprint = "2501.06294",
    archivePrefix = "arXiv",
    primaryClass = "hep-ph",
    doi = "10.1103/27kn-ly9l",
    journal = "Phys. Rev. D",
    volume = "112",
    number = "11",
    pages = "115013",
    year = "2025"
}

@article{NA64:2019auh,
    author = "Banerjee, D. and others",
    collaboration = "NA64",
    title = "{Improved limits on a hypothetical X(16.7) boson and a dark photon decaying into $e^+e^-$ pairs}",
    eprint = "1912.11389",
    archivePrefix = "arXiv",
    primaryClass = "hep-ex",
    reportNumber = "CERN-EP-2019-284",
    doi = "10.1103/PhysRevD.101.071101",
    journal = "Phys. Rev. D",
    volume = "101",
    number = "7",
    pages = "071101",
    year = "2020"
}

@article{Ferreira:2025qui,
    author = "Ferreira, Ricardo Z. and Marsh, M. C. David and Ravensburg, Eike",
    title = "{ALP couplings to muons and electrons: a comprehensive analysis of supernova bounds}",
    eprint = "2510.14469",
    archivePrefix = "arXiv",
    primaryClass = "hep-ph",
    month = "10",
    year = "2025"
}

@article{Caputo:2021rux,
    author = "Caputo, Andrea and Raffelt, Georg and Vitagliano, Edoardo",
    title = "{Muonic boson limits: Supernova redux}",
    eprint = "2109.03244",
    archivePrefix = "arXiv",
    primaryClass = "hep-ph",
    reportNumber = "MPP-2021-154",
    doi = "10.1103/PhysRevD.105.035022",
    journal = "Phys. Rev. D",
    volume = "105",
    number = "3",
    pages = "035022",
    year = "2022"
}

@article{Bollig:2020xdr,
    author = "Bollig, Robert and DeRocco, William and Graham, Peter W. and Janka, Hans-Thomas",
    title = "{Muons in Supernovae: Implications for the Axion-Muon Coupling}",
    eprint = "2005.07141",
    archivePrefix = "arXiv",
    primaryClass = "hep-ph",
    doi = "10.1103/PhysRevLett.125.051104",
    journal = "Phys. Rev. Lett.",
    volume = "125",
    number = "5",
    pages = "051104",
    year = "2020",
    note = "[Erratum: Phys.Rev.Lett. 126, 189901 (2021)]"
}

@article{Croon:2020lrf,
    author = "Croon, Djuna and Elor, Gilly and Leane, Rebecca K. and McDermott, Samuel D.",
    title = "{Supernova Muons: New Constraints on $Z$' Bosons, Axions and ALPs}",
    eprint = "2006.13942",
    archivePrefix = "arXiv",
    primaryClass = "hep-ph",
    reportNumber = "MIT-CTP/5214, FERMILAB-PUB-20-246-A-T",
    doi = "10.1007/JHEP01(2021)107",
    journal = "JHEP",
    volume = "01",
    pages = "107",
    year = "2021"
}

@misc{AxionLimits,
  author       = {Ciaran O'Hare},
  title        = {cajohare/AxionLimits: AxionLimits},
  month        = jul,
  year         = 2020,
  publisher    = {Zenodo},
  version      = {v1.0},
  doi          = {10.5281/zenodo.3932430},
  howpublished = {\url{https://cajohare.github.io/AxionLimits/}}
}

@article{Marciano:1977wx,
    author = "Marciano, W. J. and Sanda, A. I.",
    title = "{Exotic Decays of the Muon and Heavy Leptons in Gauge Theories}",
    reportNumber = "COO-2232B-116",
    doi = "10.1016/0370-2693(77)90377-X",
    journal = "Phys. Lett. B",
    volume = "67",
    pages = "303--305",
    year = "1977"
}

@article{Adler:1969gk,
    author = "Adler, Stephen L.",
    title = "{Axial vector vertex in spinor electrodynamics}",
    doi = "10.1103/PhysRev.177.2426",
    journal = "Phys. Rev.",
    volume = "177",
    pages = "2426--2438",
    year = "1969"
}

@article{Georgi:1986df,
    author = "Georgi, Howard and Kaplan, David B. and Randall, Lisa",
    title = "{Manifesting the Invisible Axion at Low-energies}",
    reportNumber = "HUTP-86/A004",
    doi = "10.1016/0370-2693(86)90688-X",
    journal = "Phys. Lett. B",
    volume = "169",
    pages = "73--78",
    year = "1986"
}

@article{Srednicki:1985xd,
    author = "Srednicki, Mark",
    title = "{Axion Couplings to Matter. 1. CP Conserving Parts}",
    reportNumber = "Print-85-0247 (UC,SANTA BARBARA)",
    doi = "10.1016/0550-3213(85)90054-9",
    journal = "Nucl. Phys. B",
    volume = "260",
    pages = "689--700",
    year = "1985"
}

@article{Feng:1997tn,
    author = "Feng, Jonathan L. and Moroi, Takeo and Murayama, Hitoshi and Schnapka, Erhard",
    title = "{Third generation familons, b factories, and neutrino cosmology}",
    eprint = "hep-ph/9709411",
    archivePrefix = "arXiv",
    reportNumber = "LBL-40822-REV, LBNL-40822-REV, UCB-PTH-97-47-REV, LBNL-40822, UCB-PTH-97-47",
    doi = "10.1103/PhysRevD.57.5875",
    journal = "Phys. Rev. D",
    volume = "57",
    pages = "5875--5892",
    year = "1998"
}

@article{Rashkovetskyi:2021rwg,
    author = "Rashkovetskyi, Michael and Mu{\~n}oz, Julian B. and Eisenstein, Daniel J. and Dvorkin, Cora",
    title = "{Small-scale clumping at recombination and the Hubble tension}",
    eprint = "2108.02747",
    archivePrefix = "arXiv",
    primaryClass = "astro-ph.CO",
    doi = "10.1103/PhysRevD.104.103517",
    journal = "Phys. Rev. D",
    volume = "104",
    number = "10",
    pages = "103517",
    year = "2021"
}

@article{Arias:2012az,
    author = "Arias, Paola and Cadamuro, Davide and Goodsell, Mark and Jaeckel, Joerg and Redondo, Javier and Ringwald, Andreas",
    title = "{WISPy Cold Dark Matter}",
    eprint = "1201.5902",
    archivePrefix = "arXiv",
    primaryClass = "hep-ph",
    reportNumber = "DESY-11-226, MPP-2011-140, CERN-PH-TH-2011-323, IPPP-11-80, DCPT-11-160",
    doi = "10.1088/1475-7516/2012/06/013",
    journal = "JCAP",
    volume = "06",
    pages = "013",
    year = "2012"
}

@article{deSalas:2020pgw,
    author = "de Salas, P. F. and Forero, D. V. and Gariazzo, S. and Mart{\'\i}nez-Mirav{\'e}, P. and Mena, O. and Ternes, C. A. and T{\'o}rtola, M. and Valle, J. W. F.",
    title = "{2020 global reassessment of the neutrino oscillation picture}",
    eprint = "2006.11237",
    archivePrefix = "arXiv",
    primaryClass = "hep-ph",
    doi = "10.1007/JHEP02(2021)071",
    journal = "JHEP",
    volume = "02",
    pages = "071",
    year = "2021"
}

@article{Capozzi:2021fjo,
    author = "Capozzi, Francesco and Di Valentino, Eleonora and Lisi, Eligio and Marrone, Antonio and Melchiorri, Alessandro and Palazzo, Antonio",
    title = "{Unfinished fabric of the three neutrino paradigm}",
    eprint = "2107.00532",
    archivePrefix = "arXiv",
    primaryClass = "hep-ph",
    doi = "10.1103/PhysRevD.104.083031",
    journal = "Phys. Rev. D",
    volume = "104",
    number = "8",
    pages = "083031",
    year = "2021"
}

@article{Esteban:2020cvm,
    author = "Esteban, Ivan and Gonzalez-Garcia, M. C. and Maltoni, Michele and Schwetz, Thomas and Zhou, Albert",
    title = "{The fate of hints: updated global analysis of three-flavor neutrino oscillations}",
    eprint = "2007.14792",
    archivePrefix = "arXiv",
    primaryClass = "hep-ph",
    reportNumber = "IFT-UAM/CSIC-112, YITP-SB-2020-21",
    doi = "10.1007/JHEP09(2020)178",
    journal = "JHEP",
    volume = "09",
    pages = "178",
    year = "2020"
}

@article{Badziak:2025mkt,
    author = "Badziak, Marcin and Gomu{\l}ka, Adam and Laletin, Maxim and Szafra{\'n}ski, Krzysztof",
    title = "{Improved cosmological constraints on axion-lepton interactions}",
    eprint = "2511.14864",
    archivePrefix = "arXiv",
    primaryClass = "hep-ph",
    month = "11",
    year = "2025"
}

@article{Barbieri:2025moq,
    author = "Barbieri, Nicola and Brinckmann, Thejs and Gariazzo, Stefano and Lattanzi, Massimiliano and Pastor, Sergio and Pisanti, Ofelia",
    title = "{Current Constraints on Cosmological Scenarios with Very Low Reheating Temperatures}",
    eprint = "2501.01369",
    archivePrefix = "arXiv",
    primaryClass = "astro-ph.CO",
    doi = "10.1103/j5rj-dz1k",
    journal = "Phys. Rev. Lett.",
    volume = "135",
    number = "18",
    pages = "181003",
    year = "2025"
}

@article{Green:2021hjh,
    author = "Green, Daniel and Guo, Yi and Wallisch, Benjamin",
    title = "{Cosmological implications of axion-matter couplings}",
    eprint = "2109.12088",
    archivePrefix = "arXiv",
    primaryClass = "astro-ph.CO",
    doi = "10.1088/1475-7516/2022/02/019",
    journal = "JCAP",
    volume = "02",
    number = "02",
    pages = "019",
    year = "2022"
}

@article{Micheli:2024hfe,
    author = "Micheli, Silvia and others",
    title = "{Systematic effects induced by half-wave plate differential optical load and TES nonlinearity for LiteBIRD}",
    eprint = "2407.15294",
    archivePrefix = "arXiv",
    primaryClass = "astro-ph.IM",
    doi = "10.1117/12.3018553",
    journal = "Proc. SPIE Int. Soc. Opt. Eng.",
    volume = "13102",
    pages = "131022R",
    year = "2024"
}

@Manual{mcmcse,
    title = {mcmcse: Monte Carlo Standard Errors for MCMC},
    author = {James M. Flegal and John Hughes and Dootika Vats and Ning
      Dai and Kushagra Gupta and Uttiya Maji},
    year = {2025},
    address = {Riverside, CA, and Kanpur, India},
    note = {R package version 1.5-1},
  }

@article{Elbers:2025vlz,
    author = "Elbers, W. and others",
    title = "{Constraints on neutrino physics from DESI DR2 BAO and DR1 full shape}",
    eprint = "2503.14744",
    archivePrefix = "arXiv",
    primaryClass = "astro-ph.CO",
    reportNumber = "FERMILAB-PUB-25-0168-PPD",
    doi = "10.1103/w9pk-xsk7",
    journal = "Phys. Rev. D",
    volume = "112",
    number = "8",
    pages = "083513",
    year = "2025"
}

@article{Badziak:2024qjg,
    author = "Badziak, Marcin and Laletin, Maxim",
    title = "{Precise predictions for the QCD axion contribution to dark radiation with full phase-space evolution}",
    eprint = "2410.18186",
    archivePrefix = "arXiv",
    primaryClass = "hep-ph",
    doi = "10.1007/JHEP02(2025)108",
    journal = "JHEP",
    volume = "02",
    pages = "108",
    year = "2025"
}

@article{Planck:2018jri,
    author = "Akrami, Y. and others",
    collaboration = "Planck",
    title = "{Planck 2018 results. X. Constraints on inflation}",
    eprint = "1807.06211",
    archivePrefix = "arXiv",
    primaryClass = "astro-ph.CO",
    doi = "10.1051/0004-6361/201833887",
    journal = "Astron. Astrophys.",
    volume = "641",
    pages = "A10",
    year = "2020"
}

@article{Kullback:1951zyt,
    author = "Kullback, S. and Leibler, R. A.",
    title = "{On Information and Sufficiency}",
    doi = "10.1214/aoms/1177729694",
    journal = "The Annals of Mathematical Statistics",
    volume = "22",
    number = "1",
    pages = "79--86",
    year = "1951"
}

@article{Bayes:1764vd,
    author = "Bayes, Rev., Thomas",
    title = "{An essay toward solving a problem in the doctrine of chances}",
    doi = "10.1098/rstl.1763.0053",
    journal = "Phil. Trans. Roy. Soc. Lond.",
    volume = "53",
    pages = "370--418",
    year = "1764"
}

@article{ParticleDataGroup:2024cfk,
    author = "Navas, S. and others",
    collaboration = "Particle Data Group",
    title = "{Review of particle physics}",
    doi = "10.1103/PhysRevD.110.030001",
    journal = "Phys. Rev. D",
    volume = "110",
    number = "3",
    pages = "030001",
    year = "2024"
}

\end{document}